\documentclass[11pt]{article}

\usepackage{latexsym,amsmath,amssymb,theorem,epsfig,multirow}
\usepackage[noconfig]{refstyle}
\usepackage[dvipsnames]{xcolor}
\usepackage[hyperfootnotes=false, linktocpage=true, colorlinks, citecolor=blue, linkcolor=blue, urlcolor=Maroon]{hyperref}

\newlength{\xtrawidth}
\newlength{\xtraheight}
\DeclareFontFamily{OT1}{rsfs10}{}
\DeclareFontShape{OT1}{rsfs10}{m}{n}{ <-> rsfs10 }{}
\DeclareMathAlphabet{\mathscript}{OT1}{rsfs10}{m}{n}

\numberwithin{equation}{section}

\newcommand{\pt}{\partial}

\def\fnote#1#2{\begingroup\def\thefootnote{#1}\footnote{#2}
     \addtocounter{footnote}{-1}\endgroup}

\def\a{\alpha}
\def\b{\beta}
\def\g{\gamma}

\def\d{\delta}
\def\e{\epsilon}

\def\k{\kappa}
\def\l{\lambda}
\def\m{\mu}

\def\o{\omega}

\def\r{\rho}
\def\s{\sigma}

\def\x{\xi}

\def\D{\Delta}

\def\L{\Lambda}
\def\O{\Omega}

\def\gsim{ \lower .75ex \hbox{$\sim$} \llap{\raise .27ex \hbox{$>$}} }
\def\lsim{ \lower .75ex \hbox{$\sim$} \llap{\raise .27ex \hbox{$<$}} }
\def\be{\begin{equation}}
\def\ee{\end{equation}}
\def\bea{\begin{eqnarray}}
\def\eea{\end{eqnarray}}

\def \a {\alpha}

\def \a {\alpha }
\def \eps {\epsilon}

\def \l {\lambda}

\def \ed {\end{document}}

\begin{document}

\begin{titlepage}

\title{{\LARGE\bf   Holomorphic Yukawa Couplings in Heterotic String Theory}\\[1em] }
\author{
Stefan Blesneag${}^{1}$,
Evgeny I. Buchbinder${}^{2}$,
Philip Candelas${}^{3}$,
Andre Lukas${}^{1}$
}

\date{}
\maketitle
\begin{center} { 
${}^1${\it Rudolf Peierls Centre for Theoretical Physics, Oxford University,\\
1 Keble Road, Oxford, OX1 3NP, U.K.\\[3mm]
${}^2$ The University of Western Australia, \\
35 Stirling Highway, Crawley WA 6009, Australia\\[3mm]
${}^3$Mathematical   Institute,   University   of   Oxford,\\   
Andrew   Wiles   Building,   Radcliffe   Observatory   Quarter,\\
Woodstock Road, Oxford, OX2 6GG, U.K.}}\\
\end{center}

\fnote{}{stefan.blesneag@wadh.ox.ac.uk}
\fnote{}{evgeny.buchbinder@uwa.edu.au}
\fnote{}{candelas@maths.ox.ac.uk}
\fnote{}{lukas@physics.ox.ac.uk} 

\vskip 1cm

\begin{abstract}\noindent
We develop techniques, based on differential geometry, to compute holomorphic Yukawa couplings for heterotic line bundle models on Calabi-Yau manifolds defined as complete intersections in projective spaces. It is shown explicitly how these techniques relate to algebraic methods for computing holomorphic Yukawa couplings. We apply our methods to various examples and evaluate the holomorphic Yukawa couplings explicitly as functions of the complex structure moduli. It is shown that the rank of the Yukawa matrix can decrease at specific loci in complex structure moduli space. In particular, we compute the up Yukawa coupling and the singlet-Higgs-lepton trilinear coupling in the heterotic  standard model described in Ref.~\cite{Buchbinder:2014qda}.
\end{abstract}

\thispagestyle{empty}

\end{titlepage}

\tableofcontents

\section{Introduction}
String model building based on heterotic Calabi-Yau compactifications~\cite{Candelas:1985en,Strominger:1985it,Witten:1985xc} has seen considerable progress 
over the past ten years~\cite{Braun:2005ux}--\cite{Braun:2011ni} and large classes of models with the MSSM spectrum can now be constructed 
using algorithmic approaches~\cite{Anderson:2011ns,Anderson:2012yf,Anderson:2013xka}. Given that compactifications with the 
correct spectrum can now be readily engineered, 
one of the most pressing problems is the calculation of Yukawa couplings for such models. Remarkably little is known about this problem, both in terms of general 
techniques and actual specific results. In this paper, we will attempt to make some progress in this direction and develop new methods, mainly based on 
differential geometry, to calculate holomorphic Yukawa couplings for heterotic line bundle models. 

Calculating the physical Yukawa couplings of a supersymmetric string compactification comes in two parts: the calculation 
of the holomorphic Yukawa couplings, that is, the couplings in the superpotential, and the calculation of the matter field Kahler metric, in order to 
work out the field normalisation. The holomorphic Yukawa couplings are quasi-topological - they do not depend on the Calabi-Yau metric or the hermitian Yang-Mill 
connection on the bundle - and they can, therefore, in principle, be calculated with algebraic methods. 
The situation is very different for the Kahler metric which does depend on the metric and the bundle connection. 
It is unlikely that an algebraic method for its calculation can be found and, hence,  methods of differential geometry will be required. 

At present, a full calculation of the physical (perturbative) Yukawa couplings is only understood for heterotic Calabi-Yau models with standard embedding. In this case, the holomorphic Yukawa couplings for the $(1,1)$ matter fields are given by the Calabi-Yau triple intersection numbers~\cite{Strominger:1985ks} while the holomorphic $(2,1)$ Yukawa couplings have been worked out in Ref.~\cite{Candelas:1987se}. The matter field Kahler metrics are known and basically given by the corresponding moduli space metrics as given in Ref.~\cite{Candelas:1990pi}. Further, in Ref.~\cite{Candelas:1987se}, the relation between the analytic calculation of $(2,1)$ holomorphic Yukawa couplings and the algebraic approach has been worked out in detail and we will review this discussion in the appendix of the present paper. Yukawa couplings for the early three-generation model of Refs.~\cite{Greene:1986bm,Greene:1986jb} have been presented in Ref.~\cite{Greene:1987xh}.

Much less is known for heterotic Calabi-Yau models with general vector bundles. Holomorphic Yukawa couplings for specific quasi-realistic models have been computed in Refs.~\cite{Braun:2006me,Bouchard:2006dn}. An algebraic approach for the calculation of holomorphic Yukawa couplings for such ``non-standard embedding" models has been outlined and applied to examples in Ref.~\cite{Anderson:2009ge}. However, the matter field Kahler metric has not been computed for any non-standard embedding model on a Calabi-Yau manifold and no clear method for its computation has been formulated. 

The purpose of the present paper is two-fold. First, we would like to develop explicit methods based on differential geometry to compute the holomorphic Yukawa couplings for heterotic models with non-standard embedding. Secondly, we would like to understand how these methods relate to the algebraic ones pioneered in Ref.~\cite{Candelas:1987se} and further developed in Ref.~\cite{Anderson:2009ge}. Apart from occasional remarks we will not be concerned with the matter field Kahler metric in the present paper. However, we hope that the differential geometry methods which we develop will eventually be of help for its calculation. For ease of terminology,  the term ``Yukawa couplings" refers to the holomorphic Yukawa couplings in the remainder of the paper.

The present work will be carried out within the context of heterotic line bundle models~\cite{Anderson:2011ns,Anderson:2012yf,Anderson:2013xka}, perhaps the simplest class of heterotic Calabi-Yau models with non-standard embedding. For those models, the gauge bundle has an Abelian structure group and is realised by a sum of line bundles, a feature which makes explicit calculations of bundle properties significantly more accessible. Yukawa textures due to the additional $U(1)$-symmetries in line bundle models have been studied in Ref.~\cite{Anderson:2010tc}.  Furthermore, we will work within perhaps the simplest class of Calabi-Yau manifolds, namely complete intersections in products of projective spaces~\cite{Green:1986ck,Candelas:1987kf,Hubsch:1992nu} (Cicys). More specifically, we focus on hyper-surfaces in products of projective spaces and the tetra-quadric in the ambient space ${\cal A}=\mathbb{P}^1\times\mathbb{P}^1\times\mathbb{P}^1\times\mathbb{P}^1$ in particular. On the one hand, the simplicity of the set-up facilitates developing new and explicit methods to calculate Yukawa couplings. On the other hand, it is known~\cite{Anderson:2011ns,Anderson:2012yf} that this class contains interesting models with a low-energy MSSM spectrum, so that we will be able to apply our methods to quasi-realistic examples. 

The plan of the paper is as follows. In the next section, we will lay the ground by reviewing some of the basics, including the general structure of heterotic Yukawa couplings, heterotic line bundle models and complete intersection Calabi-Yau manifolds. Since our main focus will be on the tetra-quadric Calabi-Yau manifold we need to understand in some detail the differential geometry of $\mathbb{P}^1$ and its line bundles. This will be developed in Section~\ref{forms}. General results for Yukawa couplings on the tetra-quadric and some toy examples are given in Section~\ref{toyex}. Section~\ref{realex} presents a complete calculation of the Yukawa couplings for a quasi-realistic model~\cite{Anderson:2011ns,Anderson:2012yf, Buchbinder:2013dna, Buchbinder:2014qda,Buchbinder:2014sya,Buchbinder:2014qca} with MMSM spectrum on the tetra-quadric. We conclude in Section~\ref{conclusions}.

Some related matters and technical issues have been deferred to the Appendices. Appendix~\ref{app:21} contains a review of holomorphic $(2,1)$ Yukawa couplings for standard embedding models, following Ref.~\cite{Candelas:1987se} and, in particular, elaborates on the algebraic approach for their computation. The vanishing of a certain boundary integral which is crucial for our calculation of Yukawa couplings is demonstrated in Appendix~\ref{app:bound}. Appendix~\ref{app:Kbundle} provides a concise review of bundles on Kahler manifolds, as required in the main text, largely following Ref.~\cite{H}. Finally, Appendix~\ref{app:proof} proofs a crucial but somewhat technical property for bundle-valued harmonic forms on $\mathbb{P}^1$ which is the key to establishing the relation between the analytic and the algebraic calculation of Yukawa couplings. 


\section{Yukawa couplings for line bundle models}


\subsection{General properties of Yukawa couplings in heterotic Calabi--Yau compactifications}


We will start with a review of holomorphic Yukawa couplings in the context of the $E_8 \times E_8$ heterotic string theory on a 
Calabi--Yau manifold (see, for example, Ref.~\cite{GSW}). The matter fields originate from the $E_8\times E_8$ gauge fields $A$ and 
the associated gauginos. Here we focus on one $E_8$ factor (``the visible sector") and assume that the Calabi-Yau manifold, $X$, carries a 
principal bundle with structure group $G$ embedded into $E_8$. The (visible) low-energy gauge group, $H$, is then the 
commutant of $G$ within $E_8$ and the types of matter multiplets can be read off from the branching
\begin{equation} 
 {\bf 248}\rightarrow \left[({\rm Adj}_G,{\bf 1})\oplus ({\bf 1},{\rm Adj}_H)\oplus\bigoplus ({\cal R}_G, {\cal R}_H)\right]_{G\times H} \label{genbranch}
\end{equation}
of the ${\bf 248}$ adjoint representation of $E_8$ under $G\times H$. Specifically, for the above branching, the low-energy theory can contain  matter multiplets transforming as representations ${\cal R}_H$ under $H$. These multiplets descend from harmonic bundle valued (0,1)-forms $\nu\in H^1(X,V)$, where $V$ is a vector bundle associated to the principal bundle via the $G$ representations ${\cal R}_G$. Consider three representations $({\cal R}_G^i,{\cal R}_H^i)$, where $i=1,2,3$, which appear in the decomposition~\eqref{genbranch}, such that ${\cal R}_G^1\otimes {\cal R}_G^2\otimes {\cal R}_G^3$ contains a singlet. The three associated vector bundles are denoted as $V_i$ with harmonic bundle-valued (0,1)-forms $\nu_i\in H^1(X,V_i)$. Then, the associated holomorphic Yukawa couplings can be computed from
\begin{equation}
\lambda(\nu_1,\nu_2,\nu_3)=\int_X\Omega\wedge\nu_1\wedge\nu_2\wedge\nu_3\; , \label{Yukgen}
\end{equation}
where $\Omega$ is the holomorphic $(3,0)$ form on $X$ and an appropriate contraction over the bundle indices in $\nu_i$ onto the singlet direction is implied. Let us introduce sets of basis forms, $\nu_{i,r}$, where $r=1,\ldots ,h^1(X,V_i)$, for the cohomologies $H^1(X,V_i)$ and define $\lambda_{rst}=\lambda(\nu_{1,r},\nu_{2,s},\nu_{3,t})$. The four-dimensional $N=1$ chiral superfields associated to $\nu_{i,r}$ are denoted $C_i^r$ and these fields transform as ${\cal R}_H^i$ under the gauge group $H$. The superpotential for these fields can be written as
\begin{equation}
 W=\lambda_{rst}C_1^rC_2^sC_3^t\; .
\end{equation} 
Here, we are mainly interested in the phenomenologically promising structure groups $G=SU(3)$, $SU(4)$, $SU(5)$ (and their maximal rank sub-groups), which lead to the low-energy gauge groups $H=E_6$, $SO(10)$, $SU(5)$ (times possible $U(1)$ factors), respectively. For these three groups, the decomposition~\eqref{genbranch} takes the form
\begin{eqnarray}
  {\bf 248}&\rightarrow& \left[({\bf 8},{\bf 1})\oplus ({\bf 1},{\bf 78})\oplus ({\bf 3},{\bf 27})\oplus (\overline{\bf 3},\overline{\bf 27})\right]_{SU(3)\times E_6}\label{E6dec}\\
   {\bf 248}&\rightarrow& \left[({\bf 15},{\bf 1})\oplus ({\bf 1},{\bf 45})\oplus ({\bf 4},{\bf 16})\oplus (\overline{\bf 4},\overline{\bf 16})\oplus ({\bf 6},{\bf 10})\right]_{SU(4)\times SO(10)}\\
   {\bf 248}&\rightarrow& \left[({\bf 24},{\bf 1})\oplus ({\bf 1},{\bf 24})\oplus ({\bf 5},{\bf 10})\oplus (\overline{\bf 5},\overline{\bf 10})\oplus ({\bf 10},\overline{\bf 5})\oplus (\overline{\bf 10},{\bf 5})\right]_{SU(5)\times SU(5)} \label{SU5dec}
\end{eqnarray}  
For $G=SU(3)$ we have matter multiplets in representations ${\bf 27}$, $\overline{\bf 27}$ and ${\bf 1}$ of the low-energy gauge group $H=E_6$ and possible Yukawa couplings of type ${\bf 27}^3$, $\overline{\bf 27}^3$, ${\bf 1}\,{\bf 27}^2$ and ${\bf 1}\,\overline{\bf 27}^2$.

For $H=SU(4)$, the families come in ${\bf 16}$ representations and the anti-families  in $\overline{\bf 16}$ representations of $SO(10)$. Higgs multiplets reside in ${\bf 10}$ representations and bundle moduli in singlets, ${\bf 1}$. Possible Yukawa couplings are of type ${\bf 10}\,{\bf 16}^2$, ${\bf 10}\,\overline{\bf 16}^2$, ${\bf 1}\,{\bf 16}\,\overline{\bf 16}$ and ${\bf 1}\,{\bf 10}^2$.

Finally, for $G=SU(5)$ and low-energy gauge group $H=SU(5)$ we have families in $\overline{\bf 5}\oplus {\bf 10}$, anti-families in ${\bf 5}\oplus\overline{\bf 10}$ and bundle moduli singlets, ${\bf 1}$. Allowed Yukawa couplings include the up-type Yukawa couplings ${\bf 5}\,{\bf 10}^2$, the down-type Yukawa couplings $\overline{\bf 5}\,\overline{\bf 5}\,{\bf 10}$ as well as the singlet couplings ${\bf 1}\,{\bf 5}\,\overline{\bf 5}$, ${\bf 1}\,{\bf 10}\,\overline{\bf 10}$.

While Eq.~\eqref{Yukgen} has been, initially, written down in terms of the harmonic representatives $\nu_i$ of the cohomologies $H^1(X,V_i)$ is it important to note that the expression is, in fact, independent of the choice of representatives. 
To see this, perform the transformation\footnote{Here and in the following, we will often denote the derivative $\bar{\partial}_E$ on differential forms taking values in the vector bundle $E$ simply by $\bar{\partial}$ to avoid cluttering the notation.} $\nu_i\rightarrow \nu_i+\bar{\partial}\xi_i$ on Eq.~\eqref{Yukgen}, where $\xi_i$ are sections of $V_i$. Then, integrating by parts and using $\bar{\partial}\nu_i=0$, $\bar{\partial}\Omega=0$ and $\bar{\partial}^2=0$ it follows immediately that
\begin{equation}
 \lambda(\nu_1+\bar{\partial}\xi_1,\nu_2+\bar{\partial}\xi_2,\nu_3+\bar{\partial}\xi_3)=\lambda(\nu_1,\nu_2,\nu_3)\; .
\end{equation}
This quasi-topological property of the holomorphic Yukawa couplings means that they can, in principle, be computed purely algebraically, as has been noted in Refs.~\cite{Candelas:1987se,Anderson:2009ge}. To recall how this works we focus on the case $G=SU(3)$ and low-energy gauge group $H=E_6$. The families in ${\bf 27}$ descend from bundle-valued (0,1)-forms $\nu,\mu,\rho \in H^1(X,V)$, where $V$ is the associated vector bundle in the fundamental representation, ${\bf 3}$, of $SU(3)$. Since $c_1(V)=0$ it follows that $\wedge^3 V\cong{\cal O}_X$ and we have a map
\be 
H^1 (X, V) \times H^1 (X, V) \times H^1 (X, V) \to H^3 (X, \wedge^3 V)\simeq H^3 (X, {\cal O}_X) \simeq {\mathbb C}\,. 
\label{1.4}
\ee
More explicitly, this can be expressed by the cup product
\begin{equation}
 \nu\wedge\mu\wedge\rho=\kappa(\nu,\mu,\rho)\,\overline{\Omega}\; ,\label{cup}
\end{equation}
Inserting into Eq.~\eqref{Yukgen}, it follows that the complex number $\kappa(\nu,\mu,\rho)$ is proportional to the Yukawa couplings via
\begin{equation}
 \lambda(\mu,\nu,\rho)=\kappa(\nu,\mu,\rho)\int_X\Omega\wedge\overline{\Omega}\; .
\end{equation} 
This means that the ${\bf 27}^3$ Yukawa couplings, up to an overall constant, can be computed algebraically, by performing a (cup) product between three cohomology representatives. Similar arguments can be made for the other Yukawa couplings in the $SU(3)$ case and indeed for other bundle structure groups $G$.

Such an algebraic calculation has been carried out for certain examples in Refs.~\cite{Candelas:1987se,Anderson:2009ge}. While it is elegant and avoids the evaluation of integrals it also has a number of drawbacks. As a practical matter, the relevant cohomologies are not always directly known but are merely represented by certain isomorphic cohomologies. In this case, it is not always obvious how the cup product should be carried out. Perhaps more significantly, computing the physical (rather than just the holomorphic) Yukawa couplings also requires knowledge of the matter field K\"ahler potential which is proportional to the inner product
\begin{equation}
 (\nu,\omega)=\int_X\nu\wedge \bar{\star}_E\,\omega
\end{equation}
between two harmonic $(0,1)$ forms $\nu$, $\omega$ representing cohomologies in $H^1(X,V)$. Unlike the holomorphic Yukawa couplings, this expression is not independent of the choice of representatives due to the presence of the complex conjugation, as can be seen by performing a transformation $\nu\rightarrow\nu+\bar{\partial}\alpha$, $\omega\rightarrow\omega+\bar{\partial}\beta$. It needs to be computed with the harmonic $(0,1)$-forms and requires knowledge of the Ricci-flat Calabi-Yau metric.  Consequently, a full calculation of the physical Yukawa couplings  will have to rely on differential geometry. One purpose of the present paper is to develop such differential geometry methods, for the immediate purpose of calculating the holomorphic Yukawa couplings, but in view of a full calculation of the physical couplings in the future.


\subsection{A review  of line bundle models}


Perhaps the simplest heterotic compactifications for which to calculate Yukawa couplings, 
apart from models with standard embedding, are line bundle models. In the remainder of this paper, we will focus on calculating holomorphic Yukawa 
couplings for such line bundle models and, in the present sub-section, we begin by reviewing their general structure, following 
Refs.~\cite{Anderson:2011ns, Anderson:2012yf}. 

Heterotic line bundle models rely on a gauge bundle with (visible) Abelian structure group $G=S(U(1)^n)$ which can be described by a line bundle sum
\be 
V = \bigoplus_{a=1}^n L_a\quad\mbox{with}\quad c_1(V)=0\; ,
\label{Vdef}
\ee
where $L_a\rightarrow X$ are line bundles over the Calabi-Yau manifold $X$. Here, the condition $c_1(V)=0$ ensures that the structure group of $V$ is indeed special unitary, rather than merely unitary. 

As every heterotic model, line bundle models need to satisfy two basic consistency conditions. Firstly, the bundle $V$ needs to be supersymmetric which is equivalent to requiring vanishing slopes
\begin{equation}
 \mu(L_a)\equiv\int_X c_1(L_a)\wedge J\wedge J\stackrel{!}{=}0
\end{equation}
for all line bundles $L_a$, where $J$ is the Kahler form of the Calabi-Yau manifold $X$. The slope-zero conditions 
are constraints in K\"ahler moduli space which have to be solved simultaneously for all line bundles in order for the bundle $V$ to preserve supersymmetry. 
Secondly, we need to be able to satisfy the heterotic anomaly condition which is guaranteed if we require that
\begin{equation}
 c_2(TX)-c_2(V)\in \mbox{Mori cone of }X\; .
\end{equation} 
In this case, the anomaly condition can always be satisfied by adding five-branes to the 
model (although other completions involving a non-trivial hidden bundle or a combination of hidden bundle and five-branes are usually possible).\\[2mm]
Of particular interest are line bundle sums with rank $n=3,4,5$ for which the associated (visible) 
low-energy gauge groups are $H=E_6\times S(U(1)^3)$, $H=SO(10)\times S(U(1)^4)$ and $SU(5)\times S(U(1)^5)$, respectively. 
For the non-Abelian part of these gauge groups, the multiplet structure of the low-energy theory can be read off 
from Eqs.~\eqref{E6dec}--\eqref{SU5dec}. In addition, multiplets carry charges under the Abelian part, $S(U(1)^n)$, of the gauge group. 
It is convenient to describe these charges by an integer vector  ${\bf q} =(q_1, q_2, \dots, q_n)$. Since we would like to label 
representations of $S (U(1)^n)$, rather than of $U(1)^n$, two such vectors ${\bf q}$ and ${\bf \tilde{q}}$ have to be identified 
if ${\bf q} - {\bf \tilde{q}} \in {\mathbb Z} (1, 1, \dots, 1)$. This charge vector will be attached as a subscript to the 
representation 
of the non-Abelian part. The number of each type of multiplet equals the dimension of the cohomology $H^1(X, K)$ for a certain 
line bundle $K$, which is either one of the line bundles $L_a$ or a tensor product thereof. 
The precise list of multiplets for the three cases $n=3,4,5$, together with the associated line bundles $K$ is provided in 
Tables~\ref{tab:e6}, \ref{tab:so10} and \ref{tab:su5}.
\begin{table}[h]
\begin{center}
\begin{tabular}{|l|l|l|l|}\hline
multiplet&indices&line bundle $K$&intepretation\\\hline\hline
${\bf 27}_{{\bf e}_a}$&$a=1,2,3$&$L_a$&families/Higgs\\\hline
$\overline{\bf 27}_{-{\bf e}_a}$&$a=1,2,3$&$L_a^*$&mirror-families/Higgs\\\hline
${\bf 1}_{{\bf e}_a-{\bf e}_b}$&$a,b=1,2,3\,,\;a\neq b$&$L_a\otimes L_b^*$&bundle moduli\\\hline
\end{tabular}
\end{center}
\caption{\it Multiplets and associated line bundles for bundle structure group $G=S(U(1)^3)$ and low-energy gauge group $H=E_6\times S(U(1)^3)$.}\label{tab:e6}
\end{table}
\begin{table}[h]
\begin{center}
\begin{tabular}{|l|l|l|l|}\hline
multiplet&indices&line bundle $K$&intepretation\\\hline\hline
${\bf 16}_{{\bf e}_a}$&$a=1,2,3,4$&$L_a$&families\\\hline
$\overline{\bf 16}_{-{\bf e}_a}$&$a=1,2,3,4$&$L_a^*$&mirror-families\\\hline
${\bf 10}_{{\bf e}_a+{\bf e}_b}$&$a=1,2,3,4\,,\;a<b$&$L_a\otimes L_b$&Higgs\\\hline
${\bf 1}_{{\bf e}_a-{\bf e}_b}$&$a,b=1,2,3,4\,,\;a\neq b$&$L_a\otimes L_b^*$&bundle moduli\\\hline
\end{tabular}
\end{center}
\caption{\it Multiplets and associated line bundles for bundle structure group $G=S(U(1)^4)$ and low-energy gauge group $H=SO(10)\times S(U(1)^4)$.}\label{tab:so10}
\end{table}
\begin{table}[h]
\begin{center}
\begin{tabular}{|l|l|l|l|}\hline
multiplet&indices&line bundle $K$&intepretation\\\hline\hline
${\bf 10}_{{\bf e}_a}$&$a=1,2,3,4,5$&$L_a$&$(Q,u,e)$ families\\\hline
$\overline{\bf 10}_{-{\bf e}_a}$&$a=1,2,3,4,5$&$L_a^*$&$(\tilde{Q},\tilde{u},\tilde{e})$ mirror-families\\\hline
$\overline{\bf 5}_{{\bf e}_a+{\bf e}_b}$&$a,b=1,2,3,4,5\,,\;a<b$&$L_a\otimes L_b$&$(L,d)$ families\\\hline
${\bf 5}_{-{\bf e}_a-{\bf e}_b}$&$a,b=1,2,3,4,5\,,\;a<b$&$L_a^*\otimes L_b^*$&$(\tilde{L},\tilde{d})$ mirror-families\\\hline
${\bf 1}_{{\bf e}_a-{\bf e}_b}$&$a,b=1,2,3,4,5\,,\;a\neq b$&$L_a\otimes L_b^*$&bundle moduli\\\hline
\end{tabular}
\end{center}
\caption{\it Multiplets and associated line bundles for bundle structure group $G=S(U(1)^5)$ and low-energy gauge group 
$H=SU(5)\times S(U(1)^5)$. 
}\label{tab:su5}
\end{table}
As is clear from the tables, all relevant $S(U(1)^n)$ charges can be expressed easily in terms of the $n$-dimensional standard unit vectors ${\bf e}_a$. 
Frequently, in order to simplify the notation for multiplets, we will replace the subscripts ${\bf e}_a$ simply by $a$.  
For example, in the $SO(10)\times S(U(1)^4)$ case, the multiplet ${\bf 16}_{{\bf e}_a}$ becomes ${\bf 16}_a$ or the multiplet ${\bf 10}_{{\bf e}_a+{\bf e}_b}$ becomes ${\bf 10}_{a,b}$.

For all three cases, the low-energy spectrum contains fields ${\bf 1}_{a,b}$ which are singlets under the non-Abelian part of the 
gauge group but are charged under $S(U(1)^n)$. These fields should be interpreted as bundle moduli which parameterise 
deformations away from a line bundle sum to bundles with non-Abelian structure group. For many models of interest these bundle moduli 
are present in the low-energy spectrum and, in such cases, the Abelian bundle is embedded in a moduli space of generically non-Abelian bundles. 
Much can be learned about non-Abelian bundles by such deformations away from the Abelian locus. This is one of the reasons why studying Yukawa 
couplings for line bundle models can yield insights into the structure of Yukawa couplings for non-Abelian bundles. Another reason is more technical. 
In practice, non-Abelian bundles are often constructed from line bundles, for example via extension or monad sequences, and, hence, some of the 
methods developed for line bundles will be useful to tackle the non-Abelian case.
\vskip 2mm
So far, we have considered the ``upstairs" theory with a GUT-type gauge group. In order to break this theory to the 
standard-model group we require a freely-acting symmetry $\Gamma$ on the Calabi-Yau manifold $X$. The line bundle sum $V$ should descend to the 
quotient Calabi-Yau $X/\Gamma$, that is, it should have a $\Gamma$-equivariant structure. 
Downstairs, on the manifold $X/\Gamma$, we should include a Wilson line, defined by a representation $W$ of $\Gamma$ into 
the (hypercharge direction of the) GUT group. As a result, each downstairs multiplet, $\psi$, acquires an induces $\Gamma$-representation 
denoted $\chi_\psi$. Luckily, the resulting downstairs spectrum can be computed in a simple group-theoretical fashion from the upstairs spectrum. 
Consider a certain type of upstairs multiplet with associated line bundle $K$. 
By virtue of the $\Gamma$-equivariant structure of $V$, the cohomology $H^1(X, K)$, associated to the upstairs multiplet, 
becomes a $\Gamma$-representation~\footnote{In more complicated cases line bundles might not be equivariant individually 
but several line bundles may form an equivariant block. However, the computation of downstairs cohomology for such 
cases proceeds in a similar group-theoretical fashion.}. To compute the spectrum of a certain type, $\psi$, of downstairs multiplet 
contained in $H^1(X, K)$ we should determine the $\Gamma$-singlet part of 
\begin{equation}
 H^1(X, K)\otimes \chi_\psi\; . \label{equivcoh}
\end{equation} 
Fortunately, the computation of Yukawa couplings relates to this Wilson line breaking mechanism in a straightforward way. 
We can obtain the downstairs (holomorphic) Yukawa couplings by basically extracting the relevant $\Gamma$-singlet directions of the upstairs 
Yukawa couplings.

In our later examples, we will consider Wilson line breaking for the gauge group $SU(5)$. In this case, the Wilson line can be conveniently described in terms of two one-dimensional $\Gamma$-representations $\chi_2$, $\chi_3$, satisfying $\chi_2^2\otimes\chi_3^3=1$ and with at least one of them non-trivial. Such a Wilson line, embedded into the hypercharge direction, breaks $SU(5)$ to the standard model group. The $\Gamma$-representations $\chi_\psi$ of the various standard model multiplets, which enter Eq.~\eqref{equivcoh}, are then explicitly given by
\begin{equation}
 \chi_Q=\chi_2\otimes\chi_3\,,\quad \chi_u= \chi_3^2\,,\quad \chi_e=\chi_2^2\,,\quad \chi_d=\chi_3^*\,,\quad \chi_L=\chi_2^*\,,\quad
 \chi_H=\chi_2^*\,,\quad \chi_{\bar{H}}=\chi_2\;. \label{WLcharges}
\end{equation} 


\subsection{Holomorphic Yukawa couplings for line bundle models}
\label{examples}

For heterotic line bundle models, the $(0,1)$-forms $\nu_1$, $\nu_2$ and $\nu_3$, contained in the general expression~\eqref{Yukgen} for the 
Yukawa couplings, represent the first cohomologies of certain line bundles, denoted by $K_1$, $K_2$ and $K_3$, so that $\nu_i\in H^1(X,K_i)$. 
The structure of the integral~\eqref{Yukgen} 
\begin{table}[h]
\begin{center}
\begin{tabular}{|l||c|c|c|c|c|}\hline
Gauge group&Yukawa coupling&$K_1$&$K_2$&$K_3$&index constraint\\\hline\hline
\multirow{3}{*}{$E_6\times S(U(1)^3)$}
&${\bf 27}_a\,{\bf 27}_b\,{\bf 27}_c$&$L_a$&$L_b$&$L_c$&$a,b,c$ all different\\\cline{2-6}
&$\overline{\bf 27}_a\,\overline{\bf 27}_b\,\overline{\bf 27}_c$&$L_a^*$&$L_b^*$&$L_c^*$&$a,b,c$ all different\\\cline{2-6}
&${\bf 1}_{a,b}\,{\bf 27}_b\,\overline{\bf 27}_a$&$L_a\otimes L_b^*$&$L_b$&$L_a^*$&$a\neq b$\\\hline\hline
\multirow{3}{*}{$SO(10)\times S(U(1)^4)$}
&${\bf 10}_{a,b}\,{\bf 16}_a\,{\bf 16}_b$&$L_a\otimes L_b$&$L_a$&$L_b$&$a\neq b$\\\cline{2-6}
&${\bf 10}_{a,b}\,\overline{\bf 16}_a\, \overline{\bf 16}_b$&$L_a\otimes L_b$&$L_a^*$&$L_b^*$&$a\neq b$\\\cline{2-6}
&${\bf 1}_{a,b}\,{\bf 16}_b\, \overline{\bf 16}_a$&$L_a\otimes L_b^*$&$L_b$&$L_a^*$&$a\neq b$\\\hline\hline
\multirow{6}{*}{$SU(5)\times S(U(1)^5)$}
&$\overline{\bf 5}_{a,b}\,\overline{\bf 5}_{c,d}\,{\bf 10}_e$&$L_a\otimes L_b$&$L_c\otimes L_d$&$L_e$&$a,b,c,d,e$ all different\\\cline{2-6}
&${\bf 5}_{a,b}\,{\bf 10}_a\,{\bf 10}_b$&$L_a^*\otimes L_b^*$&$L_a$&$L_b$&$a\neq b$\\\cline{2-6}
&${\bf 5}_{a,b}\,{\bf 5}_{c,d}\,\overline{\bf 10}_e$&$L_a^*\otimes L_b^*$&$L_c^*\otimes L_d^*$&$L_e^*$&$a,b,c,d,e$ all different\\\cline{2-6}
&$\overline{\bf 5}_{a,b}\,\overline{\bf 10}_a\,\overline{\bf 10}_b$&$L_a\otimes L_b$&$L_a^*$&$L_b^*$&$a\neq b$\\\cline{2-6}
&${\bf 1}_{a,b}\,{\bf 5}_{a,c}\,\overline{\bf 5}_{b,c}$&$L_a\otimes L_b^*$&$L_a^*\otimes L_c^*$&$L_b\otimes L_c$&$a\neq b\,,a\neq c\,,b\neq c$\\\cline{2-6}
&${\bf 1}_{a,b}\,{\bf 10}_b\,\overline{\bf 10}_a$&$L_a\otimes L_b^*$&$L_b$&$L_a^*$&$a\neq b$\\\hline
\end{tabular}
\caption{Relation between the line bundles $K_i$ which enter the expression~\eqref{Yuklb} for the Yukawa couplings and the line bundles $L_a$ which define the vector bundle $V$ in Eq.~\eqref{Vdef}. Note that $K_1\otimes K_2\otimes K_3={\cal O}_X$ always follows, 
in some cases due to $c_1(V)=0$ which imples $L_1\otimes\cdots\otimes L_n={\cal O}_X$.}\label{tab:KLrel}
\end{center}
\end{table}
(or, equivalently, four-dimensional gauge symmetry) means that such a line bundle Yukawa coupling can be non-zero only if
\begin{equation}
 K_1 \otimes K_2 \otimes K_3 = {\cal O}_X\; .
\end{equation}
Provided this is the case, the Yukawa coupling is given by 
\begin{equation}
\lambda(\nu_1,\nu_2,\nu_3)=\int_X\Omega\wedge\nu_1\wedge\nu_2\wedge\nu_3\; , \label{Yuklb}
\end{equation}
an expression similar to Eq.~\eqref{Yukgen}, but with the $(0,1)$-forms $\nu_i$ now taking values in the line bundles $K_i$. The precise relation between the line bundles $K_i$ and the line bundles $L_a$ in Eq.~\eqref{Vdef} which define the vector bundle $V$ depends on the low-energy gauge group and the type of Yukawa coupling under consideration. For the three gauge groups of interest and the relevant types of Yukawa couplings these relations are summarised in Table~\ref{tab:KLrel}.
From Eq.~\eqref{Yuklb} it is clear that the Yukawa couplings can depend on the complex structure moduli of the Calabi-Yau manifold $X$. Later, we will see examples with and without explicit complex structure dependence. Given that individual line bundles have no moduli, line bundle Yukawa couplings do not depend on bundle moduli. However, as discussed earlier, line bundle models often reside in a larger moduli space of non-Abelian bundles and Yukawa couplings on this larger moduli space will, in general, display bundle moduli dependence. In this context, our results for line bundle models can be interpreted as a leading-order expressions which are exact at the line bundle locus and provide a good approximation for small deformations away from the line bundle locus.
 

\subsection{Projective ambient spaces}
\label{coboundary}


So far our discussion applies to line bundle models on any Calabi-Yau manifold. In this sub-section and from now on we will specialise to what is perhaps the simplest class of Calabi-Yau manifolds, namely, Calabi-Yau hyper-surfaces in products of projective spaces. Restricting to this class allows us to take the first steps towards evaluating the Yukawa integral~\eqref{Yuklb} and, later on, to explicitly construct the relevant cohomology representatives and compute the integral.

Concretely, we will consider ambient spaces of the form
\be 
{\cal A}= {\mathbb P}^{n_1} \times {\mathbb P}^{n_2} \times \dots \times {\mathbb P}^{n_m} \,, 
\label{Adef}
\ee
where $n_1+n_2+\cdots n_m=4$. The Calabi-Yau hyper-surface $X$ in ${\cal A}$ is defined as the zero-locus of a homogeneous polynomial $p$ with multi-degree $(n_1+1,n_2+1,\ldots ,n_m+1)$ which can be thought of as a section of the line bundle ${\cal N}={\cal O}_{\cal A}(n_1+1,n_2+1,\ldots ,n_m+1)$. Examples in this class include  the quintic in ${\mathbb P}^4$, the bi-cubic in ${\mathbb P}^2 \times {\mathbb P}^2 $ and the tetra-quadric in ${\mathbb P}^1 \times {\mathbb P}^1 \times {\mathbb P}^1 \times {\mathbb P}^1$. 

To  evaluate the Yukawa couplings for such Calabi-Yau hyper-surfaces we first assume that the relevant $(0,1)$-forms $\nu_i$ and the $(3,0)$-form $\Omega$ on $X$ can be obtained as restrictions of ambient space counterparts $\hat{\nu}_i$ and $\hat{\Omega}$. Under this assumption and by inserting an appropriate delta-function current~\cite{Candelas:1987se} we can re-write Eq.~\eqref{Yuklb} as the ambient space integral
\be 
\l(\nu_1,\nu_2,\nu_3) =- \frac{1}{2 i}\int_{{\cal A}} \hat{\Omega} \wedge  \hat{\nu}_{1}  \wedge  \hat{\nu}_{2}   \wedge  \hat{\nu}_{3} \wedge  \d^2 (p) dp \wedge d {\bar p}\,. 
\label{Yukamb}
\ee
The construction of $\Omega$ and $\hat{\Omega}$ for Calabi--Yau hyper-surfaces in products of projective spaces is 
well known~\cite{Witten:1985xc, Strominger:1985it, Candelas:1987se, Candelas:1987kf} and we will simply present the result. To this end, we introduce the forms 
\be 
\mu_i = \frac{1}{n_i!} \eps_{\a_0 \a_1 \dots \a_{n_i}} x^{\a_0}_i d x^{\a_1}_i \wedge \dots \wedge dx^{\a_{n_i}}_i\;,\quad
\mu =\mu_1 \wedge \mu_2 \wedge \dots \wedge \mu_m\,. 
\label{10.2}
\ee
where $x^\a_i$ are the homogeneous coordinates on $\mathbb{P}^{n_i}$. With these definitions, the form $\hat{\Omega}$ satisfies 
\be 
\hat{\Omega} \wedge d p = \mu\,. 
\label{10.4}
\ee
Combining this relation with the current identity
\be 
\delta^2 (p) d \bar p = \frac{1}{\pi} {\bar \pt} \Big( \frac{1}{p}\Big)
\label{10.5}
\ee
leads to the following expression
\be 
\l(\nu_1,\nu_2,\nu_3) =- \frac{1}{2 \pi i}\int_{{\cal A}} \frac{\mu}{p} \wedge \Big[ {\bar \pt} \hat{\nu}_1 \wedge \hat{\nu}_2 \wedge \hat{\nu}_3-
 \hat{\nu}_1 \wedge {\bar \pt} \hat{\nu}_2 \wedge \hat{\nu}_3 +\hat{\nu}_1 \wedge  \hat{\nu}_2 \wedge {\bar \pt} \hat{\nu}_3
 \Big]\,. 
 \label{Yukamb1}
 \ee
for the Yukawa couplings. In deriving this expression, we have performed an integration by parts and ignored the boundary term. This boundary term will be more closely examined in Appendix B and we will show that it vanishes in all cases of interest. 

To understand the implications of this result we need to analyse the relation between the ambient space forms $\hat{\nu}_i$ and their restrictions, $\nu_i$, to the Calabi-Yau manifold $X$. Let $K$ be any of the line bundles $K_1$, $K_2$, $K_3$ and ${\cal K}$ its ambient space counterpart, so that $K={\cal K}|_X$. For a given cohomology representative $\nu\in H^1(X,K)$ we would like to construct an ambient space form $\hat{\nu}$ with $\nu=\hat{\nu}|_X$. The line bundles $K$ and ${\cal K}$ are related by the Koszul sequence
\be 
0 \longrightarrow {\cal N}^* \otimes {\cal K} \stackrel{p}{\longrightarrow} {\cal K}  \stackrel{r}{\longrightarrow} K  \longrightarrow 0\; , 
\label{10.8}
\ee
a short exact sequence with $p$ the defining polynomial of the Calabi-Yau manifold and $r$ the restriction map. This short exact sequence leads to an associated long exact sequence in cohomology whose relevant part is given by
\bea
 \cdots&\longrightarrow &  H^1 ({\cal A}, {\cal N}^* \otimes {\cal K}) \stackrel{p}{\longrightarrow} H^1 ({\cal A}, {\cal K}) \stackrel{r}{\longrightarrow} H^1 (X, K)\nonumber \\
&\stackrel{\delta}{\longrightarrow} &H^2 ({\cal A}, {\cal N}^* \otimes {\cal K}) \stackrel{p}{\longrightarrow} H^2 ({\cal A}, {\cal K}) \stackrel{r}{\longrightarrow} H^2 (X, K) 
\longrightarrow \dots\; ,
\label{longex}
\eea
where $\delta$ is the co-boundary map. This sequence allows us to relate the cohomology $H^1(X,K)$ to ambient space cohomologies, namely
\be 
H^1 (X, K) =  r \Big( {\rm Coker} \Big( H^1 ({\cal A}, {\cal N}^* \otimes {\cal K}) \stackrel{p}{\rightarrow}
H^1 ({\cal A}, {\cal K})\Big) \Big) \oplus 
\d^{-1} \Big( {\rm Ker} \Big( H^2 ({\cal A}, {\cal N}^* \otimes {\cal K}) \stackrel{p}{\rightarrow}
H^2 ({\cal A}, {\cal K})\Big) \Big) \; .
\label{H1eq}
\ee
Evidently, $H^1(X,K)$ can receive two contributions, one from $H^1({\cal A},{\cal K})$ (modulo identifications) and the other from (the kernel in) $H^2({\cal A},{\cal N}^*\otimes{\cal K})$. Let us discuss these two contributions separately, keeping in mind that the general case is a sum of these.\\[2mm]
{\bf Type 1}: If $\nu$ descends from $H^1({\cal A},{\cal K})$ we refer to it as ``type 1". In this case we have a $(0,1)$-form $\hat{\nu}\in H^1({\cal A},{\cal K})$ which, under the map $r$, restricts to $\nu\in H^1(X,K)$. What is more, since $\hat{\nu}$ represents an ambient space cohomology it is closed, so
\begin{equation}
 \bar{\partial}\hat{\nu}=0\; .
\end{equation} 
{\bf Type 2:} The situation is somewhat more involved if $\nu$ descends from $H^2({\cal A},{\cal N}^*\otimes{\cal K})$, a situation we refer to as ``type 2". In this case, we can start with an ambient space $(0,2)$-form $\hat{\omega}=\delta(\nu)\in H^2({\cal A},{\cal N}^*\otimes{\cal K})$ which is the image of $\nu$ under the co-boundary map. The definition of the co-boundary map tells us that, in this case, $\nu$ can be obtained as the restriction to $X$ of an ambient space $(0,1)$-form $\hat{\nu}$ which is related to $\hat{\omega}$ by
\be
{\bar \partial} \hat{\nu}=p \hat{\omega}  \,. 
\label{coboundarymap}
\ee
Unlikely in the previous case, the form $\hat{\nu}$ is no longer closed.\\[2mm]
The Yukawa coupling~\eqref{Yukamb1} involves three $(0,1)$-forms, $\hat{\nu}_1$, $\hat{\nu}_2$ and $\hat{\nu}_3$, each of which can be either of type 1 or type 2 (or a combination of both types) so that a variety of possibilities ensues. Perhaps the simplest possibility arises when all three forms are of type 1, so that $\bar{\partial}\hat{\nu}_i=0$ for $i=1,2,3$. Then, Eq.~\eqref{Yukamb1} shows that the Yukawa coupling vanishes,
\begin{equation}
 \l(\nu_1,\nu_2,\nu_3)=0\;.
\end{equation} 
 This vanishing is quasi-topological and related to the cohomology structure for $K_1$, $K_2$ and $K_3$ in the sequence~\eqref{longex} - there is no expectation that it can be explained in terms of a symmetry in the four-dimensional theory. An explicit example of this case will be presented later.
 
The next simplest possibility is for two of the forms, say $\hat{\nu}_1$ and $\hat{\nu}_2$, to be of type 1, so that $\bar{\partial}\hat{\nu}_1= \bar{\partial}\hat{\nu}_2=0$ while $\hat{\nu}_3$ is of type 2, so that $\bar{\partial}\hat{\nu}_3=p\hat{\omega}_3$ for some $(0,2)$-form $\hat{\omega}_3$. Inserting into Eq.~\eqref{Yukamb1}, the Yukawa coupling now reduces to the simple expression
\be
\l(\nu_1,\nu_2,\nu_3) = - \frac{1}{2 \pi i}\int_{{\cal A}} \mu  \wedge \hat{\nu}_1 \wedge  \hat{\nu}_2 \wedge  \hat{\omega}_3\,. 
\label{Yuk112}
\ee
As we will see, this formula is very useful since it is expressed in terms of ambient space forms which can often be written down explicitly. When more than one of the forms is of type 2, the general formula~\eqref{Yukamb1} needs to 
be used and working out all the required forms becomes more complicated. We will study  examples for all these cases later on.
 

\section{Line bundle valued harmonic forms } 
\label{forms}
Henceforth we will focus on tetra-quadric Calabi-Yau manifolds in the ambient space ${\cal A}={\mathbb P}^1\times{\mathbb P}^1\times{\mathbb P}^1\times{\mathbb P}^1$. Besides the general usefulness of working with a concrete example, the tetra-quadric offers a number of additional advantages. Firstly, the ambient space consists of ${\mathbb P}^1$ factors only and is, therefore, particularly simple to handle. Moreoever, it is known~\cite{Anderson:2011ns,Anderson:2012yf} that quasi-realistic line bundle standard models exist on the tetra-quadric, so we will be able to apply our methods for calculating Yukawa couplings to physically relevant models. However, the methods we develop in the context of the tetra-quadric can be generalised to other Calabi-Yau hypersurfaces in products of projective spaces and, presumably, with some more effort, to complete intersection Calabi-Yau manifolds in products of projective spaces.

The main purpose of this chapter is to set out the relevant differential geometry for $\mathbb{P}^1$, find the harmonic 
bundle-values forms for all line bundles on $\mathbb{P}^1$ and apply the results to the full ambient space ${\cal A}$. In particular, 
we will work out a multiplication rule for bundle-valued harmonic forms which will be crucial in order to establish the relation between the 
algebraic and analytic methods for calculating holomorphic Yukawa couplings. Since Yukawa couplings depend only on the cohomology classes of the corresponding forms, we are free to use any non-trivial representatives. For our calculation we will rely on forms which are harmonic relative to the Fubini-Study metric on ${\cal A}$. As we will see, these can be explicitly constructed. For easier accessibility, this chapter is kept somewhat informal. A review of some relevant mathematical background, mostly following Ref.~\cite{H}, can be found in Appendix~\ref{app:Kbundle}. The proof of the multiplication rule for harmonic forms on $\mathbb{P}^1$ is contained in Appendix~\ref{app:proof}.


\subsection{Construction of line bundle valued harmonic forms on ${\mathbb P}^1$}
\label{p1}
We begin by collecting some well-known properties of ${\mathbb P}^1$. Homogeneous coordinates on ${\mathbb P}^1$ are denoted by $x^\a$, where $\a=0,1$, and we introduce the standard open patches $U_{(\a)}=\{[x^0:x^1]\,|\, x^\a\neq 0\}$ with affine coordinates $z=x^1/x^0$ on $U_{(0)}$ and $w=x^0/x^1$ on $U_{(1)}$.  The transition function on the overlap is given by $w=1/z$. For convenience, subsequent formulae will usually be written on the patch $U_{(0)}$ and in terms of the coordinate $z$.

The Kahler potential for the Fubini--Study metric on ${\mathbb P}^1$ reads
\be 
\mathfrak{K}= \frac{i}{2 \pi} \log \kappa\,, \qquad \kappa= 1+ |z|^2\,,
\label{2.1}
\ee
with associated Kahler form and Kahler metric given by
\be
J= \pt {\bar \pt}\mathfrak{K}= \frac{i}{2 \pi \kappa^2} dz \wedge d {\bar z}\,, \qquad  g_{z \bar z}= -i J_{z \bar z} =\frac{1}{2 \pi \kappa^2}\,. 
\label{2.2}
\ee
Note that the normalisation of $\mathfrak{K}$ has been chosen such that $\int_{{\mathbb P}^1} J=1$.

Line bundles on ${\mathbb P}^1$ are classified by an integer $k$ and are denoted by ${\cal O}_{\mathbb{P}^1}(k)$. They can be explicitly constructed by dualising and taking tensor powers of the universal bundle ${\cal O}_{\mathbb{P}^1}(-1)$. With  the above covering of ${\mathbb P}^1$ and the fiber coordinate $v$, the transition function of $ {\cal O}_{{\mathbb P}^1} (k)$ can be written as
\be 
\phi_{01} (z, v)=  (1/z, z^{k} v)\,. 
\label{transfct}
\ee
This means that a section of $ {\cal O}_{{\mathbb P}^1} (k)$  given by $s_{(0)}$ on  $U_{(0)}$ and  $s_{(1)}$ on $U_{(1)}$  transforms as $s_{(0)}(z)= z^k s_{(1)}(1/z)$. 

A hermitian structure $H$ on ${\cal L}={\cal O}_{\mathbb{P}^1}(k)$ can be introduced by
\be 
H= \kappa^{-k}\; , 
\label{2.5}
\ee
and the associated Chern connection, $\nabla^{0,1}= \bar \pt$ and $\nabla^{1,0}= \pt+ A$, with gauge potential $A= {\bar H}^{-1} \pt {\bar H} = \pt \log {\bar H}$ and curvature $F= d A= {\bar \pt} {\pt} \log {\bar H}$ is explicitly specified by
\be 
A= -\frac{k \bar z}{\kappa} dz\,, \quad  F=- 2 \pi i k J\,. 
\label{2.6}
\ee
The last result for the field strength allows the calculation of the first Chern class of ${\cal L}$ which is given by
\be
c_1 ({\cal L})= \frac{i}{2 \pi}  F =k J\,, \quad \int_{{\mathbb P}^1} c_1 ({\cal L}) =k\,. 
\label{2.8}
\ee
Having introduced a hermitian structure and a connection on the line bundles ${\cal L}$, we can now turn to a discussion of their cohomology and their associated harmonic bundle-values forms. As explained in Appendix~\ref{app:Kbundle}, an  ${\cal L}$-valued harmonic form $\a$ is characterised by the equations
\be 
{\bar \pt} \a =0\,, \quad \pt ({\bar H} \star \a)=0\,, 
\label{harmeqs}
\ee
where $\star$ is the Hodge star on $\mathbb{P}^1$ with respect to the Fubini-Study metric. The first of these equations simply asserts the $\bar{\partial}$-closure of $\a$, which is already sufficient to obtain representatives for cohomology. However, $\bar{\partial}$-closed forms which differ by a $\bar{\partial}$-exact form describe the same cohomology class and such a redundant description of cohomology is not convenient for our purposes. For this reason, we will solve both equations~\eqref{harmeqs} and work with the resulting harmonic representatives which are in one-to-one correspondence with the relevant cohomology. 

The cohomology of ${\cal L}={\cal O}_{\mathbb{P}^1}(k)$ is obtained from the Bott formula and we should distinguish three qualitatively different cases. 
For $k\geq 0$ only the zeroth cohomology is non-vanishing, while for $k\leq -2$ only the first cohomology is non-vanishing. For $k=-1$ the cohomology is entirely trivial. We will now discuss these three cases in turn and explicitly compute the bundle-valued harmonic forms by solving Eqs.~\eqref{harmeqs}.\\[2mm]
{\bf Case 1)} $k\geq 0$: In this case, the Bott formula implies that $h^0(\mathbb{P}^1,{\cal L})=k+1$ and $h^1(\mathbb{P}^1,{\cal L})=0$. Hence, we are looking for sections or bundle-valued $(0,0)$-forms of ${\cal L}$. In this case, the second equation~\eqref{harmeqs} is automatically satisfied while the first one implies that the section is holomorphic, so $\alpha=\alpha(z)$. For a monomial $\alpha=z^l$ a transformation to the other patch gives $z^l=w^{-l}=z^kw^{k-l}$ with the $z^k$ factor the desired transition function. This means that the section is holomorphic in both patches only if $l=0,\ldots ,k$. This leads to the well-known result that the sections are given by degree $k$ polynomials, that is,
\begin{equation}
 \alpha=P_{(k)}(z)\; .
 \end{equation}
Note that the space of these polynomials is indeed $k+1$-dimensional, as required.\\[2mm]
{\bf Case 2)} $k=-1$: In this case, all cohomologies of ${\cal L}$ vanish and there are no forms to be determined.\\[2mm]
{\bf Case 3)} $k\leq -2$: Now, $h^1(\mathbb{P}^1,{\cal L})=-k-1$ and $h^0(\mathbb{P}^1,{\cal L})=0$. Hence, we are looking for harmonic $(0,1)$-forms $\alpha=f(z,\bar{z})d\bar{z}$. Clearly, the first equation~\eqref{harmeqs} is automatically satisfied for such $\alpha$. Using $\star d\bar{z}=-id\bar{z}$ and $\star\alpha =-i\alpha$, the second equation can be written as $\partial(\bar{H}\alpha)=0$ which leads to the general solution $\alpha = \kappa^kg(\bar{z})d\bar{z}$, with a general anti-holomorphic function $g(\bar{z})$. For a monomial $g(\bar{z})=\bar{z}^l$, this transforms to the other patch as
\begin{equation}
 \alpha=(1+|z|^2)^k\bar{z}^ld\bar{z}=-z^k(1+|w|^2)^k\bar{w}^{-k-l-2}d\bar{w}\; . \label{harm01}
\end{equation} 
For holomorphy in both patches we should therefore have $l=0,\ldots ,-k-2$, so $g(\bar{z})$ is a general polynomial of degree $-k-2$ in $\bar{z}$. It will be convenient to denote such a polynomial of degree $-k-2$ by $P_{(k)}$ with the understanding that the negative degree subscript implies a dependence on $\bar{z}$, rather than $z$. With this notation, the full solution takes the form
\begin{equation}
 \alpha=\kappa^kP_{(k)}(\bar{z})d\bar{z}\; . \label{1forms}
\end{equation} 
Note that the space of degree $-k-2$ polynomials has indeed dimension $-k-1$, as required.\\[2mm]
 

\subsection{Maps between line bundle cohomology on ${\mathbb P}^1$}
\label{maps}
Calculating Yukawa couplings requires performing a wedge product of bundle-valued forms. It is, therefore, natural to study how the harmonic forms on $\mathbb{P}^1$ found in the previous sub-section multiply. Recall that we have harmonic $(0,0)$-forms taking values in ${\cal O}_{\mathbb{P}^1}(k)$ for $k\geq 0$ and harmonic $(0,1)$-forms taking values in ${\cal O}_{\mathbb{P}^1}(k)$ for $k\leq -2$. 

Multiplying two harmonic $(0,0)$-forms, representing classes in $H^0({\mathbb P}^1,{\cal O}_{{\mathbb P}^1}(k))$ and $H^0({\mathbb P}^0,{\cal O}_{{\mathbb P}^1}(l))$ respectively, is straightforward and it leads to another harmonic $(0,0)$-form which represents a class in $H^0({\mathbb P}^1,{\cal O}_{{\mathbb P}^1}(k+l))$. 

The only other non-trivial case - the multiplication of a harmonic $(0,0)$-form with a harmonic $(0,1)$-form - is less straightforward. To be concrete, for $k\leq -2$ and $\delta>0$, we consider a harmonic $(0,1)$-form $\a_{(k-\d)} \in H^1 ({\mathbb P}^1,  {\cal O}_{{\mathbb P}^1} (k-\d))$ and a degree $\delta$ polynomial $p_{(\d)}$, representing a class in $H^0 ({\mathbb P}^1,  {\cal O}_{{\mathbb P}^1} (\d))$. The product $p_{(\d)}\a_{(k-\d)}$ is a $(0,1)$-form which represent a class in $H^1 ({\mathbb P}^1,  {\cal O}_{{\mathbb P}^1} (k))$ but it is not of the form~\eqref{harm01} and, hence, is not harmonic. We would, therefore, like to work out the harmonic representative, denoted $\a_{(k)}\in H^1 ({\mathbb P}^1,  {\cal O}_{{\mathbb P}^1} (k))$, which is equivalent in cohomology to this product $p_{(\d)}\a_{(k-\d)}$. This means we should solve the equation
\be 
p_{(\d)} \a_{(k-\d)} + {\bar \pt} s= \a_{(k)}\; ,
\label{prodeqgen}
\ee
where $s$ is a suitable section of ${\cal O}_{\mathbb{P}^1}(k)$. In general, the section $s$ an be cast into the form
\be 
s= \sum_{m \geq -k} S_{(k+m, -m-2)} (z, \bar z) \kappa^{-m}\,, 
\label{2.14}
\ee
where $S_{(k+m, -m-2)} (z, \bar z)$ is a polynomial of degree $k+m$ in $z$ and of degree $m$ in $\bar z$.  This can be seen be demanding the correct transformation under the transition function~\eqref{transfct}. It turns out that in order to solve Eq.~\eqref{prodeqgen} we only require the single term with $m=-k+\delta-1$ in this sum for $s$. Using this observation and the general formula~\eqref{1forms} for harmonic $(0,1)$-forms, we insert the following expressions
\begin{equation}
\a_{k-\d} =\kappa^{k-\d} P_{(k- \d)} (\bar z) d \bar z\;, \quad   \a_k =\kappa^{k} Q_{(k)} (\bar z) d \bar z
\;,\quad s= \kappa^{k-\d+1} S_{(\d-1, k -\d-1)} (z, \bar z)\,. 
\end{equation}
into Eq.~\eqref{prodeqgen} to cast it into the more explicit form
\be
p P +\k \pt_{\bar z} S - (-k+\d-1) z S = \k^{\d} Q\,.
\label{prodeq}
\ee
Here, for simplicity of notation, we have dropped the subscripts indicating degrees. Eq.~\eqref{prodeq} determines the polynomials $Q$ and $S$ for given $p$ and $P$ and can be solved by comparing monomial coefficients. This is relatively easy to do for low degrees and we will discuss a few explicit examples below. For arbitrary degrees Eq.~\eqref{prodeq} seems surprisingly complicated and it is, therefore, remarkable that a closed solution for $Q$ can be written down. To formulate this solution, we introduce the homogeneous counterparts of the polynomials $p$, $P$, $Q$ and $S$ which we denote as $\tilde{p}, \tilde{P}$, $\tilde{Q}$ and $\tilde{S}$. They depend on the homogeneous coordinates $x^0$, $x^1$ and are obtained from the original polynomials by replacing $z=x^1/x^0$ and multiplying with the appropriate powers of $x^0$ and $\bar{x}^0$. Then, the polynomial $\tilde{Q}$  which solves Eq.~\eqref{prodeq} can be written as
\be 
\tilde{Q} ({\bar x}^0, {\bar x}^1) = c_{k-\d, \d} \ \tilde{p} (\pt_{ {\bar x}^0}, \pt_{ {\bar x}^1})  \tilde{P}  ({\bar x}^0, {\bar x}^1)\,, \quad
c_{k-\d, \d} =\frac{(-k-1)!}{(\d-k-1)!}\,. 
\label{prodsol}
\ee
Here  $\tilde{p} (\pt_{ {\bar x}^0}, \pt_{ {\bar x}^1})$ denotes the polynomial $\tilde{p}$ with the coordinates replaced by the 
corresponding partial derivatives. These derivatives act on the polynomial $\tilde{P}$ in the usual way and thereby lower the degree to the one expected for $\tilde{Q}$. The proof of Eq.~\eqref{prodsol} is given in Appendix D. Unfortunately, we are not aware at present of a similar closed solution for the polynomial $S$.\\[2mm]
While this discussion may have been somewhat technical the final result is relatively simple and can be summarised as follows. For $k\geq 0$ the harmonic $(0,0)$-forms representing the cohomology $H^0(\mathbb{P}^1,{\cal O}_{\mathbb{P}^1}(k))$ are given by degree $k$ polynomials $P_{(k)}(z)$ which depend on the coordinate $z$. For $k\leq -2$ the harmonic $(0,1)$-forms representing the cohomology $H^1(\mathbb{P}^1,{\cal O}_{\mathbb{P}^1}(k))$ can be identified with degree $-k-2$ polynomials, denoted as $P_{(k)}(\bar{z})$, which depend on $\bar{z}$. The product of two $(0,0)$-forms is simply given by polynomial multiplication while the product of a $(0,0)$-form and a $(0,1)$-form is performed by using the homogeneous versions of these polynomials and converting the coordinates in the former to partial derivatives which act on the latter.  Let us finish this subsection by illustrating the above discussion with two explicit example.\\[2mm]
{\bf Example 1:} Consider the case $k=-3$ and $\d=1$ so that the relevant forms and associated polynomials are explicitly given by
\begin{equation}
\begin{array}{lllllll}
 \alpha_{(-4)}&=&\kappa^{-4}P_{(-4)}(\bar{z})d\bar{z}&\qquad& P_{(-4)}&=&a_0+a_1\bar{z}+a_2\bar{z}^2\\
 \alpha_{(-3)}&=&\kappa^{-3}Q_{(-3)}d\bar{z}&\qquad& Q_{(-3)}&=&b_0+b_1\bar{z}\\
 s&=&\k^{-3}S_{(0,-5)}&\qquad&S_{(0,-5)}&=&c_{0,0}+c_{0,1}\bar{z}+c_{0,2}\bar{z}^2+c_{0,3}\bar{z}^3\\
 p_{(1)}&=&f_0+f_1 z\; ,
\end{array} 
\end{equation} 
where $a_i$, $b_i$, $f_i$ and $c_{i,j}$ are constants.  Inserting these polynomials into Eq.~\eqref{prodeq}, comparing coefficients for same monomials and solving for the $b_i$ and $c_{i,j}$ in terms of the $a_i$ and $f_i$ results in
\begin{eqnarray}
Q_{(-3)}&=&\frac{1}{3} \left( 2 a_0 f_0+a_1  f_1+ \left(a_1 f_0+2 a_2 f_1\right)\bar{z}\right)\label{Qres2}\\
S_{(0,-5)}&=&\frac{1}{3}\left( -a_2 f_0 \bar{z}^3+  \left(a_2 f_1-a_1   f_0\right)\bar{z}^2+  \left(a_1 f_1-a_0 f_0\right)\bar{z}+a_0 f_1\right)
\end{eqnarray}
For the calculation based on Eq.~\eqref{prodsol}, we start with the homogenous polynomials
\begin{equation}
\tilde{p}=f_0 x_0+f_1 x_1\;,\quad \tilde{P}=a_0\bar{x}_0^2+a_1 \bar{x}_0\bar{x}_1+a_2\bar{x}_1^2\;,\quad \tilde{S}=c_{0,0}\bar{x}_0^3+c_{0,1}\bar{x}_0^2\bar{x}_1+c_{0,2}\bar{x}_0\bar{x}_1^2+c_{0,3}\bar{x}_1^3\; .
\end{equation}
Inserting these into Eq.~\eqref{prodsol} gives
\begin{equation}
\tilde{Q}=\frac{1}{3} \left( (2 a_0 f_0+a_1  f_1)\bar{x}_0+ \left(a_1 f_0+2 a_2 f_1\right)\bar{x}_1\right)\; ,
\end{equation}
which is indeed the homogeneous version of the polynomial $Q_{(-3)}$ in Eq.~\eqref{Qres2}.\\[2mm]
{\bf Example 2:} Let us choose $k=-1$ and $\d=2$. Since there are no harmonic forms for $k=-1$ we have $Q=0$, while the other forms and polynomials are given by
\be
\begin{array}{lllllll}
\a_{(-3)} &=& \k^{-3} P_{(-3)} (\bar z) d \bar z&\quad& P_{(-3)}&=& a_0 + a_1 \bar z\\
s &=&\k^{-2}S_{(1,-4)}&\quad&S_{(1,-4)}&=&c_{0,0} +c_{0,1} \bar z + c_{0,2} {\bar z}^2 + c_{1,0} z + c_{1,1} |z|^2 + c_{1,2} {\bar z} |z|^2\\
p_{(2)}&=&p_0 +p_1 z +p_2 z^2\; .
\end{array}
\ee
We note that, from~\eqref{prodeqgen}, we now need to solve the equation $p_{(2)} \a_{(-3)} =-{\bar \pt} s$ which is similar in structure to Eq.~\eqref{coboundary} which determines the co-boundary map. Indeed, we will later find the present example useful to explicitly work out a co-boundary map. Inserting the above polynomials into Eq.~\eqref{prodeq} and comparing coefficients as before leads to
\be 
S_{(2,-4)} =\frac{1}{2} (p_1 a_0 + p_2 a_1) - p_0 a_0 \bar z -\frac{1}{2} p_0 a_1 {\bar z}^2 +\frac{1}{2} p_2 a_0 z
+p_2 a_1 |z|^2 -\frac{1}{2} (p_0 a_0 + p_1 a_1){ \bar z} |z|^2\,.  \label{coboundres}
\ee


\subsection{Line bundle valued harmonic forms on ${\mathbb P}^1 \times {\mathbb P}^1\times {\mathbb P}^1  \times {\mathbb P}^1 $}
\label{maps1}


In this sub-section, we generalise the above results for $\mathbb{P}^1$ to the ambient space ${\cal A}= {\mathbb P}^1 \times {\mathbb P}^1\times {\mathbb P}^1  \times {\mathbb P}^1 $. On each ${\mathbb P}^1$ we introduce homogeneous coordinates $(x^0_i, x^1_i)$, where $i=1, \dots, 4$, and cover each ${\mathbb P}^1$ with two standard open sets $U_{(i,\a)}=\{[x_i^0:x_i^1]\,|\,x_i^\a\neq 0\}$ . Further, we introduce affine coordinates $z_i= x^1_i/x^0_i$ on $U_{(i,0)}$ and $w_i= x^0_i/x^1_i$ on $U_{(i,1)}$. On the intersection of $U_{(i,0)}$ and $U_{(i,1)}$
we have $z_i= 1/w_i$. An open cover for the entire space ${\cal A}$ is given by the $16$ sets $U_{(1,\a_1)}\times \cdots \times U_{(4,\a_4)}$. For practical purposes we will usually work on the set $U_{(1,0)}\times \cdots \times U_{(4,0)}$ with coordinates $z_1,\ldots,z_4$.

For each $\mathbb{P}^1$ we have a Fubini--Study Kahler potential and Kahler form given by
\be 
\mathfrak{K}_i=\frac{i}{2 \pi} \log \kappa_i \,, \qquad \kappa_i= 1+ |z_i|^2\,,\qquad J_i = \frac{i}{2 \pi \kappa_i^2} dz_i \wedge d {\bar z}_i\
\ee
and the Kahler cone of ${\cal A}$ is parametrised by $J=\sum_{i=1}^4 t^iJ_i$, with all $t^i>0$. 

The line bundles on ${\cal A}$ are obtained as the tensor products
\begin{equation}
 {\cal O}_{\cal A}({\bf k})={\cal O}_{\mathbb{P}^1}(k^1)\otimes\cdots\otimes {\cal O}_{\mathbb{P}^1}(k^4)
\end{equation} 
and are, hence, labeled by a four-dimensional integer vector ${\bf k} =  (k^1, k^2, k^3, k^4)$. Straightforwardly generalising Eq.~\eqref{2.5}, we can introduce a Hermitian structure
\be 
H= \prod_{i=1}^4 \kappa_i^{-k^i}\,. 
\label{20.4}
\ee
on these line bundles. The gauge field and gauge field strength for the associated Chern connection
\be 
A=- {\bar H}^{-1} \pt {\bar H}=- \sum_{i=1}^4 k^i \pt \log \kappa_i\;,\quad F ={\bar \pt} A =- 2 \pi i  \sum_{i=1}^4 k^i J_i\,,
\label{20.5}
\ee
lead to the first Chern class
\be
c_1 \left( {\cal O}_{\cal A}({\bf k})\right)= \frac{i}{2 \pi} F = \sum_{i=1}^4 k^i J_i\,.
\label{20.5.2}
\ee
The cohomology for ${\cal K}={\cal O}_{\cal A}({\bf k})$ can be obtained by combining the Bott formula for cohomology on $\mathbb{P}^1$ with the Kunneth formula. If any of the integers $k^i$ equals $-1$ all cohomologies of ${\cal K}$ vanish. In all other cases, precisely one cohomology, $H^q({\cal A},{\cal K})$, is non-zero, and $q$ equals the number of negative integers $k^i$. The dimension of this non-vanishing cohomology is given by
\begin{equation}
 h^q({\cal A},{\cal K})=\prod_{i:k^i\geq 0}(k^i+1)\prod_{i:k^i\leq -2}(-k^i-1)\; .
\end{equation} 
Generalising our results for $\mathbb{P}^1$, the harmonic $(0,q)$-forms representing this cohomology can be written as
\be 
\a_{({\bf k})} = P_{({\bf k})} \prod_{i: k^i \leq -2} \k_i^{k^i} d {\bar z}_i\,,
\label{akres}
\ee
where $P_{({\bf k})} $ is a polynomial of degree $k^i$ in $z_i$ provided $k^i \geq 0$ and of degree $-k^i-2$ in ${\bar z}_i$ if $k^i \leq -2$. It is also useful to write down a homogenous version of these forms which is given by
\be 
\a_{({\bf k})} = \tilde{P}_{({\bf k})} \prod_{i: k^i \leq -2} \s_i^{k^i}  {\bar \m}_i\,,
\label{akreshom}
\ee
where
\be
\s_i=|x_i^0|^2+|x_i^1|^2\,,\qquad \mu_i=\e_{\a\b}x_i^\a x_i^\b\;.
\ee
and $\tilde{P}_{\bf k}$ denotes the homogenous counterpart of $P_{\bf k}$. 

We would now like to generalise our rule for the multiplication of forms obtained on $\mathbb{P}^1$. In general, we have a map
\be
H^q ({\cal A}, {\cal O}_{{\cal A}} ({\bf k}))\times 
H^p ({\cal A}, {\cal O}_{{\cal A}} ({\bf l}))  \to  H^{q+p} ({\cal A}, {\cal O}_{{\cal A}} ({\bf k}+{\bf l}))
\label{20.10.1}
\ee
between cohomologies induced by the wedge product and we would like to work out this map for the above harmonic representatives. For a harmonic $(0,q)$-form $\a_{({\bf k})}\in H^q({\cal A},{\cal O}_{\cal A}({\bf k}))$ with associated polynomial $P_{({\bf k})}$ and a harmonic $(0,p)$-form $\b_{({\bf l})}$ with associated polynomial $R_{({\bf l})}$ the wedge product $\a_{({\bf k})}\wedge\b_{({\bf l})}$ is equivalent in cohomology to a harmonic $(0,q+p)$-form which we denote by $\g_{({\bf k}+{\bf l})}\in H^{q+p} ({\cal A}, {\cal O}_{{\cal A}} ({\bf k}+{\bf l}))$ with associated polynomial $Q_{({\bf k}+{\bf l})}$. In general, the relation between those forms can be written as
\be 
\a_{{\bf k}}  \wedge \b_{{\bf l}}  +{\bar \pt} s = \g_{{\bf k} +{\bf l}} 
\label{20.12}
\ee
for a suitable $(0,p+q-1)$-form $s$ taking values in ${\cal O}_{\cal A}({\bf k}+{\bf l})$. Our earlier results for $\mathbb{P}^1$ show that the polynomial
$Q_{({\bf k}+{\bf l})}$ which determines $\g_{{\bf k} +{\bf l}}$ can be directly obtained from $P_{({\bf k})}$ and $R_{({\bf l})}$ by the formula
\be 
\tilde{Q}= c_{{\bf k}, {\bf l}} \tilde{ P} \tilde {R}\,,
\ee
where, as before, $\tilde{ P}, \tilde{R}, \tilde{Q}$ are the  homogeneous counterparts of $P, R, Q$ and $c_{{\bf k}, {\bf l}}$ is the appropriate product of numerical factors in Eq.~\eqref{prodsol}. The understanding is that positive degrees in a particular $\mathbb{P}^1$, represented by powers of $x_i^\a$ should be converted into derivatives $\partial_{\bar{x}^i_\a}$ whenever they act on negative degrees in the same $\mathbb{P}^1$, represented by $\bar{x}_i^\a$. When both degrees in $\tilde{P}$ and $\tilde{R}$ are positive for a given $\mathbb{P}^1$  a simple polynomial multiplication should be carried out. Finally, for two negative degrees in the same $\mathbb{P}^1$ the resulting $\tilde{Q}$ vanishes (since there will be a term $d\bar{z}^i\wedge d\bar{z}^i$ in the corresponding wedge product of the forms).


\subsection{Line bundles and cohomology on the tetra-quadric}
\label{relations}


As the final step in our discussion of line bundles and harmonic forms we need to consider line bundles on the tetra-quadric $X$. Recall that a tetra-quadric resides in the ambient space ${\cal A}=\mathbb{P}^1\times\mathbb{P}^1\times \mathbb{P}^1\times \mathbb{P}^1$  and is defined as the zero locus of a polynomial $p$ of multi-degree $(2,2,2,2)$, which can be seen as a section of the line bundle 
\be 
{\cal N}= {\cal O}_{{\cal A}} ({\bf q})\,, \quad  {\bf q} = (2, 2, 2, 2)\,. 
\label{20.14}
\ee
The tetra-quadric has Hodge numbers $h^{1,1}(X)=4$ and $h^{2,1}(X)=68$. Later, we will use the freely-acting $\Gamma=\mathbb{Z}_2\times\mathbb{Z}_2$ symmetry of the tetra-quadric whose generators are given by
\begin{equation}
 g_1=\left(\begin{array}{cc}1&0\\0&-1\end{array}\right)\;,\quad g_2=\left(\begin{array}{cc}0&1\\1&0\end{array}\right)\; . \label{g1g2}
 \end{equation}
These matrices act simultaneously on all four pairs of homogeneous coordinates. The quotient $\tilde{X}=X/\Gamma$ is a Calabi-Yau manifold with Hodge numbers $h^{1,1}(\tilde{X})=4$ (since all four Kahler forms $J_i$ are $\Gamma$-invariant) and $h^{2,1}(\tilde{X})=20$ (using divisibility of the Euler number).

All line bundles on the tetra-quadric can be obtained as restriction of line bundles on ${\cal A}$, that is
\begin{equation}
 {\cal O}_X({\bf k})={\cal O}_{{\cal A}}({\bf k})|_X\; .
\end{equation} 
As discussed in Section~\ref{coboundary}, the Koszul sequence and its associated long exact sequence provide a close relationship between line bundle cohomology on ${\cal A}$ and $X$ which is summarised by Eq.~\eqref{H1eq}. This equation shows that the cohomology of a line bundle $K={\cal O}_X({\bf k})$  depends on the first and second cohomologies of the ambient space line bundles ${\cal K}={\cal O}_{\cal A}({\bf k})$ and ${\cal N}^*\otimes{\cal K}={\cal O}_{\cal A}({\bf k}-{\bf q})$. As discussed earlier, line bundles on ${\cal A}$ have at most one non-vanishing cohomology and, hence, ${\cal K}$ and ${\cal N}^*\otimes{\cal K}$ have at most one non-zero cohomology each. This leads to the following four cases:
\begin{enumerate}
\item[1)]\underline{$H^2 ({\cal A}, {\cal N}^* \otimes {\cal K})=0$  and $H^2 ({\cal A}, {\cal K})=0$}\\
In this case, $H^1 (X, K)$ is given by $(0,1)$-forms $\a_{{\bf k}}$, as in Eq.~\eqref{akres}, with associated polynomials
$P_{({\bf k})}$ and, in the terminology of Section~\ref{coboundary}, the cohomology representatives are of type 1. If $H^1({\cal A},{\cal N}^*\otimes {\cal K})$ is non-trivial we have to compute the co-kernel in Eq.~\eqref{H1eq} which amounts to imposing the identification $\tilde{P}_{({\bf k})} \sim \tilde{P}_{({\bf k})} +\tilde{p} \tilde{Q}_{({\bf k}- {\bf q})}$ for arbitrary polynomials $\tilde{Q}_{({\bf k}- {\bf q})}$ of multi-degree ${\bf k}-{\bf q}$. Recall that the tilde denotes the homogeneous version of the polynomials and that coordinates appearing with positive degree have to be converted into derivatives whenever they act on negative degree coordinates, as discussed at the end of the last sub-section. Since the coefficients of $p$ depend on the complex structure, this identification leads to complex structure dependence of the representatives.
\item[2)] \underline{$H^1 ({\cal A}, {\cal N}^* \otimes {\cal K})=0$  and $H^1 ({\cal A}, {\cal K})=0$}\\
In this case, $H^1 (X, K)$ is represented by $(0,2)$-forms $\a_{{\bf k}- {\bf q}}$, with  associated polynomials $P_{({\bf k}-{\bf q})}$, satisfying 
$p  \a_{{\bf k}- {\bf q}}= {\bar \pt} \b_{{\bf k}}$ for a suitable $(0,1)$-form $\b_{{\bf k}}$. Using the terminology of Section~\ref{coboundary}, this corresponds to type 2 representatives. If $H^2 ({\cal A}, {\cal K})\neq 0$, we have to work out the kernel in Eq.~\eqref{H1eq} which amounts to imposing the condition $\tilde{p}\tilde{P}_{({\bf k}-{\bf q})}=0$. This leads to explicit complex structure dependence of the representatives.
\item[3)] \underline{$H^1 ({\cal A}, {\cal N}^* \otimes {\cal K})=0$  and $H^2 ({\cal A}, {\cal K})=0$}\\
This is a combination of the previous two cases where $H^1 (X, K)$ is a direct sum of type 1 and type 2 contributions.
\item[4)] \underline{$H^2 ({\cal A}, {\cal N}^* \otimes {\cal K})=0$   and $H^1 ({\cal A}, {\cal K})=0$}\\
In this case $H^1(X, K)=0$. 
\end{enumerate}


\section{Yukawa couplings on the tetra-quadric and some toy examples}\label{toyex}


We have now collected all relevant technical details on line bundles and harmonic bundle-valued forms on the tetra-quadric and are ready to apply these to concrete calculations of Yukawa couplings. To begin we derive some general statements on Yukawa couplings on the tetra-quadric - including the precise relation between the analytic calculation of the integral and a corresponding algebraic calculation - and then move on to work out Yukawa couplings for a number of toy examples. In the next section, we compute the Yukawa couplings for a quasi-realistic standard model on the tetra-quadric.


\subsection{General properties of Yukawa couplings}
\label{comments}
As we have discussed earlier, we can distinguish two types of harmonic bundle-valued $(0,1)$-forms on the tetra-quadric: forms of type 1 which descend from harmonic $(0,1)$-form on the ambient space and forms of type 2 which descend from harmonic $(0,2)$-forms on the ambient space. 
The Yukawa couplings involve three harmonic $(0,1)$-forms and, as shown in Section~\ref{coboundary}, their structure depends on the types of these $(0,1)$-forms.

Let us consider a line bundle model on the tetra-quadric, specified by line bundles $L_a$, where $a=1,\ldots ,n$, and a 
Yukawa coupling with three associated line bundles $K_1={\cal O}_X({\bf k}_1)$, $K_2={\cal O}_X({\bf k}_2)$ and $K_3={\cal O}_X({\bf k}_3)$, 
which are related to $L_a$ as in Table~\ref{tab:KLrel}. Consider three harmonic $(0,1)$-forms $\nu_i\in H^1(X,K_i)$. 
We have seen that the Yukawa coupling vanishes if these three forms are of type 1. 
The next simplest case, when two of the forms, say $\nu_1$ and $\nu_2$, are of type 1 and 
descend from ambient space harmonic $(0,1)$-forms $\hat{\nu}_1\in H^1({\cal A},{\cal O}_{\cal A}({\bf k}_1))$ 
and $\hat{\nu}_2\in H^1({\cal A},{\cal O}_{\cal A}({\bf k}_2))$ while $\nu_3$ is of type 2 and descends from a harmonic 
ambient space $(0,2)$-form $\hat{\omega}_3\in H^2 ({\cal A}, {\cal O}_X ({\bf k}_3 -{\bf q}))$, leads to the particularly simple formula
\be 
\l(\nu_1,\nu_2,\nu_3) = - \frac{1}{2 \pi i}\int_{{\mathbb C}^4} d^4 z  \wedge \hat{\nu}_1 \wedge  \hat{\nu}_2 \wedge  \hat{\omega}_3\,, 
\label{Yuk112copy}
\ee
for the Yukawa coupling. This follows from Eq.~\eqref{Yuk112} together with Eqs.~\eqref{10.2} which shows that the form $\mu$ is given by
\be
\mu = d z_1 \wedge dz_2 \wedge dz_3 \wedge dz_4=d^4z\, . 
\label{muform}
\ee
The integral over ${\cal A}$ can then be thought of as the integral over ${\mathbb C}^4$ provided the forms
$ \hat{\nu}_1,   \hat{\nu}_2,  \hat{\omega}_3$ transform to the other patches as sections of the appropriate line bundles. 
Since $\hat{\nu}_1$ and $\hat{\nu}_2$ are $(0,1)$-forms the vectors ${\bf k}_1$ and ${\bf k}_2$ should contain 
precisely one entry $\leq -2$ each while the vector ${\bf k}_3$ contains precisely two entries $\leq 0$, in line with $\hat{\o}_3$ 
being a $(0,2)$-form. Further, recall from Table~\ref{tab:KLrel} that ${\cal K}_1\otimes{\cal K}_2\otimes{\cal K}_3={\cal O}_{\cal A}$ and, 
hence, ${\bf k}_1+{\bf k}_2+{\bf k}_3=0$. This means that the four non-positive entries in these vector must all arise in different $\mathbb{P}^1$ directions. 
Hence, we can assume, possibly after re-ordering, that $k_1^1 \leq -2$, $k_2^2 \leq -2$ and $k_3^3, k_3^4 \leq 0$ while all other entries are non-negative. 
With these conventions, we can apply Eq.~\eqref{akres} to write down the relevant forms as
\be 
\hat{\nu}_1 = \k_1^{k_1^1} P_{({\bf k}_1)} d {\bar z}_1\,, \qquad 
\hat{\nu}_2 = \k_2^{k_2^2} R_{({\bf k}_2)} d {\bar z}_2\,, \qquad 
\hat{\o}_3 = \k_3^{k_3^3 -2 } \k_4^{k_3^4 -2 } T_{({\bf k}_3  -{\bf q} )} d {\bar z}_3 \wedge  d {\bar z}_4\,. 
\label{3.9}
\ee
Inserting these forms into Eq.~\eqref{Yuk112copy} leads to the integral
\be 
\l(\nu_1,\nu_2,\nu_3)  = - \frac{1}{2 \pi i}\int_{{\mathbb C}^4} d^4 z \ d^4 {\bar z}  \ \k_1^{k_1^1} \k_2^{k_2^2}\k_3^{k_3^3 -2 } \k_4^{k_3^4 -2 } 
P_{({\bf k}_1)} R_{({\bf k}_2)} T_{({\bf k}_3  -{\bf q} )}  \,. 
\label{Yuk112spec}
\ee
There are two ways of evaluating this integral. Firstly, we can explicitly insert the factors $\kappa_i=1+|z_i|^2$ and the polynomials and simply integrate, using polar coordinates in each $\mathbb{C}$ plane. All terms with non-matching powers of $z_i$ and $\bar{z}_i$ vanish due to the angular integration. The remaining terms all reduce to the standard integrals
\begin{equation}
 \int_{\mathbb{C}}\frac{|z|^{2q}}{\kappa^p}dz\,d\bar{z}=2\pi i I_{p,q}\;,\qquad I_{p,q}=2\int_0^\infty dr\frac{r^{2q+1}}{(1+r^2)^p}=\frac{q!}{(p-1)\cdots (p-q-1)}\; .\label{stdint}
\end{equation}
Alternatively, we can work out the integral~\eqref{Yuk112spec} ``algebraically". To do this we first note that the integrand $\hat{\nu}_1 \wedge  \hat{\nu}_2 \wedge  \hat{\omega}_3$ represents an element of the one-dimensional cohomology $H^4({\cal A},{\cal N}^*)$. It can, therefore, be written as  $\mu (P, R, T)\k_1^{-2} \k_2^{-2} \k_3^{-2} \k_4^{-2} d^4 {\bar z}$ where
\be
 \mu (P, R, T)= \tilde{P} \tilde{R} \tilde{T}  \label{Yukalg}
\ee
is the product of the three associated polynomials (carried out as discussed in Section~\ref{maps1}) and simply a complex number. Inserting this into Eq.~\eqref{Yuk112copy} shows that 
\be 
\l(\nu_1,\nu_2,\nu_3) =  8 i \pi^3  c \mu (P, R, T)\; , \label{lmurel}
\ee
where the numerical factor $c$ follows from Eq.~\eqref{prodsol} and is explicitly given by
\be 
c=c_{k_1^1, -k_1^1-2} \ c_{k_2^2, -k_2^2-2}\  c_{k_3^3-2, -k_3^3} \ c_{k_3^4-2, -k_3^4}\, . \label{cgen}
\ee
In conclusion, up to an overall numerical (and explicitly computed) factor, the Yukawa couplings are simply given by Eq.~\eqref{Yukalg} and can, hence, be obtained by a multiplication of the associated polynomials.

In the general case,  the Yukawa couplings are given by the integral~\eqref{Yukamb1} which can be written as
\be 
\l(\nu_1,\nu_2,\nu_3) =- \frac{1}{2 \pi i}\int_{{\mathbb C}^4} d^4 z \wedge [\hat{\o}_1 \wedge \hat{\nu}_2 \wedge \hat{\nu}_3-
 \hat{\nu}_1 \wedge \hat{\o}_2 \wedge \hat{\nu}_3 +\hat{\nu}_1 \wedge  \hat{\nu}_2 \wedge  \hat{\o}_3 ]\,. 
 \label{Yukgen4}
 \ee
with the $(0,1)$-forms $\hat{\nu}_i$ and the $(0,2)$-forms $\hat{\o}_i$ in this expression related by
\be
{\bar \pt} \hat{\nu}_i=p \hat{\o}_i \,. 
\ee
If the Yukawa coupling depends on more than one form of type 2 we have to solve this last equation for some of the $\hat{\nu}_i$ in terms of $\hat{\omega}_i$. This can be done explicitly for specific examples, as we will demonstrate later, but as discussed in Section~\ref{maps}, we are currently not aware of a general solution.


\subsection{An example with vanishing Yukawa couplings}
\label{vanishing}


We would like to consider a rank four line bundle sum on the tetra-quadric specified by the line bundles 
\be
L_1= {\cal O}_X  (-1, 0, 0, 1)\,, \quad L_2= {\cal O}_X  (0, -2, 1, 3)\,, \quad L_3= {\cal O}_X  (0, 0, 1, -3)\,. \quad
L_4= {\cal O}_X  (1, 2, -2, -1)\,, 
\ee
This bundle leads to a four-dimensional theory with gauge group $SO(10)\times S(U(1)^4)$. Table~\ref{tab:so10} contains the basic information required to determine the multiplet content of such a theory and together with the cohomology results
\be
\begin{array}{lll}
h^{\bullet} (X, L_2)= (0, 8, 0, 0)&h^{\bullet} (X, L_3)= (0, 4, 0, 0)&h^{\bullet} (X, L_1 \otimes L_4)= (0, 3, 3, 0)\\
h^{\bullet} (X, L_2 \otimes L_3)= (0, 3, 3, 0)&h^{\bullet} (X, L_1 \otimes L_2^*)= (0, 0, 12, 0)&h^{\bullet} (X, L_1 \otimes L_3^*)= (0, 0, 12, 0)\\ 
h^{\bullet} (X, L_2 \otimes L_3^*)= (0, 7, 15, 0)&h^{\bullet} (X, L_2 \otimes L_4^*)= (0, 60, 0, 0)&h^{\bullet} (X, L_3 \otimes L_4^*)= (0, 0, 36, 0)
\end{array}
\ee
we find the upstairs spectrum
\be 
8 \ {\bf 16}_2\,, \  4 \ {\bf 16}_3\,, \ 3 \ {\bf 10}_{1,4}\,, \ 3 \ {\bf 10}_{2,3}\,,  \ 12 \ {\bf 1}_{2,-1}\,, \ 12 \ {\bf 1}_{3,-1}\,, \ 
7 \ {\bf 1}_{2,-3}\,, \ 15 \ {\bf 1}_{3,-2}\,, \ 60 \ {\bf 1}_{2,-4}\,, \ 36 \ {\bf 1}_{4,-3}\,.
\ee
This spectrum is designed to produce a standard-model with three families upon dividing by a freely-acting symmetry of order four. Such symmetries are indeed available for the tetra-quadric however, unfortunately, for group-theoretical reasons these symmetries cannot break the $SO(10)$ gauge group to the standard model group. For this reason, the above model should be considered a toy example.

Nevertheless, it is useful to calculate the Yukawa couplings for this model, in order to gain some experience with our formalism. Specifically, we are interested in couplings of the type
\be 
\l_{IJK} {\bf 10}^{(I)}_{1, 4} {\bf 16}^{(J)}_{2} {\bf 16}^{(K)}_{3} \,. 
\ee
which are allowed by the $SO(10)\times S(U(1)^4)$ gauge symmetry. Following Table~\ref{tab:KLrel}, the required harmonic forms are contained in the first cohomologies of the line bundles
\be
K_1= L_1 \otimes L_4 = {\cal O}_X (0, 2, -2, 0)\,,\quad K_2= L_2 ={\cal O}_X  (0, -2, 1, 3)\,, \quad
K_3= L_3 ={\cal O}_X  (0, 0, 1, -3)\,. 
\ee
These line bundles satisfy $H^1 (X, K_i)\cong H^1 ({\cal A}, {\cal K}_i)$ and $H^2 ({\cal A}, {\cal N}^* \otimes {\cal K}_i)=0$, where ${\cal  K}_i$ are the corresponding ambient space line bundles with $K_i = {\cal K}_i|_X$. This shows (see Section~\ref{coboundary}) that all three harmonic forms which enter the Yukawa integral are of type 1. From our general arguments this means that the Yukawa couplings vanish, so
\be
 \l_{IJK} =0\; .
\ee 
Note that this vanishing is, apparently, not caused by a symmetry in the low-energy theory but happens due to quasi-topological reasons related to the cohomology of the line bundles involved. (However, we do not rule out that a symmetry which explains this vanishing result may be found.)

\subsection{An $E_6$ example} \label{E6example}
For a simple example with gauge group $E_6\times S(U(1)^3)$ consider the following choice of line bundles
\be
 L_1=K_1={\cal O}_X(-2,0,1,0)\,,\quad L_2=K_2={\cal O}_X(0,-2,0,1)\,,\quad L_3=K_3={\cal O}_X(2,2,-1,-1)\; .
\ee
These line bundles $K_i$ may also arise as appropriate tensor products for other gauge groups, see Table~\ref{tab:KLrel}, and the subsequent calculation also applies to these cases. However, for definiteness we will focus on $E_6\times S(U(1)^3$ and the corresponding multiplets, as summarised in Table~\ref{tab:e6}. The cohomology results
\be
h^\bullet(K_1)=(0,2,0,0)\,,\quad h^\bullet(K_2)= (0,2,0,0)\,,\quad h^\bullet(K_3)=(0,4,0,0)
\ee
show that we have a spectrum
\be
2\; {\bf 27}_1\,,\;2\;{\bf 27}_2\,,\;4\;{\bf 27}_3
\ee
plus $E_6$ singlets which are irrelevant to the present discussion. We are interested in the Yukawa couplings
\be
 \l_{IJK}{\bf 27}_1^{(I)}\,{\bf 27}_2^{(J)}\,{\bf 27}_3^{(K)}\; .
\ee 
Clearly, the first two line bundles are of type 1 with the corresponding harmonic $(0,1)$-forms contained in $H^1({\cal A},{\cal K}_1)$ and $H^1({\cal A},{\cal K}_2)$. However, $K_3$ is of type two and the associated harmonic $(0,2)$-forms represent the cohomology $H^2({\cal A},{\cal N}^*\otimes{\cal K}_2)$. Altogether, using Eq.~\eqref{akres}, this means the relevant harmonic forms and polynomials are
\begin{equation}
\begin{array}{lllllll}
 \hat{\nu}_1&=&\kappa_1^{-2}P_{(-2,0,1,0)}d\bar{z}_1&\quad&P_{(-2,0,1,0)}&=&p_0+p_1z_3\\
 \hat{\nu}_2&=&\kappa_2^{-2}Q_{(0,-2,0,1)}d\bar{z}_2&\quad&Q_{(0,-2,0,1)}&=&q_0+q_1z_4\\
 \hat{\omega}_3&=&\kappa_3^{-3}\kappa_4^{-3}R_{(0,0,-3,-3)}d\bar{z}_3\wedge d\bar{z}_4&\quad&
 R_{(0,0,-3,-3)}&=&r_0+r_1\bar{z}_3+r_2\bar{z}_4+r_3\bar{z}_3\bar{z}_4
 \end{array}
 \end{equation}
where $p_I$, $q_I$ and $r_I$ are complex coefficients parametrising the various ${\bf 27}$ multiplets. Multiplying the three polynomials and discarding terms with different powers of $z_i$ and $\bar{z}_i$ gives
\begin{equation}
 PQR=p_0q_0r_0+p_0q_1r_2|z_4|^2+p_1q_0r_1|z_3|^2+p_1q_1r_3|z_3|^2|z_4|^2+\mbox{ non-matching terms}\; .
\end{equation}
This can be directly inserted into the integral~\eqref{Yuk112spec} and together with the standard integrals~\eqref{stdint} (specifically, $I_{2,0}=1$, $I_{3,0}=1/2$, $I_{3,1}=1/2$) we find
\begin{equation}
 \lambda(P,Q,R)=2 i \pi^3\left(p_0q_0r_0+p_0q_1r_2+p_1q_0r_1+p_1q_1r_3\right)\; . \label{exresint1}
\end{equation} 
Alternatively, we can use the algebraic calculation method based on Eq.~\eqref{Yukalg}. For simplicity of notation, we denote the four sets of homogenous ambient space coordinates by
\begin{equation}
 (x_i^\a)=((x_0,x_1),(y_0,y_1),(u_0,u_1),(v_0,v_1))\; .
\end{equation} 
Then, the homogenous versions of the three polynomials read explicitly
\begin{equation}
 \tilde{P}=p_0u_{0}+p_1u_{1}\;,\quad \tilde{Q}=q_0v_{0}+q_1v_{1}\;,\quad \tilde{R}=r_0\bar{u}_{0}\bar{v}_{0}+r_1\bar{v}_{0}\bar{u}_{1}+r_2\bar{u}_{0}\bar{v}_{1}+r_3\bar{u}_{1}\bar{v}_{1}\; .
\end{equation} 
Their product is given by
\begin{eqnarray}
 \mu(P,Q,R)&=&\left(p_0{\partial}_{\bar{u}_0}+p_1{\partial}_{\bar{u}_1}\right)\left(q_0{\partial}_{\bar{v}_0}+q_1{\partial}_{\bar{v}_1}\right)
 \left(r_0\bar{u}_{0}\bar{v}_{0}+r_1\bar{v}_{0}\bar{u}_{1}+r_2\bar{u}_{0}\bar{v}_{1}+r_3\bar{u}_{1}\bar{v}_{1}\right)\\
 &=&p_0q_0r_0+p_0q_1r_2+p_1q_0r_1+p_1q_1r_3\; ,
\end{eqnarray} 
where we have converted the coordinates in $\tilde{P}$ and $\tilde{Q}$ into derivatives, as required by our general rules. Inserting the correct numerical coefficient from Eqs.~\eqref{lmurel} and \eqref{cgen} this indeed coincides with the result~\eqref{exresint1} from direct evaluation of the integral. If we choose a standard basis where one each of the coefficients $p_I$, $q_I$ and $r_I$ equals one while all others vanish we can write down the explicit Yukawa matrices
\begin{equation}
( \l_{1JK})=2 i \pi^3 \left(\begin{array}{llll}1&0&0&0\\0&0&1&0\end{array}\right)\,,\quad
( \l_{2JK})=2 i \pi^3 \left(\begin{array}{llll}0&1&0&0\\0&0&0&1\end{array}\right)\; .
\end{equation}
Both matrices have maximal rank and are independent of complex structure.

\subsection{An example with complex structure dependence}\label{csexample}
We would like to discuss the Yukawa couplings related to the three line bundles
\be
 K_1={\cal O}_X(0,-2,1,1)\,,\quad K_2={\cal O}_X(-4,0,1,1)\,,\quad K_3={\cal O}_X(4,2,-2,-2)\; ,
\ee 
with cohomologies
\be
 h^\bullet(K_1)=(0,4,0,0)\,,\quad h^\bullet(K_2)=(0,12,0,0)\,,\quad h^\bullet(K_3)=(0,12,0,0)\; .
\ee 
It will be convenient to think about this situation as arising from an $SU(5)\times S(U(1)^5)$ model, defined by five line bundles $L_a$, with $K_1=L_1\otimes L_2$ and $K_2=L_3\otimes L_4$ and $K_3=L_5$. Then, using the correspondence from Table~\ref{tab:KLrel} the $SU(5)\times S(U(1))$ spectrum related to $K_1$, $K_2$ and $K_3$ is
\be
 4\;\overline{\bf 5}_{1,2}\;,\quad12\;\overline{\bf 5}_{3,4}\;,\quad 12\;{\bf 10}_5\; .
\ee 
We will later introduce a $\mathbb{Z}_2\times\mathbb{Z}_2$ Wilson line to break to the standard model group in which case, as we will see, the above spectrum reduces to
\be
 H_{1,2}\;,\quad 3\;d_{3,4}\;,\quad 3\;Q_5\; .
\ee
We are interested in computing the d-quark Yukawa couplings
\begin{equation}
 \l^{(d)}_{JK}H_{1,2}d_{3,4}^JQ_5^K\; .
\end{equation}   
However, for now we construct the relevant bundle-valued forms in the upstairs theory and restrict to the $\mathbb{Z}_2\times\mathbb{Z}_2$-quotient later. The line bundles $K_1$ and $K_2$ are both of type one with $H^1(X,K_1)\cong H^1(X,{\cal K}_1)$ and $H^1(X,K_2)\cong H^1(X,{\cal K}_2)$ while $K_3$ is of type 2 and
\be 
H^1(X,K_3)\cong {\rm Ker}(H^2({\cal A},{\cal N}^*\otimes{\cal K}_3)\stackrel{p}{\rightarrow }H^2({\cal A},{\cal K}_3) )\; . \label{ex3ker}
\ee
Hence, following Eq.~\eqref{akreshom}, the relevant ambient space forms and polynomials can be written in terms of homogenous coordinates as
\begin{equation}
\arraycolsep=1.2pt\def\arraystretch{1.2}
\begin{array}{lllllllll}
4\;\overline{\bf 5}_{1,2}&\;\longrightarrow\;&\hat{\nu}_1&=&\sigma_2^{-2}\tilde{Q}_{(0,-2,1,1)}\bar{\mu}_2&\quad& \tilde{Q}&\in&{\rm Span}(u_0v_0,u_0v_1,u_1v_0,u_1v_1)\\
 12\;\overline{\bf 5}_{3,4}&\;\longrightarrow\;&\hat{\nu}_2&=&\sigma_1^{-4}\tilde{R}_{(-4,0,1,1)}\bar{\mu}_1&\quad& \tilde{R}&\in&{\rm Span}(\bar{x}_0^2,\bar{x}_0\bar{x}_1,\bar{x}_1^2)\,{\rm Span}(u_0,u_1)\,{\rm Span}(v_0,v_1)\\
 12\;{\bf 10}_5&\;\longrightarrow\;&\hat{\omega}_3&=&\sigma_3^{-2}\sigma_4^{-2}\tilde{S}_{(2,0,-4,-4)}\bar{\mu}_3\wedge \bar{\mu}_4&\quad&\tilde{S}&\in&{\rm Span}(x_0^2,x_0x_1,x_1^2)\,{\rm Span}(\bar{u}_0^2,\bar{u}_0\bar{u}_1,\bar{u}_1^2)\\
 &&&&&&&&{\rm Span}(\bar{v}_0^2,\bar{v}_0\bar{v}_1,\bar{v}_1^2)\; .
 \end{array}
 \end{equation}
The polynomial $\tilde{S}$ lies in a $27$-dimensional space which, in line with Eq.~\eqref{ex3ker}, is mapped into the $15$-dimensional space
\begin{equation}
 {\rm Span}(x_0^4,x_0^3x_1,x_0^2x_1^2,x_0x_1^3,x_1^4)\,{\rm Span}(y_0^2,y_0y_1,y_1^2)\; .
\end{equation}
We have to ensure that $\tilde{S}$ resides in the kernel of this map which amounts to imposing the condition
\be
 \tilde{p}\tilde{S}=0\; . \label{pS0}
\ee
This leads to a $12$-dimensional space, as expected.

These results are quite complicated due to the large number of multiplets. To simplify matters, it is useful to quotient by the freely-acting $\Gamma=\mathbb{Z}_2\times\mathbb{Z}_2$ symmetry with generators~\eqref{g1g2}. Representations of this symmetry are denoted by a pair of charges, $(q_1,q_2)$, where $q_i\in\{0,1\}$.  We choose a trivial equivariant structure for all line bundles and, following the discussion around Eq.~\eqref{WLcharges}, a Wilson line specified by $\chi_2=(1,1)$, $\chi_3=(0,0)$ with associated multiplet charges
\begin{equation}
 \chi_H=\chi_2^*=(1,1)\;,\quad \chi_d=\chi_3^*=(0,0)\;,\quad \chi_Q=\chi_2\otimes \chi_3=(1,1)\; .
\end{equation}
Taking into account that the differentials $\mu_i$ carry charge $(1,1)$ under the $\mathbb{Z}_2\times\mathbb{Z}_2$ symmetry, this choice means we should project onto the $(0,0)$ states for $\tilde{Q}$, and the $(1,1)$ states for $\tilde{R}$ and $\tilde{S}$. This leads to to the explicit $\mathbb{Z}_2\times\mathbb{Z}_2$-equivariant polynomials
\begin{eqnarray}
 \tilde{Q}&=&u_0v_0+u_1v_1\\
 \tilde{R}&=&a_3 \left(u_0 v_0 \bar{x}_0 \bar{x}_1-u_1 v_1 \bar{x}_0 \bar{x}_1\right)+a_1 \left(u_0
   v_1 \bar{x}_0^2-u_1 v_0 \bar{x}_1^2\right)+a_2 \left(u_1 v_0 \bar{x}_0^2-u_0 v_1
   \bar{x}_1^2\right)\\
   \tilde{S}&=&b_4 \left(x_0^2 \bar{u}_1^2 \bar{v}_0 \bar{v}_1-x_1^2 \bar{u}_0^2 \bar{v}_0
   \bar{v}_1\right)+b_1 \left(x_0^2 \bar{u}_0^2 \bar{v}_0 \bar{v}_1-x_1^2 \bar{u}_1^2
   \bar{v}_0 \bar{v}_1\right)+b_6 \left(x_0 x_1 \bar{u}_0^2 \bar{v}_1^2-x_0 x_1
   \bar{u}_1^2 \bar{v}_0^2\right)+\nonumber\\
   &&b_3 \left(x_0^2 \bar{u}_0 \bar{u}_1 \bar{v}_1^2-x_1^2
   \bar{u}_0 \bar{u}_1 \bar{v}_0^2\right)+b_2 \left(x_0^2 \bar{u}_0 \bar{u}_1
   \bar{v}_0^2-x_1^2 \bar{u}_0 \bar{u}_1 \bar{v}_1^2\right)+b_5 \left(x_0 x_1 \bar{u}_0^2
   \bar{v}_0^2-x_0 x_1 \bar{u}_1^2 \bar{v}_1^2\right)\; .
\end{eqnarray} 
Hence, we are left with a single Higgs multiplet, $H_{1,2}$, three d-quarks, $d_{3,4}^J$, with parameters ${\bf a}=(a_I)$ and six left-handed quarks $Q_5^J$ with parameters ${\bf b}=(b_J)$. In terms of these parameters, the Yukawa couplings are given by
\begin{equation}
 \mu(Q,R,S)=\tilde{Q}\tilde{R}\tilde{S}=8 \left(a_1 \left(b_1+b_3\right)+a_2 \left(b_2+b_4\right)+a_3 b_5\right)\; . \label{yukex3}
\end{equation} 
However, for the ``physical" result we still have to find the kernel~\eqref{ex3ker}, that is, compute the vectors ${\bf b}$ which satisfy Eq.~\eqref{pS0}. To this end, we write down the most general tetra-quadric polynomial consistent with the $\Gamma=\mathbb{Z}_2\times \mathbb{Z}_2$ symmetry.
\begin{eqnarray}
\tilde{p}&=&C_1 u_0 u_1 v_0 v_1 x_0 x_1 y_0 y_1+C_2 (u_1^2 x_0 x_1 y_0 y_1 v_0^2+u_0^2 v_1^2 x_0x_1 y_0 y_1)+\nonumber\\
   &&C_3 (u_0^2 x_0 x_1 y_0 y_1 v_0^2+u_1^2 v_1^2 x_0 x_1 y_0 y_1)+C_{14}(u_0 u_1 v_1^2 y_0 y_1 x_0^2+u_0 u_1 v_0^2 x_1^2 y_0 y_1)+\nonumber\\
   &&C_{13} (u_1^2 v_0 v_1 y_0 y_1 x_0^2+u_0^2 v_0 v_1 x_1^2 y_0y_1)+C_{16} (u_0^2 v_0 v_1 y_0 y_1 x_0^2+u_1^2 v_0 v_1 x_1^2 y_0 y_1)+\nonumber\\
   &&C_{15} (u_0 u_1 v_0^2 y_0 y_1 x_0^2+u_0 u_1 v_1^2 x_1^2 y_0 y_1)+C_{12} (u_1^2 v_1^2 x_1^2 y_0^2+u_0^2 v_0^2 x_0^2 y_1^2)+\nonumber\\
   &&C_9(u_0^2 v_1^2 x_1^2 y_0^2+u_1^2 v_0^2 x_0^2 y_1^2)+C_{10} (u_0 u_1 v_0 v_1 x_1^2 y_0^2+u_0 u_1 v_0 v_1 x_0^2 y_1^2)+\nonumber\\
   &&C_{11}(u_1^2 v_0^2 x_1^2 y_0^2+u_0^2 v_1^2 x_0^2 y_1^2)+ C_8 (u_0^2 v_0^2 x_1^2 y_0^2+u_1^2 v_1^2  x_0^2 y_1^2)+\nonumber\\
   &&C_5(u_0 u_1 v_1^2 x_0 x_1 y_0^2+u_0 u_1 v_0^2 x_0 x_1 y_1^2)+C_4 (u_1^2 v_0 v_1 x_0 x_1 y_0^2+u_0^2 v_0 v_1 x_0 x_1y_1^2)+\nonumber\\
   &&C_7(u_0^2 v_0 v_1 x_0 x_1 y_0^2+u_1^2 v_0 v_1 x_0 x_1 y_1^2)+C_6 (u_0 u_1 v_0^2 x_0 x_1 y_0^2+u_0 u_1 v_1^2 x_0 x_1y_1^2)+\nonumber\\
   &&C_{17} (u_1^2 v_1^2 x_0^2 y_0^2+u_0^2 v_0^2 x_1^2 y_1^2)+C_{20} (u_0^2 v_1^2 x_0^2 y_0^2+u_1^2 v_0^2 x_1^2y_1^2)+\nonumber\\
   &&C_{19}(u_0 u_1 v_0 v_1 x_0^2 y_0^2+u_0 u_1 v_0 v_1 x_1^2 y_1^2)+C_{18} (u_1^2 v_0^2 x_0^2 y_0^2+u_0^2 v_1^2 x_1^2 y_1^2)+\nonumber\\
   &&C_{21} (u_0^2 v_0^2 x_0^2 y_0^2+u_1^2 v_1^2 x_1^2 y_1^2)\label{pgen}
\end{eqnarray}
The dimension of the complex structure moduli space for $\tilde{X}=X/\mathbb{Z}_2\times \mathbb{Z}_2$ is given by $h^{2,1}(\tilde{X})=20$. The $21$ coefficients $C_i$ in the above polynomial provide projective (local) coordinates on this moduli space. Using this polynomial, Eq.~\eqref{pS0} is solved by vectors ${\bf b}$ satisfying
\be
 M{\bf b}=0\,,\qquad 
 M=\left(
\begin{array}{cccccc}
 \frac{C_{16}}{2} & \frac{C_{15}}{2} & \frac{C_{14}}{2} & \frac{C_{13}}{2} & 0 & 0 \\
 \frac{C_7}{2} & \frac{C_6}{2} & \frac{C_5}{2} & \frac{C_4}{2} & C_{21}-C_{17} &  C_{20}-C_{18} \\
 \frac{C_4}{2} & \frac{C_5}{2} & \frac{C_6}{2} & \frac{C_7}{2} & C_{12}-C_8 & C_{11}-C_9  \\
\end{array}
\right)\; .
\ee
The matrix $M$ has indeed a (generically) three-dimensional kernel but its basis vectors ${\bf v}_I$, where $I=1,2,3$, are very complicated functions of the complex structure moduli. In principle, this basis can be computed and ${\bf b}$ can then be written as
\begin{equation}
 {\bf b}=\sum_{I}\beta_I{\bf v}_I \label{bspec}
\end{equation}
where the three $\beta_I$ now parametrize  the three left-handed quark families. Inserting this result into Eq.~\eqref{yukex3} gives the desired result for the Yukawa couplings and it can be shown that the rank of the Yukawa matrix $\l_{IJ}^{(d)}$ is three at generic loci in the complex structure moduli space.

In order to obtain a more explicit result, we restrict to a five-dimensional sub-locus of our $20$-dimensional complex structure moduli space, described by polynomials of the form
\begin{eqnarray}
\tilde{p}_s&=&c_1 u_0 u_1 v_0 v_1 x_0 x_1 y_0 y_1+c_2 (u_0^2 v_0 v_1 x_0^2 y_0 y_1+u_1^2 v_0 v_1x_0^2 y_0 y_1+u_0 u_1 v_0^2 x_1 x_0 y_0^2+\nonumber\\
   &&u_0 u_1 v_1^2 x_1 x_0 y_0^2+u_0 u_1 v_0^2   x_1 x_0 y_1^2+u_0 u_1 v_1^2 x_1 x_0 y_1^2+u_0^2 v_0 v_1 x_1^2 y_0 y_1+u_1^2 v_0 v_1
   x_1^2 y_0 y_1)+\nonumber\\
   &&c_5 (u_0^2 v_1^2 x_0^2 y_0^2+u_0^2 v_0^2 x_1^2 y_0^2+u_1^2
   v_1^2 x_0^2 y_1^2+u_1^2 v_0^2 x_1^2 y_1^2)+c_4 (u_0^2 v_0^2 x_0 x_1 y_0
   y_1-u_1^2 v_0^2 x_0 x_1 y_0 y_1+\nonumber\\
   &&u_0 u_1 v_1 v_0 x_0^2 y_0^2-u_0 u_1 v_1 v_0 x_1^2y_0^2-u_0 u_1 v_1 v_0 x_0^2 y_1^2+u_0 u_1 v_1 v_0 x_1^2 y_1^2-u_0^2 v_1^2 x_0 x_1 y_0
   y_1+\nonumber\\
   &&u_1^2 v_1^2 x_0 x_1 y_0 y_1)+c_3(u_1^2 v_0^2 x_0^2 y_0^2+u_1^2 v_1^2 x_1^2 y_0^2+u_0^2 v_0^2 x_0^2 y_1^2+u_0^2 v_1^2 x_1^2 y_1^2)+c_6 (u_0^2
   v_0^2 x_0^2 y_0^2+\nonumber\\
   &&u_1^2 v_1^2 x_0^2 y_0^2+u_1^2 v_0^2 x_1^2 y_0^2+u_0^2 v_1^2 x_1^2 y_0^2+u_1^2 v_0^2 x_0^2 y_1^2+u_0^2 v_1^2 x_0^2 y_1^2+u_0^2 v_0^2 x_1^2 y_1^2+u_1^2v_1^2 x_1^2 y_1^2)\label{pspec}\; .
 \end{eqnarray}  
In fact, this polynomial is the most general consistent with the freely-acting $\mathbb{Z}_4\times\mathbb{Z}_4$ symmetry of the tetra-quadric which contains the $\mathbb{Z}_2\times \mathbb{Z}_2$ symmetry used previously as a sub-group. The equation $\tilde{p}_s\tilde{S}=0$ for the kernel now reads
\begin{equation}\label{Mex3}
 M{\bf b}=0\,,\qquad 
 M=\left(
\begin{array}{cccccc}
 c_2 & 0 & 0 & c_2 & 0 & 0 \\
 0 & c_2 & c_2 & 0 & 0 & 2 c_5-2 c_3 \\
 0 & c_2 & c_2 & 0 & 2 c_3-2 c_5 & 0 \\
\end{array}
\right)\; .
\end{equation}
Generically, the dimension of this kernel is three and a basis can be readily found as
\be
{\bf v}_1=\frac{1}{8}\left(0,2 \left(c_3-c_5\right),0,0,-c_2,c_2\right)^T\,,\;
{\bf v}_2=\frac{1}{8}\left(-c_2,0,0,c_2,0,0\right)^T\,,\;
{\bf v}_3=\frac{1}{8}\left(0,-c_2,c_2,0,0,0\right)^T\; .
\ee
Inserting these vectors into Eq.~\eqref{bspec} and \eqref{yukex3} and choosing a standard basis for the coefficients ${\bf a}$ and ${\boldsymbol\beta}$ then gives the Yukawa couplings
\begin{equation}
\lambda^{(d)}=i \pi^3c
\left(
\begin{array}{ccc}
 0 & -c_2 & c_2 \\
 2 c_3-2 c_5 & c_2 & -c_2 \\
 -c_2 & 0 & 0 \\
\end{array}
\right)\label{yukex4}
\end{equation}
where $c$ is the numerical factor from Eq.~\eqref{lmurel}. Evidently, the generic rank of this matrix is two. This shows that the rank of the Yukawa matrix can vary in complex structure moduli space and can reduce at specific loci. In the present case, it is generically of rank three in the $20$-dimensional complex structure moduli space described by the polynomials~\eqref{pgen}. On the five-dimensional sub-locus, described by the polynomials~\eqref{pspec}, the rank reduces to two.

If we specialise further to the four-dimensional locus where $c_2=0$ the rank of \eqref{yukex4} reduces to one. It turns out that the tetra-quadric~\eqref{pspec} remains generically smooth on this sub-locus. However, we have to be careful since the rank of the matrix M in Eq.~\eqref{Mex3} also depends on the complex structure. In fact, for $c_2=0$ the rank of $M$ reduces to two so that the dimension of the kernel increases from three to four. Hence, on this sub-locus the spectrum in the low-energy theory enhances from three left-handed quark multiplets to four (plus one mirror left-handed quark multiplet since the index remains unchanged). A basis of the kernel is then given by ${\bf v}_I={\bf e}_I/8$, where $I=1,\ldots ,4$ and ${\bf e}_I$ are the six-dimensional standard unit vectors. From Eq.~\eqref{bspec} and \eqref{yukex3} this leads to the Yukawa couplings
\be
\l^{(d)}=i \pi^3 c
 \left(
\begin{array}{cccc}
 1 & 0 & 1 & 0 \\
 0 & 1 & 0 & 1 \\
 0 & 0 & 0 & 0 \\
\end{array}
\right)
\ee
Hence, after properly including the additional multiplet the rank of the Yukawa matrix remains two.



\section{Yukawa couplings in a quasi-realistic model on the tetra-quadric}\label{realex}
In the previous section, we have applied our methods to a number of toy examples and we have seen cases with vanishing and non-vanishing Yukawa couplings, both with and without complex-structure dependence. We would now like to calculate Yukawa couplings in a quasi-realistic model on the tetra-quadric, that is, a model with gauge group $SU(3)\times SU(2)\times U(1)$ (plus additional $U(1)$ symmetries which are Green-Schwarz anomalous or can be spontaneously broken) and the exact MSSM spectrum (plus moduli fields uncharged under the standard model group, including bundle moduli singlets). 
This model appears in the standard model data base~\cite{Anderson:2011ns,Anderson:2012yf} and has been further analysed in 
Refs.~\cite{Buchbinder:2013dna, Buchbinder:2014qda,Buchbinder:2014sya,Buchbinder:2014qca}. 
We begin by reviewing the basic structure of this model and then calculate the two types of non-vanishing Yukawa couplings which arise, that is, the standard up-quark Yukawa couplings and the singlet Yukawa couplings of the form $SL\overline{H}$, with bundle moduli singlets $S$. 


\subsection{The model}
The upstairs model is based on a rank five line bundle sum, $V=\bigoplus_{a=1}^5L_a$, on the tetra-quadric, with the five line bundles explicitly given by
\begin{equation}\label{lbs5}
 \begin{array}{lllllllllll}
 L_1&=&{\cal O}_X(-1,0,0,1)&,&L_2&=&{\cal O}_X(-1,-3,2,2)&,&L_3&=&{\cal O}_X(0,1,-1,0)\\
 L_4&=&{\cal O}_X(1,1,-1,-1)&,&L_5&=&{\cal O}_X(1,1,0,-2)\; .
 \end{array}
\end{equation}
Hence, the low-energy GUT group is $SU(5)\times S(U(1)^5)$. The non-zero cohomologies of line bundles appearing in $V$, $\wedge^2V$ and $V\otimes V^*$ are
\begin{equation}
\begin{array}{lllllll}
 h^{^{\!\bullet}}(X,L_2)&=&(0,8,0,0)&,&h^{^{\!\bullet}}(X,L_5)&=&(0,4,0,0)\\[4pt]
 h^{^{\!\bullet}}(X,L_2\otimes L_4)&=&(0,4,0,0)&,&h^{^{\!\bullet}}(X,L_2\otimes L_5)&=&(0,3,3,0)\\[4pt]
 h^{^{\!\bullet}}(X,L_4\otimes L_5)&=&(0,8,0,0)&,&h^{^{\!\bullet}}(X,L_1\otimes L_2^*)&=&(0,0,12,0)\\[4pt]
 h^{^{\!\bullet}}(X,L_1\otimes L_5^*)&=&(0,0,12,0)&,&h^{^{\!\bullet}}(X,L_2\otimes L_3^*)&=&(0,20,0,0)\\[4pt]
 h^{^{\!\bullet}}(X,L_2\otimes L_4^*)&=&(0,12,0,0)&,&h^{^{\!\bullet}}(X,L_3\otimes L_5^*)&=&(0,0,4,0)\; .
\end{array} 
\end{equation}
Following Table~\ref{tab:su5}, these cohomologies give rise to the GUT spectrum
\begin{equation}
 8\, {\bf 10}_2\,,\; 4\,{\bf 10}_5\,,\;4\,\overline{\bf 5}_{2,4}\,,\;3\,\overline{\bf 5}_{2,5}^H\,,\;8\,\overline{\bf 5}_{4,5}\,,\; 3{\bf 5}_{2,5}^{\overline{H}}\,,\;
 12\,{\bf 1}_{2,1}\,,\;12\,{\bf 1}_{5,1}\,,\;20\,{\bf 1}_{2,3}\,,\;12\,{\bf 1}_{2,4}\,,\;4\,{\bf 1}_{5,3}\; .\label{gutspec}
\end{equation}
At the GUT level the only superpotential terms allowed by the gauge symmetry are
\begin{equation}
 W=\lambda_{IJK} {\bf 5}_{2,5}^{(I)}{\bf 10}_2^{(J)}{\bf 10}_5^{(K)}+\rho_{IJK} {\bf 1}_{2,4}^{(I)}\overline{\bf 5}_{4,5}^{(J)}{\bf 5}_{2,5}^{(K)}\; ,
 \label{Wgut}
\end{equation} 
where the indices $I,J,K\ldots $ run over various ranges, as indicated by the multiplicities in the spectrum~\eqref{gutspec} and $\lambda_{IJK}$ and $\rho_{IJK}$ are the couplings we would like to calculate.

Evidently, the above GUT model has $12$ families of quarks and leptons, three vector-like ${\bf 5}^{\overline{H}}$--$\overline{\bf 5}^H$ pairs, which can account for the Higgs multiplets, and a spectrum of bundle moduli singlets. This is a promising upstairs spectrum which may lead to a downstairs standard model upon dividing by a freely-acting symmetry of order four. This can indeed be accomplished using the $\mathbb{Z}_2\times\mathbb{Z}_2$ symmetry with generators~\eqref{g1g2}, a choice of Wilson line specified by $\chi_2=(0,1)$ and $\chi_3=(0,0)$ and a trivial equivariant structure for all line bundles. The relevant GUT multiplets branch as ${\bf 10}\rightarrow (Q,u,e)$, $\overline{\bf 5}\rightarrow (d,L)$, $\overline{\bf 5}^H\rightarrow (T,H)$ and  ${\bf 5}^{\overline{H}}\rightarrow (\bar{T},\bar{H})$ (where $T$ and $\bar{T}$ are the Higgs triplets, to be projected out). From Eq.~\eqref{WLcharges}, these standard model multiplets carry the Wilson line charges
 \be
 \begin{array}{lllll}
 \chi_Q=\chi_2\otimes\chi_3=(0,1)&\quad&\chi_u=\chi_3^2=(0,0)&\quad&\chi_e=\chi_2^2=(0,0)\\
 \chi_d=\chi_3^*=(0,0)&\quad&\chi_L=\chi_2^*=(0,1)&\quad&\chi_H=\chi_2^*=(0,1)\\
 \chi_{\overline{H}}=\chi_2=(0,1)&\quad&\chi_T=\chi_3^*=(0,0)&\quad&\chi_{\overline{T}}=\chi_3=(0,0)\; .
 \end{array}
\ee 
Applying the rule~\eqref{equivcoh} for this choice of charges then leads to the downstairs spectrum
\begin{equation}
 2\, (Q,u,e)_2\,,\; (Q,u,e)_5\,,\;(d,L)_{2,4}\,,\;2\,(d,L)_{4,5}\,,\; H_{2,5}\,,\;\overline{H}_{2,5}\,,\;
 3\,{\bf 1}_{2,1}\,,\;3\,{\bf 1}_{5,1}\,,\;5{\bf 1}_{2,3}\,,\;3\,{\bf 1}_{2,4}\,,\;{\bf 1}_{5,3}\; , \label{smspec}
\end{equation} 
a perfect MSSM spectrum plus additional bundle moduli singlets. Ordering the quarks as $(Q^{(I)})=(Q_5^1,Q_5^2,Q_2)$ and $(u^{(I)})=(u_5^1,u_5^2,u_2)$, the downstairs analogue of the superpotential~\eqref{Wgut} can be written as
\be
 W=\l_{IJ}^{(u)}\overline{H}_{2,5}u^{(I)}Q^{(J)}+\rho_{IJ} {\bf 1}_{2,4}^{(I)}L_{4,5}^{(J)}\overline{H}_{2,5}\; . \label{Wsm}
\ee
The up-Yukawa matrix $\l^{(u)}$ is further constrained by the $S(U(1)^5)$ symmetry and must be of the form
\begin{equation}
 \l^{(u)}=\left(\begin{array}{lll}0&0&a\\0&0&b\\a'&b'&0\end{array}\right)\; . \label{uppattern}
\end{equation} 
However, it is not yet clear that the entries $a$, $b$, $a'$, $b'$ of this matrix are non-zero and that the rank of  the up-Yukawa matrix is indeed two, as the pattern of \eqref{uppattern} suggests. This is the question we will answer in the next sub-section. The $3\times 2$ singlet coupling matrix $\rho$ is unconstrained by gauge symmetry and evidently plays an important role for the existence of a massless Higgs doublet pair, away from the line bundle locus. More precisely, if
\be
 \langle \rho_{IJ}{\bf 1}_{2,4}^{(I)}\rangle
\ee
is non-zero then the Higgs pair (where a combination of the lepton multiplets plays the role of the down Higgs) receives a large mass and disappears 
from the spectrum. At the line bundle locus, we have $\langle  {\bf 1}_{2,4}^{(I)}\rangle=0$ and the Higgs pair is massless, consistent with 
the result of our cohomology calculation. However, once we move away from the line bundle locus \footnote{Note that we 
can turn on all the available singlets except ${\bf 1}_{2, 4}^{(I)}$ and keep the Higgs pair massless. As was shown in~Ref.~\cite{Buchbinder:2014qda} 
this deformation leads to a standard model with global $B-L$ symmetry.} such that $\langle  {\bf 1}_{2,4}^{(I)}\rangle\neq 0$,
 the Higgs pair may become massive, depending on the structure of the couplings $\rho_{IJ}$. In fact, in Ref.~\cite{Buchbinder:2014qda} we have 
verified - by performing a cohomology calculation for the associated non-Abelian bundles - that the Higgs pair does indeed become massive for generic complex 
structure, once $\langle  {\bf 1}_{2,4}^{(I)}\rangle\neq 0$. This suggests that at least some of the singlet couplings $\rho_{IJ}$ are non-zero, generically. 
Below, we will confirm this expectation by explicitly calculating the couplings $\rho_{IJ}$.


\subsection{Up Yukawa coupling}
To calculate the up Yukawa couplings we begin with the upstairs GUT model and focus on the first term in the superpotential~\eqref{Wgut}. The line bundles and ambient space harmonic forms (see Eq.~\eqref{akreshom}) for these multiplets are
\begin{equation}
 \begin{array}{lllll}
 3\;{\bf 5}^H_{2,5}&\longrightarrow&K_1=L_2^*\otimes L_5^*&\quad&\hat{\nu}_1=\s_3^{-2}\tilde{Q}_{(0,2,-2,0)}\bar{\m}_3\\
 4\;{\bf 10}_2&\longrightarrow&K_2=L_5&\quad&\hat{\nu}_2=\s_4^{-2}\tilde{R}_{(1,1,0,-2)}\bar{\m}_4\\
 8\;{\bf 10}_5&\longrightarrow&K_3=L_2&\quad&\hat{\omega}=\s_1^{-3}\s_2^{-5}\tilde{S}_{(-3,-5,0,0)}\bar{\m}_1\wedge \bar{\m}_2\; ,
 \end{array}
 \end{equation}
with associated polynomials
\begin{eqnarray}
 \tilde{Q}&=&q_0y_0^2+q_1y_0y_1+q_2y_1^2\label{Qt}\\
 \tilde{R}&=&r_0x_0y_0+r_1x_1y_0+r_2x_0y_1+r_3x_1y_1\\
 \tilde{S}&=&s_0\bar{x}_0\bar{y}_0^3+s_1\bar{x}_0\bar{y}_0^2\bar{y}_1+s_2\bar{x}_0\bar{y}_0\bar{y}_1^2+s_3\bar{x}_0\bar{y}_1^3+s_4\bar{x}_1\bar{y}_0^3+s_5\bar{x}_1\bar{y}_0^2\bar{y}_1+s_6\bar{x}_1\bar{y}_0\bar{y}_1^2+s_7\bar{x}_1\bar{y}_1^3\; ,
 \end{eqnarray} 
 and coefficients $q_I$, $r_I$ and $s_I$ parametrising the multiplets. Evidently, $K_1$ and $K_2$ are of type 1 while $K_3$ is of type 2, so we can proceed with the algebraic calculation explained in Section~\ref{comments}. Converting everything to holomorphic coordinates for simplicity of notation, we have
\begin{eqnarray}
 \mu(Q,R,S)&=&\left(q_0\partial_{y_0}^2+q_1\partial_{y_0}\partial_{y_1}+q_2\partial_{y_1}^2\right)
                          \left(r_0\partial_{x_0}\partial_{y_0}+r_1\partial_{x_1}\partial_{y_0}+r_2\partial_{x_0}\partial_{y_1}+r_3\partial_{x_1}\partial_{y_1}\right)\nonumber\\
&& \left(s_0x_0y_0^3+s_1x_0y_0^2y_1+s_2x_0y_0y_1^2+s_3x_0y_1^3+s_4x_1y_0^3+s_5x_1y_0^2y_1+s_6x_1y_0y_1^2+s_7x_1y_1^3\right)\nonumber\\
&=&2\left[3q_0r_0s_0+3q_0r_1s_4+q_0r_2s_1+q_0r_3s_5+q_1r_0s_1+q_1r_1s_5+\right.\nonumber\\
&&\left.\quad q_1r_2s_2+q_1r_3s_6+q_2r_0s_2+q_2r_1s_6+3q_2r_2s_3+3q_2r_3s_7\right]\; .
 \end{eqnarray}
Inserting standard choices for the coefficients then leads to the couplings $\l_{IJK}$ in the superpotential~\eqref{Wgut}. In particular, we see that these couplings are just numbers, that is, they are independent of complex structure.\\[2mm]

For a simpler and physically more meaningful result we should consider the downstairs theory. This means we have to extract, from the above polynomials $\tilde{Q}$, $\tilde{R}$ and $\tilde{S}$, the $\mathbb{Z}_2\times\mathbb{Z}_2$ equivariant parts. Remembering that the differentials $\mu_i$ carry charge $(1,1)$ under $\mathbb{Z}_2\times\mathbb{Z}_2$, while the $\s_i$ are invariant, this leads to
\begin{eqnarray}
\bar{H}&:&\tilde{Q}_{\bar{H}}=y_0y_1\label{b1}\\
Q_2&:&\tilde{R}_{Q_2}=y_0x_1+y_1x_0\\
u_2&:&\tilde{R}_{u_2}=y_0x_1-y_1x_0\\
Q_5^\alpha&:&\tilde{S}_{Q_5}=-x_0y_0^2+x_1y_1^3\,,\;-x_0y_0y_1^2+x_1y_1y_0^2\\
u_5^\alpha&:&\tilde{S}_{u_5}=x_0y_0^3+x_1y_1^3\,,\;x_0y_0y_1^2+x_1y_1y_0^2\label{b5}
\end{eqnarray}
To carry out the algebraic calculation, we first note that
\begin{equation}
 \lambda(Q,R,S)=\frac{i \pi^3}{24}\mu(Q,R,S)\; . 
\end{equation}  
where the additional factor of $1/4$ relative to Eq.~\eqref{lmurel} accounts for the fact that we are integrating over the upstairs manifold $X$, while the actual calculation should be carried out on the quotient $X/\Gamma$. We find
\begin{equation}
 \mu(\bar{H},u_2,Q_5^\alpha)=\left(\partial_{y_0}\partial_{y_1}\right)\left(\partial_{y_0}\partial_{x_1}-\partial_{y_1}\partial_{x_0}\right)
\left( \begin{array}{l}-x_0y_0^3+x_1y_1^3\\-x_0y_0y_1^2+x_1y_1y_0^2\end{array}\right)=\left(\begin{array}{l}0\\4\end{array}\right)
\end{equation}
and
\begin{equation}
 \mu(\bar{H},u_5^\alpha,Q_2)=\left(\partial_{y_0}\partial_{y_1}\right)\left(\partial_{y_0}\partial_{x_1}+\partial_{y_1}\partial_{x_0}\right)
\left( \begin{array}{l}x_0y_0^3+x_1y_1^3\\x_0y_0y_1^2+x_1y_1y_0^2\end{array}\right)=\left(\begin{array}{l}0\\4\end{array}\right)
\end{equation}
Combining these results leads to the up Yukawa matrix
\begin{equation}
 \lambda^{(u)}=\frac{i \pi^3}{6}\left(\begin{array}{lll}0&0&0\\0&0&1\\0&1&0\end{array}\right)\; .
\end{equation} 
We have, therefore, shown that the up Yukawa matrix has indeed rank 2, as suggested by the general structure~\eqref{uppattern}. In addition, we see that these Yukawa couplings are independent of complex structure. This happens because the cohomologies of the line bundles $K_i$ have a simple representation in terms of ambient space cohomologies without any kernel or co-kernel operations required.


\subsection{Singlet-Higgs-lepton coupling}
To calculate the singlet Yukawa coupling we start with the upstairs theory as before and focus on the second term in the superpotential~\eqref{Wgut}. The relevant line bundles and forms are
\be\label{ex4forms}
\begin{array}{lllll}
 12\;{\bf 1}_{2,4}&\longrightarrow&K_1=L_2\otimes L_4^*&\quad& \hat{\omega}_1=\kappa_1^{-4}\kappa_2^{-6}Q_{(-4,-6,1,1)}d\bar{z}_1\wedge d\bar{z}_2\\
 8\;\overline{\bf 5}_{4,5}&\longrightarrow&K_2=L_4\otimes L_5&\quad&\hat{\omega}_2=\kappa_3^{-3}\kappa_4^{-5}R_{(0,0,-3,-5)}d\bar{z}_3\wedge d\bar{z}_4\\
 4\;{\bf 5}_{2,5}^{\overline{H}}&\longrightarrow&K_3=L_2^*\otimes L_5^*&\quad& \hat{\nu}_3=\kappa_3^{-2}S_{(0,2,-2,0)}d\bar{z}_3\; .
\end{array}
\ee 
There are two additional complications, compared to the previous calculation, evident from this list of forms. First of all, the singlet space is defined as the kernel 
\be
 {\rm Ker}\left(H^2({\cal A},{\cal N}^*\otimes {\cal K}_1)\stackrel{p}{\rightarrow} H^2({\cal A},{\cal K}_1)\right)
\ee 
of a map between a $60$ and a $48$-dimensional space. These dimensions are quite large but we will improve on this shortly by taking the $\mathbb{Z}_2\times\mathbb{Z}_2$ quotient. At any rate, we should impose the constraint $\tilde{p}\tilde{Q}=0$ on the polynomials $Q$ in order to work out this kernel and this will lead to complex structure dependence.

Secondly, two line bundles, $K_1$ and $K_2$, are of type 2 which means that we will have to work with the more general Eq.~\eqref{Yukgen4} for the Yukawa couplings. Given the differentials $d\bar{z}_i$ which appear in  \eqref{ex4forms}, only the term proportional to $\hat{\omega}_1\wedge\hat{\nu}_2\wedge\hat{\nu}_3$ can contribute to the integral~\eqref{Yukgen4}. This means we need to determine the $(0,1)$-forms $\hat{\nu}_2$ satisfying
\be
\bar{\partial}\hat{\nu}_2=p\hat{\omega}_2\; .
\ee
To do this we write down the two relevant polynomials
\begin{equation}
 R_{(0,0,-3,-5)}=r_0+r_1\bar{z}_3\;,\quad p=p_0+p_1z_3+p_2z_3^2 \label{Rsplit}
\end{equation}
with the $z_3$-dependence made explicit and apply the result~\eqref{coboundres} which reads
\begin{equation}
 {\cal R}=-\frac{1}{2}(p_1r_0+p_2r_1)+p_0r_0\bar{z}_3+\frac{1}{2}p_0r_1\bar{z}^2-\frac{1}{2}p_2r_0z_3-p_2r_1|z_3|^2+\frac{1}{2}(p_0r_0+p_1r_1)\bar{z}_3|z_3|^2 \; .\label{Rdef}
\end{equation} 
Then, the desired $(0,1)$-form $\hat{\nu}_2$ can be written as
\begin{equation}
 \hat{\nu}_2=\kappa_3^{-2}\kappa_4^{-5}{\cal R}d\bar{z}_4\; .
\end{equation}
Using these results for the forms in the basic formula~\eqref{Yukgen4} for the Yukawa couplings we find
\begin{equation}
 \lambda(\nu_1,\nu_2,\nu_3)= \frac{1}{2 \pi i}\int_{\mathbb{C}^4}\frac{Q{\cal R}S}{\kappa_1^4\kappa_2^6\kappa_3^4\kappa_4^5}d^4z\,d^4\bar{z}\; . \label{yukexample3}
\end{equation}  
To simplify the calculation, we descend to the downstairs theory and divide by the $\mathbb{Z}_2\times\mathbb{Z}_2$ with generators~\eqref{g1g2}. The polynomials $Q$, $R$ and $S$ then simplify to
\begin{eqnarray}
 Q&=&a_{14} \left(z_3 \bar{z}_1 \bar{z}_2^2+z_4 \bar{z}_1 \bar{z}_2^2\right)+a_5
   \left(\bar{z}_1^2 \bar{z}_2^2+z_3 z_4 \bar{z}_2^2\right)+a_4 \left(z_3 z_4 \bar{z}_1^2
   \bar{z}_2^2+\bar{z}_2^2\right)+a_7 \left(z_3 \bar{z}_2^3+z_4 \bar{z}_1^2
   \bar{z}_2\right)+\nonumber\\
   &&a_6 \left(z_4 \bar{z}_2^3+z_3 \bar{z}_1^2 \bar{z}_2\right)+a_{13}
   \left(\bar{z}_1 \bar{z}_2^3+z_3 z_4 \bar{z}_1 \bar{z}_2\right)+a_{12} \left(z_3 z_4
   \bar{z}_1 \bar{z}_2^3+\bar{z}_1 \bar{z}_2\right)+a_2 \left(z_3 \bar{z}_1^2
   \bar{z}_2^3+z_4 \bar{z}_2\right)+\nonumber\\
   &&a_3 \left(z_4 \bar{z}_1^2 \bar{z}_2^3+z_3
   \bar{z}_2\right)+a_8 \left(\bar{z}_2^4+z_3 z_4 \bar{z}_1^2\right)+a_9 \left(z_3 z_4
   \bar{z}_2^4+\bar{z}_1^2\right)+a_{10} \left(z_3 \bar{z}_1 \bar{z}_2^4+z_4
   \bar{z}_1\right)+\nonumber\\
   &&a_{11} \left(z_4 \bar{z}_1 \bar{z}_2^4+z_3 \bar{z}_1\right)+a_1\left(\bar{z}_1^2 \bar{z}_2^4+z_3 z_4\right)+a_0 \left(z_3 z_4 \bar{z}_1^2\bar{z}_2^4+1\right)\\
   R&=&b_1 \left(\bar{z}_4^2-\bar{z}_3 \bar{z}_4\right)+b_0 \left(1-\bar{z}_3 \bar{z}_4^3\right)\\
   S&=&z_2\; .
\end{eqnarray}
We still have to impose the condition $\tilde{p}\tilde{S}=0$ which reduces the $15$ parameters ${\bf a}=(a_I)$ down to a generic number of three, corresponding to the three singlets ${\bf 1}_{2,4}$. The two coefficients ${\bf b}=(b_0,b_1)$ parametrize the leptons $L_{4,5}$ while $S=z_2$ represents the Higgs $\overline{H}_{2,5}$. From Eq.~\eqref{Rdef} and using the five-parameters $Z_4\times Z_4$-invariant family of tetra-quadrics~\eqref{pspec} in order to make the calculation manageable, we can explicitly work out the polynomial ${\cal R}$. Then, inserting into Eq.~\eqref{yukexample3}, gives
\begin{eqnarray}
\label{Yukawa_5.3}
\lambda({\bf a},{\bf b})&=&-\frac{i \pi^3}{6480}\left(2 a_{14} b_1 c_1+9 a_{12} b_0 c_2+9 a_{13} b_0 c_2-8 a_4 b_1 c_2-8 a_5 b_1 c_2+3 a_{12}
   b_1 c_2+3 a_{13} b_1 c_2-\right.\nonumber\\
   && \qquad\; 36 a_7 b_0 c_3-12 a_2 b_1 c_3-12 a_{14} b_0 c_4+6 a_2 b_1c_4+6 a_3 b_1 c_4-6 a_6 b_1 c_4-6 a_7 b_1 c_4+ \label{yukex3gen}\\
   &&\left.\qquad\; 4 a_{14} b_1 c_4-36 a_6 b_0 c_5-12 a_3 b_1 c_5-36 a_2 b_0 c_6-36 a_3 b_0 c_6-12 a_6 b_1 c_6-12 a_7 b_1 c_6\right)\nonumber
\end{eqnarray}  
We still have to impose the kernel condition on the vector ${\bf a}$, and as before, we use the five-parameter family of tetra-quadrics~\eqref{pspec}. This condition can then be written as $M{\bf a}=0$, where
\begin{align}
&M=\nonumber\\
&\tiny
\left(
\arraycolsep=1.4pt\def\arraystretch{1.5}
\begin{array}{ccccccccccccccc}
 24 c_6 & 0 & 0 & 0 & 4 c_3 & 4 c_6 & 0 & 0 & 0 & 24 c_5 & 0 & 0 & 3 c_4 & 0 & 0 \\
 24 c_5 & 0 & 6 c_2 & 0 & 4 c_6 & 4 c_3 & 0 & 6 c_2 & 0 & 24 c_6 & 0 & 0 & -3 c_4 & 0 & 0
   \\
 24 c_4 & 24 c_6 & 0 & 6 c_2 & 4 c_6-4 c_4 & 4 c_3+4 c_4 & 6 c_2 & 0 & 24 c_5 & -24 c_4 &
   12 c_2 & 0 & 3 c_1 & 3 c_4 & 2 c_2 \\
 0 & 24 c_5 & 0 & 0 & 4 c_3 & 4 c_6 & 0 & 0 & 24 c_6 & 0 & 12 c_2 & 0 & 0 & -3 c_4 & 2
   c_2 \\
 24 c_3 & 0 & 0 & 0 & 4 c_6 & 4 c_5 & 0 & 0 & 0 & 24 c_6 & 0 & 12 c_2 & -3 c_4 & 0 & 2
   c_2 \\
 24 c_6 & 24 c_4 & 6 c_2 & 0 & 4 c_4+4 c_5 & 4 c_6-4 c_4 & 0 & 6 c_2 & -24 c_4 & 24 c_3 &
   0 & 12 c_2 & 3 c_4 & 3 c_1 & 2 c_2 \\
 0 & 24 c_3 & 0 & 6 c_2 & 4 c_5 & 4 c_6 & 6 c_2 & 0 & 24 c_6 & 0 & 0 & 0 & 0 & -3 c_4 & 0
   \\
 0 & 24 c_6 & 0 & 0 & 4 c_6 & 4 c_5 & 0 & 0 & 24 c_3 & 0 & 0 & 0 & 0 & 3 c_4 & 0 \\
 0 & 0 & 12 c_6 & 12 c_6 & 8 c_2 & 8 c_2 & 12 c_3 & 12 c_5 & 0 & 0 & 0 & 0 & 0 & 0 & 4
   c_4 \\
 0 & 0 & 12 c_5 & 12 c_3 & 0 & 0 & 12 c_6 & 12 c_6 & 0 & 0 & 0 & 0 & 0 & 0 & -4 c_4 \\
 0 & 0 & 12 c_6 & 12 c_6 & 0 & 0 & 12 c_5 & 12 c_3 & 0 & 0 & 0 & 0 & 6 c_2 & 6 c_2 & 4
   c_4 \\
 0 & 0 & 12 c_3+12 c_4 & 12 c_4+12 c_5 & 8 c_2 & 8 c_2 & 12 c_6-12 c_4 & 12 c_6-12 c_4 &
   0 & 0 & 0 & 0 & 6 c_2 & 6 c_2 & 4 c_1-4 c_4 \\
\end{array}
\right)\nonumber
\end{align}
This matrix has generic rank $12$ and, hence, a three-dimensional kernel spanned by vector ${\bf v}_I$.  We can write
\begin{equation}
 {\bf a}=\sum_{I}\alpha_I{\bf v}_I\; , \label{aexp}
\end{equation}
with the three coefficients $\alpha_I$ describing the singlets $S^I$. Unfortunately, even for our 5-parameter family~\eqref{pspec} of tetra-quadrics the ${\bf v}_I$ contain very complicated functions of the complex structure moduli which make an analytic calculation impractical. Instead, we choose random numerical values for the complex structure moduli $c_1,\ldots ,c_6$, calculate a basis of ${\rm Ker}(M)$ for this choice and then work out the Yukawa matrix by inserting into Eqs.~\eqref{aexp} and \eqref{yukex3gen}. In this way we obtain an explicit numerical $3\times 2$ Yukawa matrix $\rho$, valid at this specific point in complex structure moduli space. This calculation leads to a Yukawa matrix $\rho$ with rank two and this should be considered the generic result in complex structure moduli space. 

An analytic calculation can be carried out by restricting to the 4-parameter sub-family with $c_2=0$. In this case, the kernel basis vectors are
\begin{equation}
\arraycolsep=1.2pt\def\arraystretch{1.1}
\begin{array}{lll}
{\bf v}_1&=&\left(0,0,-c_1 \left(c_3^2+c_5 c_3-2 c_6^2\right)-c_4 \left(-c_3^2+\left(3 c_4-2
   c_6\right) c_3+c_5 \left(c_5+2 c_6\right)+c_4 \left(c_5+4 c_6\right)\right),\right.\\
   &&c_4c_3^2+\left(c_4^2+2 c_6 c_4+c_1 c_5\right) c_3-c_4 c_5 \left(c_5+2 c_6\right)+c_4^2
   \left(3 c_5+4 c_6\right)+c_1 \left(c_5^2-2 c_6^2\right),0,0,\\
   && -\left(c_3-c_5\right)
   \left(c_4^2+c_3 c_4+\left(c_5+2 c_6\right) c_4-c_1 c_6\right),-\left(c_3-c_5\right)
   \left(c_4^2+c_3 c_4+\left(c_5+2 c_6\right) c_4-c_1 c_6\right),\\
   &&\left. 0,0,0,0,0,0,3\left(c_3-c_5\right) \left(c_3+c_4+c_5-2 c_6\right) \left(c_3+c_5+2 c_6\right)\right)\\
{\bf v}_2&=&\left(0,0,0,0,0,0,0,0,0,0,0,3 \left(c_3-c_5\right) \left(c_3+c_4+c_5-2 c_6\right)
   \left(c_3+c_5+2 c_6\right),0,0,0\right) \\
{\bf v}_3&=&\left(0,0,0,0,0,0,0,0,0,0,3 \left(c_3-c_5\right) \left(c_3+c_4+c_5-2 c_6\right)
   \left(c_3+c_5+2 c_6\right),0,0,0,0\right)
\end{array}   
\end{equation}   
Inserting these vectors into Eq.~\eqref{aexp} and then into the general form \eqref{yukex3gen} of the Yukawa couplings leads to
\begin{equation}
\lambda({\boldsymbol\alpha},{\bf b})=-\frac{i \pi^3}{360} \alpha _1 b_1 \left(c_3-c_5\right) \left(4 c_4^2+c_1 \left(c_3+c_5-2
   c_6\right)\right) \left(c_3+c_5+2 c_6\right)\; 
\end{equation}
For the Yukawa matrix $\rho$ in the superpotential~\eqref{Wsm} this means
\be
\rho=-\frac{i \pi^3}{360}\left(
\begin{array}{cc}0&\left(c_3-c_5\right) \left(4 c_4^2+c_1 \left(c_3+c_5-2
   c_6\right)\right) \left(c_3+c_5+2 c_6\right)\\0&0\\0&0\end{array}\right) \label{Yuksing}
\ee   
The matrix has rank one which is reduced from the generic value two which we have found for the five-dimensional family~\eqref{pspec}. Hence, we have found another example of a Yukawa coupling with rank varying as a function of complex structure. In addition, our results show that, for generic complex structure, the Higgs pair receives a mass whenever $\langle {\bf 1}_{2,4}\rangle\neq 0$, in agreement with the results in Ref.~\cite{Buchbinder:2014qda}. 

For special sub-loci of our four-parameter family of tetra-quadrics, characterised by the vanishing of one of the factors in Eq.~\eqref{Yuksing}, the Yukawa matrix vanishes entirely. However, as before, we have to be careful since the kernel of the matrix $M$ might also change in these case. Let us begin by imposing $c_3=c_5$, in addition to $c_2=0$, on the family of polynomials~\eqref{pspec}. In this case, the dimension of ${\rm Ker}(M)$ turns out to be six and a basis is given by
\be
\arraycolsep=2pt\def\arraystretch{1.1}
\begin{array}{lllllll}
\bf{v}_1&=&(0, 0, 0, 0, 0, 0, 0, 0, 0, 0, 0, 1, 0, 0, 0)^T &\quad&\bf{v}_2&=&(0, 0, 0, 0, 0, 0, 0, 0, 0, 0, 1, 0, 0, 0, 0)^T \\
\bf{v}_3&=&(0, 1, 0, 0, -6, 0, 0, 0, 0, 1, 0, 0, 0, 0, 0)^T &\quad&\bf{v}_4&=& (1, 0, 0, 0, 0, -6, 0, 0, 1, 0, 0, 0, 0, 0, 0)^T \\
\bf{v}_5&=&(0, 0, 0, 0, 0, 0, -1, 1, 0, 0, 0, 0, 0, 0, 0)^T &\quad&\bf{v}_6&=& (0, 0, -1, 1, 0, 0, 0, 0, 0, 0, 0, 0, 0, 0, 0)^T
\end{array}
\ee
Using these six vectors in Eqs.~\eqref{aexp} and \eqref{Yukawa_5.3}, leads to a $6\times 2$ Yukawa matrix which vanishes entirely. Similar results are obtained for other sub-loci of interest. If $4 c_4^2+c_1 (c_3+c_5-2c_6)=0$, in addition to $c_2=0$, the dimension of the kernel becomes four and the $4\times 2$ Yukawa matrix vanishes entirely. The same statements hold for $c_3+c_5-2c_6=0$. This shows that there are specific loci in complex structure moduli space where the Higgs pair remains massless, even in the presence of generic bundle moduli VEVs. 


\section{Conclusions}\label{conclusions}
In this paper, we have developed methods to calculate holomorphic Yukawa couplings for heterotic line bundle models, focusing on Calabi-Yau manifolds defined as hypersurfaces in products of projective spaces and the tetra-quadric in $\mathbb{P}^1\times\mathbb{P}^1\times\mathbb{P}^1\times\mathbb{P}^1$ in particular.  While our approach is based on  differential geometry, we have also made contact with the algebraic methods in Refs.~\cite{Candelas:1987se,Anderson:2009ge}.

We provide explicit rules for writing down the relevant bundle-valued harmonic forms which enter the Yukawa couplings. These forms can be identified with polynomials of certain multi-degrees which are the key players in the algebraic calculation. It turns out that these form can be of different topological types, which we have referred to as type 1 and type 2 (as well as mixed type). If all three forms involved in a Yukawa coupling are of type 1 it turns out that the Yukawa coupling vanishes. This vanishing is topological in nature and is not, apparently, due to a symmetry in the low-energy theory. Our most explicit results, see for example Eq.~\eqref{Yuk112copy}, are for Yukawa couplings which involve two forms of type 1 and one form of type 2. We also show how to compute Yukawa couplings which involve more than one form of type 2, by explicitly working out co-boundary maps. 

The various cases are illustrated with explicit toy examples on the tetra-quadric. In Section~\ref{vanishing}, we have provided an example, based on the gauge group $SO(10)$, of a ${\bf 10}\,{\bf 16}\,{\bf 16}$ Yukawa coupling with topological vanishing, due to all three relevant forms being of type 1. An example of a complex structure independent ${\bf 27}^3$ Yukawa coupling for gauge group $E_6$, with two forms of type 1 and one form of type 2 has been provided in Section~\ref{E6example}. Finally, Section~\ref{csexample} contains an example with gauge group $SU(5)$ which leads to a complex structure dependent d-quark Yukawa coupling.

In Section~\ref{realex} we have computed all Yukawa couplings allowed by the gauge symmetry for a line bundle standard model on the tetra-quadric. The up-quark Yukawa matrix turns out to be complex structure independent and of rank two while the singlet coupling to $L\overline{H}$ is complex structure dependent. The latter involves two forms of type 2 and requires an explicit calculation of a co-boundary map as well as a kernel of a map in cohomology. 

For two of our examples, we have explicitly calculated the complex structure dependence of the Yukawa matrix, if only for a sub-locus in complex structure moduli space. 
The detailed complex structure dependence of these Yukawa matrices is not necessarily physical since the matter 
field Kahler metric can be expected to depend on complex structure as well. However, the rank of the Yukawa matrices is not affected by 
the field normalisation and has to be considered a physical quantity. We have shown that this rank can vary in complex structure moduli space. 

The results of the present paper are limited to a relatively narrow class of Calabi-Yau manifolds and bundles with Abelian structure group. However, the methods we have developed point to and facilitate a number of generalisations. We expect that suitable generalisations of our approach can be used to calculate Yukawa couplings for more general classes of Calabi-Yau manifolds, notably higher co-dimension Cicys and hypersurfaces in toric varieties. Non-Abelian bundles are frequently constructed from line bundles, for example via monad or extension sequences. The results for line bundles obtained in this paper will be useful to calculate Yukawa couplings for such non-Abelian bundles. We hope to address some of these generalisations in future work. 

The most pressing problem remains the calculation of the matter field Kahler metric which is essential in order to determine the physical Yukawa couplings. While we have not addresses this problem it is clear that it requires an approach based on differential geometry. Our hope is that the methods developed in this 
paper will eventually lead to a framework for such a calculation.


\section*{Acknowledgements}
The work of E.I.B. was supported by the ARC Future Fellowship FT120100466
and in part by the ARC Discovery project DP140103925.
A.L. is partially supported by the EPSRC network grant EP/N007158/1 and by the STFC grant~ST/L000474/1.
E.I.B.~would like to thank physics department at Oxford University 
where some of this work was done for warm hospitality.
\newpage
\appendix
\section{Holomorphic Yukawa couplings for $(2,1)$-fields on the quintic}\label{app:21}
Many of the explicit methods for calculating holomorphic Yukawa couplings - including a derivation of the algebraic approach - were first presented in Ref.~\cite{Candelas:1987se}, in the context of the $(2,1)$-Yukawa couplings for standard embedding models. In this appendix, we review some of the results of this paper. In addition, in the second part, we elaborate on the algebraic approach for calculating $(2,1)$-Yukawa couplings by formulating it in the language of bundles, sequences and cohomology. 


\subsection{Explicit evaluation of $(2,1)$-Yukawa couplings}


We begin by reviewing the explicit calculation of $(2,1)$-Yukawa couplings 
for a standard embedding model on quintic Calabi-Yau manifolds, following Ref.~\cite{Candelas:1987se}. 
Quintics are defined as zero loci of polynomials $p$ which are homogeneous of degree five in the projective coordinates $Z^A$, where $A=1,\ldots 5$, 
on the ambient space ${\cal A} ={\mathbb P}^4$. Local coordinates on the quintic $X$ will be denoted as $(x^{\mu}, {\bar x}^{{\bar \mu}})$~\footnote{Here
we will follow the same notation for the coordinates as in~\cite{Candelas:1987se} which is different from our notation in
other sections. Since the material of this appendix is isolated from the rest of the paper this should not cause any confusion.}.
The Hodge numbers of the quintic are given by $h^{1,1}(X)=1$ and $h^{2,1}(X)=101$, where the latter equals 
the number of complex structure moduli on which the defining polynomials $p$ depend. 

The Yukawa couplings for the $(2, 1)$ matter fields in a standard embedding model are given by~\cite{Strominger:1985it} 
\be 
\l(a,b,c)= \int_X \Omega \wedge a^{\mu} \wedge b^{\nu} \wedge c^{\rho} \Omega_{\mu \nu \rho}\;, 
\label{A1}
\ee
a special version of the general formula~\eqref{Yukgen}. Here $a^{\mu}, b^{\nu},  c^{\rho}$ are tangent bundle valued $(0, 1)$-forms which are in one-to-one correspondence with harmonic $(2,1)$-forms, that is
\begin{equation}
 H^1(X,TX)\cong H^{2,1}(X)\; .
\end{equation} 
Following Ref.~\cite{Candelas:1987se}, these forms can be explicitly written as
\be 
a^{\mu} = q (Z^A) g^{\mu \rho} \chi_{{\bar \lambda} {\bar \rho} } d {\bar x}^{\bar \l}\,,\qquad
\chi_{\mu \nu} =\frac{\partial Z^A}{\partial x^{\mu}}\frac{\partial Z^B}{\partial x^{\nu}}\frac{\partial^2 p}{\partial Z^A  \partial Z^B}\; ,
\label{A3}
\ee
where $q(Z^A)$ are a homogeneous degree five polynomials which parametrise the 101 complex structure deformations of the quintic. The space of homogeneous polynomials of degree five in five variables has dimension $126$ but this space has to be divided by the action of $Gl(5,\mathbb{C})$ on the coordinates which reduces the dimension to the desired $101$. Of course we can choose a basis of this space (for example consisting of monomials) which is independent of the complex structure moduli. In the following, we will denote the three degree five polynomials which correspond to the three form $a$, $b$, $c$ in the Yukawa integral~\eqref{A1} by $q$, $r$ and $s$. 

On the ambient space $\mathbb{P}^4$, we can define the $(3,0)$-form
\be 
\hat{\Omega} =\frac{1}{4!} \frac{\epsilon_{ABCDE} Z^A d Z^B dZ^C dZ^D}{p_E}\,, \quad  p_E=\frac{\partial p}{\pt Z^E}\,. 
\label{A5}
\ee
whose restriction $\O= \hat{\Omega}|_X$ gives the $(3,0)$-form on the quintic $X$. It can be shown that $\O$ is non-singular as long as the derivatives $p_A$ do not all vanish simultaneously or, equivalently, if the quintic defined by $p=0$ is smooth. We also have the useful property (a special version of Eq.~\eqref{10.2})
\be 
\hat{\Omega} \wedge dp = \mu = \frac{1}{4!} \epsilon_{ABCDE} Z^A d Z^B dZ^C dZ^D dZ^E = Z^5 d^4Z\,, 
\label{A6}
\ee
where the last equality holds on the coordinate patch $Z^5 = {\rm const}$. 

To compute the integral~\eqref{A1}, we lift it to the ambient space by inserting the delta--function current
\be 
\l(a,b,c)=-\frac{1}{2 i} \int_{{\mathbb P}^4} \Omega \wedge \bar \omega \wedge \delta^2 (p) d p \wedge d {\bar p}\quad\mbox{where}\quad
\bar \omega = a^{\mu} \wedge b^{\nu}\wedge c^{\rho} \Omega_{\mu \nu \rho}\,. 
\label{A7}
\ee
With the help of Eqs.~\eqref{10.5} and \eqref{A6} and an integration by parts this integral turn into
\be 
\l(a,b,c)= -\frac{1}{2 \pi i} \int_{{\mathbb P}^4} ({\bar \partial}{\bar \omega} ) \wedge \mu \wedge d {\bar p} \frac{1}{p}\,. 
\label{A10}
\ee
To continue we introduce the definitions
\be 
||p||^2 = \sum_A p_A {\bar p}_A\,, \quad \tau = \epsilon^{ABCDE} p_A d p_B dp_C dp_D dp_E\,.
\label{A12}
\ee
and the relation~\cite{Candelas:1987se} 
\be
{\bar \partial}{\bar \omega}= \frac{5 q r s}{ ||p||^{10}} p \bar \tau\, ,
\label{A11}
\ee
which turns the Yukawa integral into
\be 
\l(q,r,s)= -\frac{5}{2 \pi i}  \int_{{\mathbb P}^4}  \frac{q r s}{||p||^{10}} Z^5 d^4 Z \bar \tau\,. 
\label{A13}
\ee
The integral can be lifted further to ${\mathbb C}^{5}$ by inserting unity in the form 
\be 
1= \frac{i}{2 \pi} \int d \eta d {\bar \eta } \ \frac{|Z^5|^2}{|\eta|^4} \delta \Big( \sigma-\frac{|Z^5|^2}{|\eta|^2} \Big)\,, 
\label{A14}
\ee
where $\sigma= \sum_A |Z^A|^2$. This leads to
\be 
\l(q,r,s)= -\frac{5}{(2 \pi )^2}  \int_{{\mathbb C}^5}  \frac{q r s}{||p||^{10}  {\bar z}^{{\bar 5}}} \delta (\s -1) d^5 z \ \bar \tau d {\bar z}^{{\bar 5}}\,. 
\label{A15}
\ee
and, with the relation $\tau dZ^5/Z^5=\tau d\s/\s$ and by integrating over $\s$, this can be re-written as
\be 
\l(q.r.s)= -\frac{5}{(2 \pi )^2}  \int_{S^9}  \frac{q r s}{||p||^{10} } d^5 z \ \bar \tau =  -
\frac{5}{(2 \pi )^2}  \int_{S^9}  \frac{q r s}{||p||^{10} }  {\rm det} \Big(  \frac{\partial Z^A}{\partial p_B} \Big)  d^5 p \ \bar \tau \,, 
\label{A17}
\ee
where $d^5 p = dp_1 dp_2 dp_3 dp_4 dp_5$. The last integral is suitable for applying the Bochner--Martinelli theorem (see, for example, Ref.~\cite{GH}), leading to
\be 
\l(q,r,s) = -\frac{5}{(2 \pi )^2}  q r s \ {\rm det} \Big(  \frac{\partial Z^A}{\partial p_B} \Big) \Big|_{p_A=0}\,. 
\label{A18}
\ee
It is convenient to re-express the last identity as a contour integral 
\be 
\l(q,r,s)= - \frac{5}{(2 \pi )^2}    \int_{\Gamma_5} \frac{ q r s \,d^5 z}{p_1p_2 p_3 p_4 p_5}\,, 
\label{A19}
\ee
with contour $\Gamma_5 =\g_1 \times \g_2 \times \g_3 \times \g_4 \times \g_5$ and $\gamma_A$ curves winding around the hyper-surfaces given by $p_A=0$. 
This last form of the integral is a suitable starting point to derive the algebraic approach for calculating $(2,1)$ Yukawa couplings. We first note that the numerator $qrs$ of the integrand is a homogeneous polynomial of degree $15$. Now assume that this polynomial can be written as
\be
q r s = E^A p_A
\label{A20}
\ee
for degree 11 polynomials $E^A$. In this case, the integral~\eqref{A19} is zero since one of the poles in the denominator of~\eqref{A19} is canceled. Hence, for the purpose of calculating Yukawa couplings, we can think of $qrs$ as an element of the quotient
 \be 
 P_{15}= A_{15}/I_{15}\,, 
 \label{A21}
\ee
which consists of the degree $15$ polynomials $A_{15}$ in the coordinate ring $A=\mathbb{C}[Z^1,\ldots ,Z^5]/\langle p\rangle$ of the quintic divided by the degree $15$ part of the ideal $I=\langle p_1,\ldots ,p_5\rangle\subset A$. By counting polynomial degrees of freedom (or, in more mathematical terms, by computing the Hilbert functions of $A$ and $I$) it can be shown that the quotient $P_{15}$ is one-dimensional. Hence, we should be able to choose a degree $15$ polynomial which represent this one -dimensional quotient space. It turns out that an appropriate choice is given by
 \be
 Q ={\rm det} \Big(  \frac{\partial^2 p}{\partial Z^A  \partial Z^B } \Big) = {\rm det} \Big(  \frac{\partial p_A}{\partial Z^B} \Big)\in A_{15} \,. 
 \label{A22}
 \ee
Indeed, replacing $qrs$ by $Q$ in the integral~\eqref{A19} gives the non-vanishing result
\be 
\l(Q) = - \frac{5}{(2 \pi )^2}   \int_{\Gamma_5} \frac{ d^5 p}{p_1p_2 p_3 p_4 p_5}  =- \frac{5}{(2 \pi )^2}  (2 \pi i )^5 \neq 0 \, ,
\label{A22.0}
\ee
which shows that $Q\notin I_{15}$ and, hence, that its associated equivalence class spans $P_{15}$. Put another way, this means that every product $qrs$ can be written as a multiple of $Q$ plus an element in the ideal $I_{15}$ or, explicitly,
\be 
q r s = \mu(q, r, s) Q + E^A p_A\,, 
\label{A23}
\ee
for some number $\mu(q, r, s)$. Inserting this expression for $qrs$ into the Yukawa integral~\eqref{A19} and using the ``normalisation"~\eqref{A22.0} for $Q$ it follows that
\be
\l(q,r,s)= -5i(2 \pi )^3 \mu(q,r,s)\,. 
\label{A24}
\ee
Hence, up to a well-defined numerical factor, the coefficient $\mu(q,r,s)$ is the desired Yukawa coupling and Eq.~\eqref{A23} provides the algebraic rule for its computation.

As an explicit example, let us consider the one-parameter family of quintics defined by the polynomials~\footnote{For the purpose of this example, we write the homogenous coordinates $Z^A$ with lower indices.}
\be 
p= Z_1^5 +  Z_2^5 + Z_3^5  + Z_4^5 +  Z_5^5 - 5 \psi Z_1  Z_2 Z_3 Z_4  Z_5\,,
\label{A25}
\ee
where $\psi$ is the complex structure modulus. We would like to compute the Yukawa coupling between the three same matter fields which correspond to the polynomials $q= r= s = Z_1 Z_2 Z_3 Z_4 Z_5 $. Using $qrs= ( Z_1 Z_2 Z_3 Z_4 Z_5)^3$ and the relation
\be 
{\rm det} \Big(  \frac{\partial^2 p}{\partial Z^A  \partial Z^B } \Big) = 5 \cdot 4^5 (1- \psi^5 ) (Z_1Z_2Z_3Z_4Z_5)^3 + E^A p_A
\label{A27}
\ee
in Eq.~\eqref{A23} we obtain the holomorphic Yukawa coupling 
\be 
\l = - \frac{i \pi^3 }{128} \frac{1}{1- \psi^5} \,. 
\label{A28}
\ee
This coupling becomes singular for $\psi^5 \to 1$ which is related to the quintic acquiring a conifold singularity in this limit.


\subsection{An algebraic approach}


It is possible to formulate the above procedure for calculating the $(2,1)$-Yukawa couplings in more algebraic terms, in analogy with the approach 
taken in Ref.~\cite{Anderson:2009ge}. Calculating $(2,1)$-Yukawa couplings can also be understood as a cup product between three 
elements of $H^1(X,TX)$ which leads to a map
\begin{equation}
 H^1(X,TX)\times  H^1(X,TX)\times  H^1(X,TX)\rightarrow H^3(X,\wedge^3 TX)=H^3(X,{\cal O}_X)\cong\mathbb{C}\; . \label{cupprod}
\end{equation} 
The target space, $H^3(X,\wedge^3 TX)$, of this map is one-dimensional as indicated and, hence, the result of the cup product can be interpreted as a number which turns out to be proportional to the Yukawa coupling. In order to turn this observation into a useful practical procedure we require polynomial 
representatives for the cohomologies involved. The tangent bundle, $T=TX$ of the quintic can be described in terms of two short exact sequences, the Euler sequence 
and the normal bundle sequence, given by
\begin{equation}\label{eulernormal}
\arraycolsep=2pt\def\arraystretch{1.1}
\begin{array}{lllclclclllllclclcll}
 &0&\rightarrow& {\cal O}_X&\stackrel{{\bf Z}}{\rightarrow}& S&\rightarrow& {\cal T}&\rightarrow& 0&\qquad& 0&\rightarrow& T&\rightarrow&{\cal T}&\stackrel{{\bf p}}{\rightarrow}&N&\rightarrow& 0\\
h^0(\cdot)&&&1&&25&&24&&&&&&0&&24&&125&&\\
h^1(\cdot)&&&0&&0&&0&&&&&&101&&0&&0&&\\
h^2(\cdot)&&&0&&0&&1&&&&&&1&&1&&0&&\\
h^3(\cdot)&&&1&&0&&0&&&&&&0&&0&&0&&
\end{array} 
\end{equation} 
where ${\cal T}=T{\cal A}|_X$ is the tangent bundle of the ambient space ${\cal A}=\mathbb{P}^4$ restricted to the quintic, $N={\cal O}_X(5)$ is the normal bundle and $S={\cal O}_X(1)^{\oplus 5}$. The two relevant maps are ${\bf Z}=(Z_1,\ldots ,Z_5)^T$ and ${\bf p}=(p_1,\ldots p_5)$, where, as before, 
$p_A=\partial p/\partial Z^A$. In Eq.~\eqref{eulernormal}, we have also indicated the dimensions of cohomologies in the associated long exact sequences. 
These show that
\begin{eqnarray}
 H^1(X,T)&\cong&{\rm Coker}\left(H^0(X,{\cal T})\stackrel{{\bf p}}{\rightarrow} H^0(X,N)\right)\\
 H^0(X,{\cal T})&\cong&{\rm Coker}\left(H^0(X,{\cal O}_X)\stackrel{{\bf Z}}{\rightarrow} H^0(X,S)\right)\; .
\end{eqnarray} 
With the coordinate ring $A=\mathbb{C}[Z^1,\ldots ,Z^5]/\langle p\rangle$ of the quintic and $H^0(X,N)\cong A_5$ and $H^0(X,S)\cong A_1^{\oplus 5}$ it follows that
\begin{equation}
 H^1(X,TX)\cong\frac{A_5}{{\bf p}(A_1^{\oplus 5})}\; .\label{famquot}
\end{equation} 
This equation provides an algebraic description for the $(2,1)$ families. They are given by quintics in $A_5$ modulo the image of five linear polynomials $(\ell^1, \dots, \ell^5)$ under the map ${\bf p}$, that is, modulo polynomials of the form $\sum_{A=1}^5p_A\ell^A$. Note that dimensions work out correctly. We have ${\rm dim}(A_5)=125$ and ${\rm dim}(A_1^{\oplus 5})=25$, however, the image of ${\bf Z}$ is ${\bf p}({\bf Z})=\sum_{A=1}^5p_AZ^A=5p$ and, hence, vanishes in $A_5$. This means ${\bf p}(A_1^{\oplus 5})\subset A_5$ has only dimension $24$ so that the entire quotient has dimensions $101$, as required. 

In order to complete the picture we should also work out an algebraic representation for the target space $H^3(X,TX)$ in Eq.~\eqref{cupprod}. To do this we consider the third wedge power sequence
\begin{equation}
 0\rightarrow \wedge^3T\rightarrow \wedge^3{\cal T}\rightarrow\wedge^2{\cal T}\otimes N\rightarrow{\cal T}\otimes S^2N\rightarrow S^3N\rightarrow 0
\end{equation}
associated to the normal bundle sequence in Eq.~\eqref{eulernormal}. By introducing suitable co-kernels $C_1$ and $C_2$, this long exact sequence can be split up into three short exact sequences
\begin{equation}
\arraycolsep=2pt\def\arraystretch{1.1}
\begin{array}{llclclclclclclclclc}
 &&\wedge^3T&\rightarrow&\wedge^3{\cal T}&\rightarrow&C_2&\qquad&C_2&\rightarrow&\wedge^2{\cal T}\otimes N&\rightarrow&C_1&\qquad&C_1&\rightarrow&{\cal T}\otimes S^2N&\rightarrow&S^3N\\
h^0(\cdot)&&1&&225&&224&&224&&2250&&2026&&2026&&4900&&2875\\
h^1(\cdot)&&0&&0&&0&&0&&0&&1&&1&&0&&0\\
h^2(\cdot)&&0&&0&&1&&1&&0&&0&&0&&0&&0\\
h^3(\cdot)&&1&&0&&0&&0&&0&&0&&0&&0&&0\\
\end{array}
\end{equation} 
For simplicity of notation we have omitted the zeros at either end of the sequences and we have added the cohomology dimensions of the associated long exact sequences. For $\wedge^3{\cal T}$, $\wedge^2{\cal T}\otimes N$ and ${\cal T}\otimes S^2N$ these dimensions follow straightforwardly from the wedge powers of the Euler sequence~\eqref{eulernormal}, multiplied with the appropriate powers of the normal bundle $N$. Chasing through these three long exact sequences we find that
\begin{equation}
 H^3(X,TX)\cong H^2(X,C_2)\cong H^1(X,C_1)\cong{\rm Coker}\left(H^0(X,{\cal T}\otimes S^2N)\stackrel{{\bf p}}{\rightarrow} H^0(X,S^3N)\right)\; .
\end{equation}
Further, the Euler sequence in Eq.~\eqref{eulernormal} tensored with $S^2N$ implies that
\begin{equation}
 H^0(X,{\cal T}\otimes S^2N)\cong{\rm Coker}\left(H^0(X,S^2N)\stackrel{{\bf Z}}{\rightarrow} H^0(X,S\otimes S^2 N)\right)\; .
\end{equation} 
Combining these last two results, together with $H^0(X,S^3 N)\cong A_{15}$ and $H^0(X,S\otimes S^3 N)\cong A_{11}^{\oplus 5}$ we learn that
\begin{equation}
 H^3(X,TX)\cong\frac{A_{15}}{{\bf p}(A_{11}^{\oplus 5})}\; .\label{Yukquot}
\end{equation} 
and this quotient space is indeed one-dimensional, as it should be. Note, since ${\bf p}\circ{\bf Z}=0$ in the coordinate ring $A$, we do not have to remove the image of ${\bf Z}$ from the denominator in Eq.~\eqref{Yukquot}. More significantly, this quotient has the right structure to serve as a target space for an algebraic computation of Yukawa couplings. Start with three quintic polynomials $q$, $r$, $s$ which represent $(2,1)$ families and classes in the quotient~\eqref{famquot}, that is, they are defined modulo
\begin{equation}
 q\sim q+{\bf p}\cdot {\boldsymbol \ell}^{(q)}\,,\quad r\sim r+{\bf p}\cdot {\boldsymbol \ell}^{(r)}\,,\quad s\sim s+{\bf p}\cdot {\boldsymbol\ell}^{(s)}\; , \label{famid}
\end{equation}
where ${\boldsymbol\ell}^{(q)}$, ${\boldsymbol\ell}^{(r)}$ and ${\boldsymbol\ell}^{(s)}$ are five-dimensional vector of linear polynomials. Then the product $qrs$ is a degree $15$ polynomial which defines an element in the one-dimensional quotient~\eqref{Yukquot}. This element is independent of the ambiguity~\eqref{famid} and, subject to choosing a basis polynomial for the quotient~\eqref{Yukquot}, provides the desired Yukawa coupling.


\section{The boundary integral}\label{app:bound}


When deriving the Yukawa coupling in the main text, in particular by converting Eq.~\eqref{Yukamb} into Eq.~\eqref{Yukamb1}, we have neglected the boundary term which arises from in the partial integration. In this appendix we show that this boundary term does indeed vanish for the cases of interest. 

Before we get to Yukawa couplings it might be useful to note that this boundary term can indeed be important for certain integrals of interest. Consider the tetra-quadric in the ambient space ${\cal A}=\mathbb{P}^1\times\mathbb{P}^1\times\mathbb{P}^1\times\mathbb{P}^1$, with the four ambient space Kahler forms $\hat{J}_i$, where $i=1,2,3,4$, normalised as $\int_{\mathbb{P}^1}\hat{J}_i=1$ and their restrictions $J_i=\hat{J}_i|_X$ to the tetra-quadric. An object of interest are the triple-intersection numbers of the tetra-quadric, for example
\begin{equation}
 d_{123}=\int_XJ_1\wedge J_2\wedge J_3\; . \label{d123}
\end{equation} 
It is well-known~\cite{Hubsch:1992nu} how to compute these intersection numbers by introducing the two-form $\mu=2\sum_{i=1}^4\hat{J}_i$ and re-writing the above expression as an ambient space integral. This leads to
\begin{equation}
 d_{123}=\int_{{\cal A}}\hat{J}_1\wedge\hat{J}_2\wedge\hat{J}_3\wedge\mu=2\; . \label{d123res}
\end{equation}
This method is applicable since the ambient space version $\hat{J}_1\wedge\hat{J}_2\wedge\hat{J}_3$ of the integrand is a closed form. However, alternatively, we may proceed  by inserting a $\delta$-function into the integral~\eqref{d123}, as we have done for Eq.~\eqref{Yukamb} and, subsequently, by using the current identity~\eqref{10.5}. This leads to
\begin{equation}
 d_{123}=\frac{1}{2\pi i}\int_{\cal A}\hat{J_1}\wedge\hat{J}_2\wedge\hat{J}_3\wedge\bar{\partial}\left(\frac{1}{p}\right)\wedge dp
 =\frac{1}{2\pi i}\int_{\cal A}\hat{J_1}\wedge\hat{J}_2\wedge\hat{J}_3\wedge\left(\bar{\partial}_{\bar{z}_4}\left(\frac{1}{p}\right)d\bar{z}_4\right)\wedge dp
\end{equation} 
Since the Kahler forms $\hat{J}_i$ are $\bar{\partial}$-closed, integration by parts and neglecting the boundary term leads to $d_{123}=0$, in contradiction with~\eqref{d123res}. Hence, in this case, the result comes entirely from the boundary term
\begin{equation}
 d_{123}=\frac{1}{2\pi i}\int_{\mathbb{P}_1\times\mathbb{P}_1\times\mathbb{P}^1\times \gamma_4}\hat{J_1}\wedge\hat{J}_2\wedge\hat{J}_3\wedge \frac{dp}{p}\; .
\end{equation} 
where $\gamma_4$ is a contour with $|z_4|\rightarrow\infty$. In this limit $p\sim z_4^2$ and $p^{-1}dp\sim 2z_4^{-1}dz_4$ which leads to the correct answer $d_{123}=2$.
 
For Yukawa integrals, the integrand is typically not a closed form so the $\delta$-function current shoud be used to re-write these as ambient space integrals. As the above example indicates, we should be careful about the boundary term in the subsequent partial integration. The basic integral we consider is of the form
\be 
\l(\nu_1,\nu_2,\nu_3)= \int_X \Omega \wedge \nu_1 \wedge \nu_2 \wedge \nu_3 =
\frac{1}{2 \pi i} \int_{{\cal A}} d^4 z \wedge \hat{\nu}_1 \wedge \hat{\nu}_2\wedge \hat{\nu}_3\wedge {\bar \pt}\Big( \frac{1}{p}\Big)\,, 
\label{B1}
\ee
where the $\nu_i$ are bundle-valued harmonic $(0,1)$-forms. We begin with the simplest case where all three 
forms are of type 1, that is, they are restrictions $\nu_i=\hat{\nu}_i|_X$ of three ambient space forms $\hat{\nu}_i$ which 
are $\bar{\partial}$-closed. Since the three associated line bundles $K_i={\cal O}_X({\bf k}_i)$ tensor to 
the trivial bundle (see Table~\ref{tab:KLrel}) the structure of line bundle cohomology on the tetra-quadric (as discussed in Section~\ref{relations}) implies 
that the vectors ${\bf k}_i$ must all vanish in one same component. For simplicity, we take this to be the fourth component. 
This means that the form $\hat{\nu}_i$ are all independent of $z_4, {\bar z}_4$ and $d {\bar z}_4$. 
Then, the boundary integral related to Eq.~\eqref{B1} becomes
\be 
-\frac{1}{2 \pi i} \int_{{\mathbb P}^1 \times {\mathbb P}^1\times {\mathbb P}^1\times \gamma_4} d^4 z \wedge 
\frac{\hat{\nu}_1 \wedge \hat{\nu}_2\wedge \hat{\nu}_3}{p} \Big|_{|z_4|\to \infty}\,, 
\label{B2}
\ee
where $\gamma_4$ is the circular contour at $|z_4|\to \infty$. Since all $\hat{\nu}_i$ are independent of $z_4$ this contour integral gives
\be 
\int_{\gamma_4} \frac{d z_4}{p} \sim \int_{\gamma_4} \frac{d z_4}{z_4^2} =0
\label{B3}
\ee
since $p$ is quadratic in $z_4$. This shows that the boundary integral~\eqref{B2} vanishes. 

Now we will consider the general case when at least one the forms $\nu_i$ is of type 2 (so that ${\bar \pt} {\hat \nu}_i \neq 0$ for these forms).
In this case, we write~\eqref{B1} as
\be
\l(\nu_1,\nu_2,\nu_3)= 
\frac{1}{2 \pi i} \int_{{\cal A}} d^4 z \wedge \hat{\nu}_1 \wedge \hat{\nu}_2\wedge \hat{\nu}_3\wedge 
\frac{\pt}{\pt {\bar z}_i}
\Big( \frac{1}{p}\Big)d {\bar z}_i\; .
\label{B4}
\ee
and integrating this by parts leads to 
\be 
\l(\nu_1,\nu_2,\nu_3) = -\sum_{i=1}^4  \int_{{\mathbb P}^1 \times {\mathbb P}^1\times {\mathbb P}^1\times \gamma_i}
\frac{\a}{p}\Big|_{|z_i|\to \infty} - \int_{{\mathbb C}^4} d^4 z \wedge \b\,. 
\label{B6}
\ee
where we have introduced the forms
\bea
&& 
\a =\frac{1}{2\pi i} d^4z \wedge  \hat{\nu}_1 \wedge \hat{\nu}_2\wedge \hat{\nu}_3\,, 
\nonumber \\
&&
\b =\frac{1}{2\pi i}   [\hat{\o}_1 \wedge \hat{\nu}_2 \wedge \hat{\nu}_3-
 \hat{\nu}_1 \wedge \hat{\o}_2 \wedge \hat{\nu}_3 +\hat{\nu}_1 \wedge  \hat{\nu}_2 \wedge  \hat{\o}_3 ]\,, 
\nonumber \\
&& 
{\bar \pt}\a = p \ d^4 z \wedge \b\,. 
\label{B5}
\eea
To evaluate the boundary term we first note that, from our discussion in Section~\ref{comments}, the form $\beta$ is a section of
$H^4 ({\cal A}, {\cal N}^*)=H^4 ({\cal A}, {\cal O}_{{\cal A}} (-2, -2, -2, -2))$ and is, hence, proportional to
\be 
\frac{d^4 {\bar z}}{\k_1^2 \k_2^2 \k_3^2 \k_4^2} = \frac{d^4 {\bar z}}{(1+ |z_1 |^2)^2 (1+ |z_2 |^2)^2(1+ |z_3 |^2)^2(1+ |z_4 |^2)^2}  \,.
\label{B7}
\ee
This means that in the limit $|z_4|\to \infty$ we get
\be 
d^4 z \wedge \b \sim \rho \wedge  \frac{d z_4 \wedge d {\bar z}_4}{z_4^2 {\bar z}_4^2}\,, \quad 
p d^4 z \wedge \b \sim \rho \wedge  \frac{d z_4 \wedge d {\bar z}_4}{ {\bar z}_4^2}\,,
\label{B8}
\ee
where $\rho$ is a $(3, 3)$-form independent of $z_4$, ${\bar z}_4$, $d z_4$ and $d {\bar z}_4$. 
Now let us solve Eq.~\eqref{B5}  for $\a$  in this limit.
The general solution for $\a$ is given by $\a$ = $\a_0 + \a_1$, 
where $\a_0$ is the general solution to the homogeneous equation and $\a_1$  is a partial solution to the inhomogeneous one.
Recall that $\hat{\nu}_1 \wedge \hat{\nu}_2 \wedge \hat{\nu}_3$ takes values in the trivial bundle. 
Since 
 $H^3 ({\cal A}, {\cal O}_{{\cal A}}) = 0$ we conclude that $\a_0=0$. 
Then from Eq.~\eqref{B8} it follows that
\be 
\a =\a_1 = \rho\wedge \frac{dz_4}{{\bar z}_4}\,.
\label{B10}
\ee
Hence, the contour integral in~\eqref{B6} becomes
\be 
\int_{\gamma_4} \frac{d z_4}{p {\bar z}_4} \sim \int_{\gamma_4} \frac{d z_4}{z_4^2 {\bar z}_4} =0\,. 
\label{B11}
\ee
This shows that the boundary contribution in~\eqref{B6} indeed vanishes. 


\section{Bundles on Kahler manifolds}\label{app:Kbundle}


In this appendix, we review some standard mathematics for Kahler manifolds and holomorphic vector bundles, which we rely on in the main part of the text. The exposition mainly follows Ref.~\cite{H}, and more details can also be found in Refs.~\cite{Candelas:1987is,GH}.

Let $M$ be a Kahler manifold of dimension $n$ and $E\rightarrow M$ be a rank $r$ holomorphic vector bundle over $M$ with fibres $E_x$, where $x \in M$. The space of $E$-valued $(p,q)$ forms on $M$ is denoted by ${\cal A}^{p, q} (E)$. The usual operator $\bar{\partial}:{\cal A}^{p, q}\rightarrow {\cal A}^{p, q+1}$ for differential forms can be generalised to $E$-valued forms
\begin{equation}
 {\bar \pt}_E : {\cal A}^{p, q} (E) \to {\cal A}^{p, q+1} (E)
\end{equation} 
mapping bundle-valued $(p,q)$-forms to bundle-valued $(p,q+1)$-forms. Explicitly, this operator is defined as follows. For a local holomorphic trivialisation $s = (s_1, s_2, \dots,  s_r)$ of $E$ we can write a vector bundle-valued $(p, q)$-form $\a \in {\cal A}^{p, q} (E)$ as $\a= \sum_{i=1}^r \a^i \otimes s_i$, where  $\a^i \in {\cal A}^{p, q}$ are regular $(p, q)$-forms. Then ${\bar \pt}_E$ acts as
\begin{equation}
 {\bar \pt}_E \a=  \sum_{i=1}^r {\bar \pt} \a^i \otimes s_i\; . \label{defbp}
\end{equation} 
Since the transition functions are holomorphic, this definition is independent of the chosen trivialisation, as it should be. It is straightforward to show from this definition that ${\bar \pt}_E^2 =0$ and that the Leibniz rule
\begin{equation}
{\bar \pt}_E (f \a) = {\bar \pt} (f) \wedge \a + f {\bar \pt}_E (\a)
\end{equation}
holds (here, $f$ is a differentiable function on $M$).\\[4mm]
A Hermitian structure on $E$ (which can also be de defined more generally over complex vector bundles) is defined
by providing  a Hermitian scalar product $h_x$ on each fibre $E_x$. Let $\s$ and $\r$ be two sections of $E$ 
which, for the aforementioned trivialisation of $E$, are expanded  as $\s= \sum_{i=1}^r \s^i s_i$ and  $\r= \sum_{i=1}^r \r^i s_i$.
Then, the Hermitian structure, acting on $\s$ and $\r$, can be written out as
\be 
h (\s, \r)= H_{ij} \s^i {\bar \r}^j  =  \s^{{\rm T}} H {\bar \rho}\,, \quad H_{i j}= h (s_i, s_j)\,. 
\label{C1}
\ee
In other words, locally, we can think of the Hermitian structure as being described by Hermitian $r \times r$ matrices $H$.
For a different local trivialisation $s' = (s'_1, s'_2, \dots,  s'_r)$ related to the original one by $s'_i = \phi^j_{\ i} s_j$ it follows that $H$ transforms as 
\be 
H'= \phi^{{\rm T}} H {\bar \phi}\,. 
\label{C2}
\ee
The Hermitian structure $h$ can also be viewed as an  isomorphism between the vector bundle $E$ and its dual $E^*$, so $h: E \stackrel{\simeq}{\rightarrow} E^*$. 
This isomorphism can be written more explicitly by introducing a ``dual" trivialisation $s_*=(s_*^1,\ldots ,s_*^r)$ of $E^*$, defined by the relations $s_*^i(s_j)=\delta^i_j$.
If we further denote the inverse map of $h$ by $h^*:E^* \stackrel{\simeq}{\rightarrow} E$ then we have
\begin{equation}
 h(s_i)=H_{ji}s^j_*\,,\qquad h^*(s^i_*)=\bar{H}^{ji}s_j\,,\qquad H^{ij}H_{jk}=\delta^i_k\; .
\end{equation}
A Hermitian structure allows one to define a generalisation of the Hodge dual operation ${\bar \star}_E : {\cal A}^{p, q} (E) \to  {\cal A}^{n-p, n-q} (E^*)$ to vector bundle-valued forms by setting
\be 
{\bar \star}_E (\a \otimes s) = \star ({\bar \a}) \otimes h(s)\,,
\label{C11}
\ee
where $\star$ the the regular Hodge star operation on forms. It follows that ${\bar \star}_E \circ {\bar \star}_E = (-1)^{p+q}$, in analogy with corresponding rule for the regular Hodge star. Using this generalised Hodge dual one can define the scalar product
\be 
(\a, \b)= \int_M \a \wedge {\bar \star}_E (\b)\,. 
\label{C12}
\ee
on  ${\cal A}^{p, q} (E)$. The adjoint operator ${\bar \pt}_E^{\dagger}: {\cal A}^{p, q} (E) \to  {\cal A}^{p, q-1} (E)$ of $\bar{\partial}_E$ relative to this scalar product satisfies
\be 
({\bar \pt}_E \a, \b)= (\a, {\bar \pt}_E^{\dagger} \b)\,, 
\label{C13}
\ee
and takes the form
\be 
{\bar \pt}_E^{\dagger}= - {\bar \star}_E \circ {\bar \pt}_{E^*} \circ {\bar \star}_E\,, 
\label{C14}
\ee
as can be seen explicitly from Eqs.~\eqref{C11}, \eqref{C12} and \eqref{C13}. Furthermore, one can define the generalised Laplacian
\be 
\D_E = {\bar \pt}_E^{\dagger} {\bar \pt}_{E} + {\bar \pt}_{E} {\bar \pt}_E^{\dagger} \,, 
\label{C15}
\ee
which is self-adjoint under the above scalar product. Bundle-valued forms  $\a \in {\cal A}^{p, q} (E)$ satisfying $\D_E \a =0$ are called harmonic with respect to the Hermitian structure $h$. For a compact manifold, the harmonic forms $\a$ are precisely the closed and co-closed forms, so the forms satisfying
\be
{\bar \pt}_E\a=0\;,\qquad {\bar \pt}_E^{\dagger} \a=0\,.
\label{C16}
\ee
These forms are in one-to-one correspondence with the cohomology groups $H^{p, q}(M, E)\cong H^q (M, E \otimes \L^p \O_M)$. Finally, there is a generalisation of the Hodge decomposition which states that every form $\a \in {\cal A}^{p, q} (E)$ can be written as a unique sum $\a =\eta +  {\bar \pt}_{E} \b +{\bar \pt}_E^{\dagger}\g$, where $\eta$ is harmonic. 
\vskip 4mm\noindent
A connection, $\nabla$, on $E$ is a map $\nabla: {\cal A}^0 (E) \to  {\cal A}^1 (E)$ which satisfies the Leibniz rule 
\be 
\nabla (f\s)= d (f) \otimes \s + f \nabla (\s)
\label{C5}
\ee
for local sections $\s$ and local functions $f$. Writing the section $\s =\sum_{i=1}^r \s^i s_i$ in terms of a local trivialisation $s=(s_1,\ldots ,s_r)$, we have 
\be 
\nabla (\s)= (d \s^i + A^{i}_{\ j} \s^j) \otimes s_i\,, \quad \nabla (s_j)= A^{i}_{\ j} s_i\,, 
\label{C6}
\ee
where $A$ is the gauge field. In short, locally, the connection can be written as $\nabla=d+A$, with the gauge field transforming as
\be 
A'= \phi^{-1} A \phi + \phi^{-1} d\phi\,. 
\label{C7}
\ee
under a change of trivialisation, $s'_i =\phi^j_{\ i} s_j$. The curvature $F_{\nabla} \in {\cal A}^2 ({\rm End} (E))$ is defined by $F_{\nabla} =\nabla \circ \nabla$. For a given trivialisation its local form is
\be 
F_{\nabla} = dA + A \wedge A\,. 
\label{C8}
\ee
A connection is called compatible with the holomorphic structure if $\nabla^{0,1}={\bar \pt}$ and it is called Hermitian if it satisfies $d (h (\s, \r))= h (\nabla (\s), \r) +
 h (\s, \nabla (\r))$ for any two sections $\s$ and $\r$. For a holomorphic vector bundle there exists a unique Hermitian connection compatible with the 
 holomorphic structure which is called the Chern connection. In a local frame, the gauge field associated to the Chern connection is given by
 \be 
 A ={\bar H}^{-1} \pt {\bar H}\,.
 \label{C9}
\ee
For a holomorphic change of the trivialisation, $s'_i =\phi^j_{\ i} s_j$,
it is straightforward to verify that Eq.~\eqref{C9} is consistent with the transformation laws~\eqref{C2} and \eqref{C7}. 
It can be shown, using Eq.~\eqref{C8}, that the curvature of the Chern connection is a $(1, 1)$-form and, locally, is explicitly given by 
\be 
F_{\nabla} = {\bar \pt} ({\bar H}^{-1} \pt {\bar H})\,. 
\label{C10}
\ee
\vskip 4mm\noindent
In the main part of the paper, we are calculating certain bundle-valued harmonic forms and it is, therefore, important to re-write the defining Eqs.~\eqref{C16} for such forms in a simple and explicit way. As before, we introduce local trivialisations $s=(s_1,\ldots ,s_r)$ and $s_*=(s_*^1,\ldots ,s_*^r)$  on $E$ and $E^*$, satisfying $s^i_* (s_j)= \d^i_{\ j}$. We start with two $(p,q)$-forms $\a =\a^i s_i$ and $\b =\b_i s^i_*$ taking values in $E$ and $E^*$, respectively. Then from the definition~\eqref{defbp} of ${\bar \pt}_E$ we have 
\be 
{\bar \pt}_E (\a)= ({\bar \pt} \a^i) \otimes s_i\,, \qquad 
{\bar \pt}_{E^*} (\b)= ({\bar \pt} \b_i) \otimes s^i_*\,.
\label{C17}
\ee
For the generalised Hodge star operation~\eqref{C11} we get 
\be
{\bar \star}_{E}  (\a) = (* {\bar \a}^i)  \otimes h(s_i)=
H_{ji} (* {\bar \a}^i) \otimes s_*^j \,, \qquad 
{\bar \star}_{E^*}  (\b) = (* {\bar \b}_i)  \otimes h^*(s^i_*)=
\bar{H}^{ji} (* {\bar \b}_i) \otimes s_j \,. 
\label{C18}
\ee
Combining these equations we obtain
\be
{\bar \pt}_E^{\dagger} \a = -\star (\d^k_{\ i} \pt + {\bar H}^{kj} \pt {\bar H}_{ji}) \star \a^i \otimes s_k
= -\star  (\d^k_{\ i} \pt + A^k_{\ i}) \star \a^i \otimes s_k\,, 
\label{C19}
\ee
where $A$ is the  Chern connection~\eqref{C9}. Hence, ${\bar \pt}_E^{\dagger}$ corresponds to the 
dual of the $\nabla^{1, 0}$ part of the Chern connection. From the above 
argument we conclude that a harmonic bundle-valued form $\a$, written as ${\boldsymbol \a}= (\a^1, \ldots ,\a^r)^{{\rm T}}$ relative to a local frame, is characterised by
\be 
{\bar \pt} \boldsymbol { \a} =0\,, \quad  (\pt + A) \star {\boldsymbol \a}=0\,, 
\label{C20}
\ee
where $A$ is the gauge field associated to the Chern connection on the bundle. Using the explicit expression~\eqref{C9} for the Chern connection, these equations can be cast into the somewhat more convenient form
\be 
{\bar \pt} \boldsymbol { \a} =0\,, \quad  \pt ({\bar H} \star {\boldsymbol \a})=0\, ,
\label{C21}
\ee
with the Hermitian structure $H$ on the bundle. 
%


\section{The solution to the map between harmonic forms on ${\mathbb P}^1$}\label{app:proof}


One of the key technical observations in the main part of the paper concerns the multiplication of harmonic bundle-valued $(0,0)$-forms with $(0,1)$-forms on $\mathbb{P}^1$. While the resulting product $(0,1)$-form represents a cohomology it is not harmonic anymore. However, the equivalent harmonic representative can be found by solving Eq.~\eqref{prodeq} which is surprisingly complicated. Remarkably, a simply solution, given by Eq.~\eqref{prodsol}, can be found for this equation. it states that the harmonic representative of the product can be obtained by converting the multiplicative action of the $(0,0)$-form into a derivative action. The purpose of this appendix is to provide a general proof for this solution.

More specifically, the set-up is as follows. On $\mathbb{P}^1$ we introduce homogeneous coordinates $x^\alpha$, where $\a=0,1$, and corresponding affine coordinates $z=x^1/x_0$, $w=x^0/x^1$ on the two standard open patches. We consider a harmonic $(0,0)$-form which represents a class in $H^0(\mathbb{P}^1,{\cal O}_{\mathbb{P}^1}(\delta))$, where $\delta\geq 0$, and, from the discussion in Section~\ref{p1}, this $(0,0)$-form is described by a holomorphic polynomial $p(z)$ of degree $\delta$ or, equivalently, by its homogeneous counterpart $\tilde{p}(x^0,x^1)$. Further, we consider a harmonic $(0,1)$-form which represents a class in $H^0(\mathbb{P}^1,{\cal O}_{\mathbb{P}^1}(k-\delta))$, where $k\leq -2$. Again, following Section~\eqref{p1}, this $(0,1)$-form is described by an anti-holomorphic polynomial $P(\bar{z})$ with degree $-k+\delta-2$ or, equivalently, by its homogeneous counterpart $\tilde{P}(\bar{x}^0,\bar{x}^1)$. The product of the two forms represent a cohomology class in $H^0(\mathbb{P}^1,{\cal O}_{\mathbb{P}^1}(k))$ but it is not harmonic. This harmonic representative, equivalent in cohomology to this product, is denoted is represented by an anti-holomorphic polynomial $Q(\bar{z})$ of degree $-k-2$ or, equivalently, by its homogenous counterpart $\tilde{Q}(\bar{x}^0,\bar{x}^1)$. 

This polynomial $Q$ can be obtained from $p$ and $P$ by the equation
\be 
p P + (1+ z {\bar z}) \pt_{{\bar z}} S - (\d-k-1) z S = (1+ z {\bar z})^{\d} Q\,, 
\label{D1}
\ee
which we have derived in Section~\ref{p1}. Here, $S=S(z,\bar{z})$ is a suitable polynomial of bi-degree $(\delta-1,\delta-k-1)$ in $(z,\bar{z})$ which, for given $p$ and $P$, is determined from the above equation along with the polynomial $Q$. Our aim is to show this equation is indeed solved by Eq.~\eqref{prodsol}.

We begin by writing the relevant polynomials out explicitly
\begin{equation}
 p=\sum_{i=0}^{\d} a_i z^i \,, \quad P= \sum_{j=0}^{\d+\ell-1} b_j {\bar z}^j\,, \quad Q= \sum_{j=0}^{\ell -1 } q_j {\bar z}^j\,,\quad
 S=  \sum_{j=0}^{\d+\ell-1} {\bar z}^j A_j + {\bar z}^{\d+\ell} A_{\d+\ell}\; .
\end{equation}
where  $\ell =-k-1 \geq 1$ and 
\be 
A_j= \sum_{i=0}^{\d-1} z^i c_{ij}\,,  \qquad   A_{\d+\ell}  =\sum_{i=0}^{\d-1} z^i d_{i}\,. 
\label{D3}
\ee
are polynomials of degree $\d-1$ in $z$. Inserting the above expressions for $p$, $P$ and $Q$ into Eq.~\eqref{D1} leads to
\be
(1+ z {\bar z}) \pt_{{\bar z}} S - (\d-k-1) z S= \sum_{j=1}^{\d+\ell-1} j {\bar z}^{j-1} A_j  +\sum_{j=0}^{\d+\ell-1}(j-\d-\ell){\bar z}^j z A_{j}  + (\d+\ell) {\bar z}^{\d+\ell-1} A_{\d+\ell} \,. 
\label{D4}
\ee
Note that the terms with the highest degrees ${\bar z}^{\d+\ell} z A_{\d +\ell}$ cancel. Now we substitute the remaining polynomials for $S$, $A_j$ and $A_{\delta+\ell}$ in order to obtain equations for the coefficients $q_j$.  Focusing on terms proportional to $z^i {\bar z}^{i+j}$, where $i=0, 1, \dots, \d$, we find the following linear system
\bea
&& 
a_0 b_j +  (j+1) c_{0,j+1} = q_j  
\nonumber \\
&&
a_1 b_{j+1} + (j+2) c_{1, j+2} + (j+1 -\d-\ell)  c_{0, j+1} =  \frac{\d!}{1 ! (\d-1)!} q_j  
\nonumber \\
&&
a_2 b_{j+2} + (j+3) c_{2, j+3} + (j+2- \d-\ell) c_{1,j+2} =\frac{\d!}{2 ! (\d-2)!} q_j 
\nonumber \\
&&
\qquad\vdots\qquad\qquad\vdots\qquad\qquad\vdots\qquad\qquad\vdots\qquad\qquad\vdots\qquad\qquad\vdots\label{D5}\\
&&
a_{\d-1} b_{\d-1 +j} + (j+\d) c_{\d-1, j+ \d} + (j-\ell -1)  c_{\d-2, j+ \d-1} =  \frac{\d!}{ (\d-1)!  1!} q_j 
\nonumber \\
&&
a_{\d} b_{\d+j } + (j-\ell)  c_{\d-1, j+ \d} =q_j \nonumber
\eea
for the coefficients $q_j$. Evidently, this system has a triangular form and we can eliminate the coefficients $c_{i j}$ step by step. 
More specifically, taking a proper linear combination of the first two equations we can eliminate $ c_{0,j+1}$, combining the 
resulting linear combination with the third equation we can eliminate $c_{1, j+2}$ and so forth. In this way, by going through all $\d+1$ equations,  
we can completely eliminate the coefficients $c_{i, j}$ and find $q_j$ in terms of $a_i$ and $b_k$. To do this explicitly, we would like to find the linear combination with coefficients $\a_0, \a_1, \dots, \a_{\d}$ of the $\d+1$ equations~\eqref{D5} for which all $c_{ij}$ cancel on the LHS. The cancellation of the terms involving  $ c_{0,j+1} ,  c_{1, j+2}, \dots,  c_{\d-1, j+ \d}$ imposes the following conditions
\bea
&& 
\a_1 = \a_0 \frac{j+1}{\d+\ell -j -1} 
\nonumber \\
&& 
\a_2= \a_1 \frac{j+2}{\d+\ell -j -2} = \a_0 \frac{ (j+1)(j+2) }{ (\d+\ell -j -1)(\d+\ell -j -1) } 
\nonumber \\
&&
\;\vdots\qquad\qquad\vdots\qquad\qquad\vdots\qquad\qquad\vdots 
\nonumber  \\
&&
\a_{\d}= \a_0 \frac{(j+1)(j+2) \dots (j+\d) }{(\d+\ell -j -1)(\d+\ell -j -1)\dots (\ell-j) }
\label{D7}
\eea
 on the ratios of these coefficients. If we choose the overall normalisation of the $\alpha_i$ by setting $\a_0= (\d+\ell-1-j) (\d+\ell-2-j)  \dots (l-j)$ it follows that
\be 
\a_i = \frac{ (\d+ \ell -j-1-i)! (i+j) !}{(\ell-j-1)! j!}\,, \quad i=0, 1, \dots \d\,.
\label{D10}
\ee
We can now work out the required linear combination of the Eqs.~\eqref{D5}, using the so-determined coefficients $\a_i$, to find
\be 
\sum_{i=0}^{\d} \a_i \  a_i b_{i+j}= \sum_{i=0}^{\d} \a_i \ \frac{\d!}{i!  (\d-i)!} \  q_j  := c_j^{-1} q_j\quad\mbox{where}\quad c_j^{-1} = \sum_{i=0}^{\d} \a_i \ \frac{\d!}{i!  (\d-i)!}= \frac{(\d+ \ell)!}{\ell!} \,. 
\label{D14}
\ee
Hence, the coefficients $c=c_j$ are independent of $j$ and our solution for $Q$ is explicitly given by
\be
 Q= \sum_{j=0}^{\ell -1 } q_j {\bar z}^j\quad\mbox{where}\quad q_j=c\sum_{i=0}^{\d} \a_i \  a_i b_{i+j}\quad\mbox{and}\quad c=\frac{\ell!}{(\d+ \ell)!}\; . \label{Qres}
\ee

We would like to compare this result for $Q$ with the expression
\be 
\tilde{p} \Big( \frac{\pt}{\pt {\bar x}^0},   \frac{\pt}{\pt {\bar x}^1}\Big) \tilde{P} ({\bar x}^0, {\bar x}^1) \;,
\label{D11}
\ee
where we recall that the tilde denotes the homogeneous counterparts of polynomials. The coefficient of the terms $({\bar x}^0)^{\ell-1-j} ({\bar x}^1)^j$ in this expression can be written as
\be
\sum_{i=0}^{\d} \b_i \ a_i b_{i+j} 
\label{D12}
\ee
for certain constants $\b_i$. It is a simple combinatorial exercise to compute $\b_i$ and to note that, in fact, $\b_i=\a_i$. Hence, with the result for $Q$ from Eq.~\eqref{Qres}, this means that
\be 
\tilde{Q}({\bar x}^0, {\bar x}^1)= c \  \tilde{p} \Big( \frac{\pt}{ \pt {\bar x}^0},   \frac{\pt}{\pt {\bar x}^1}\Big) \tilde{P} ({\bar x}^0, {\bar x}^1)\quad \mbox{where}\quad
c= \frac{\ell!}{(\delta+ \ell)!} = \frac{(-k-1)!}{(\delta -k-1)!}\; .
\label{D20}
\ee
This is the expected solution to Eq.~\eqref{D1}.



\newpage
\begin{thebibliography}{99}
\ifx\doiref\asklfhas\newcommand{\doiref}[2]{\href{http://dx.doi.org/#1}{#2}}\fi
\raggedright 
\ifx\arxivref\asklfhas\newcommand{\arxivref}[2]{\href{http://arxiv.org/abs/#1}{arXiv:#1}}\fi
\raggedright


\bibitem{Candelas:1985en}
P.~Candelas, G.~T. Horowitz, A.~Strominger, and E.~Witten, ``Vacuum Configurations for Superstrings,''
\textsf{\doiref{10.1016/0550-3213(85)90602-9}{Nucl.Phys. {\bf B258} (1985) 46--74}}.  

\bibitem{Strominger:1985it}
A.~Strominger and E.~Witten, ``New Manifolds for Superstring Compactification,''
 \textsf{\doiref{10.1007/BF01216094}{Commun.\ Math.\ Phys.\  {\bf 101} (1985) 341}}.
   
\bibitem{Witten:1985xc}
E.~Witten,  ``Symmetry Breaking Patterns in Superstring Models,''
\textsf{\doiref{10.1016/0550-3213(85)90603-0}{Nucl.\ Phys.\ B {\bf 258} (1985) 75}}.

\bibitem{Braun:2005ux}
V.~Braun, Y.-H. He, B.~A. Ovrut, and T.~Pantev, ``{A Heterotic standard model},''
\textsf{\doiref{10.1016/j.physletb.2005.05.007}{Phys.Lett. {\bf B618} (2005) 252--258}, \arxivref{hep-th/0501070}}.

\bibitem{Braun:2005bw}
V.~Braun, Y.-H. He, B.~A. Ovrut, and T.~Pantev, ``{A Standard model from the E(8) x E(8) heterotic superstring},''
\textsf{\doiref{10.1088/1126-6708/2005/06/039}{ JHEP {\bf 0506} (2005) 039}, \arxivref{hep-th/0502155}}.

\bibitem{Braun:2005nv}
V.~Braun, Y.-H. He, B.~A. Ovrut, and T.~Pantev, ``{The Exact MSSM spectrum from string theory},''
\textsf{\doiref{10.1088/1126-6708/2006/05/043}{JHEP {\bf 0605} (2006) 043}, \arxivref{hep-th/0512177}}.

\bibitem{Bouchard:2005ag}
V.~Bouchard and R.~Donagi, ``{An SU(5) heterotic standard model},''
\textsf{\doiref{10.1016/j.physletb.2005.12.042}{Phys.Lett. {\bf B633} (2006) 783--791}, \arxivref{hep-th/0512149}}.

\bibitem{Blumenhagen:2006ux} 
  R.~Blumenhagen, S.~Moster and T.~Weigand,
  ``Heterotic GUT and standard model vacua from simply connected Calabi-Yau manifolds,''
  \textsf{\doiref{10.1016/j.nuclphysb.2006.06.005}{Nucl.\ Phys.\ B {\bf 751}, 186 (2006)}, \arxivref{hep-th/0603015}}.
   
\bibitem{Blumenhagen:2006wj} 
  R.~Blumenhagen, S.~Moster, R.~Reinbacher and T.~Weigand,
  ``Massless Spectra of Three Generation U(N) Heterotic String Vacua,''
  \textsf{\doiref{10.1088/1126-6708/2007/05/041}{JHEP {\bf 0705}, 041 (2007)}, \arxivref{hep-th/0612039}}.
 
\bibitem{Anderson:2007nc}
L.~B. Anderson, Y.-H. He, and A.~Lukas, ``{Heterotic Compactification, An  Algorithmic Approach},''
\textsf{\doiref{10.1088/1126-6708/2007/07/049}{JHEP {\bf 0707} (2007) 049}, \arxivref{hep-th/0702210}}.

\bibitem{Anderson:2008uw}
L.~B. Anderson, Y.-H. He, and A.~Lukas, ``{Monad Bundles in Heterotic String Compactifications},''
\textsf{\doiref{10.1088/1126-6708/2008/07/104}{JHEP {\bf 0807} (2008) 104}, \arxivref{0805.2875}}.

\bibitem{Anderson:2009mh}
L.~B. Anderson, J.~Gray, Y.-H. He, and A.~Lukas, ``{Exploring Positive Monad Bundles And a New Heterotic Standard Model},''
\textsf{\doiref{10.1007/JHEP02(2010)054}{JHEP {\bf 1002} (2010) 054}, \arxivref{0911.1569}}.
  
\bibitem{Braun:2009qy} 
  V.~Braun, P.~Candelas and R.~Davies,
  ``A Three-Generation Calabi-Yau Manifold with Small Hodge Numbers,''
  \textsf{\doiref{10.1002/prop.200900106}{Fortsch.\ Phys.\  {\bf 58}, 467 (2010)}, \arxivref{0910.5464}}.
    
\bibitem{Braun:2011ni}
V.~Braun, P.~Candelas, R.~Davies, and R.~Donagi, ``{The MSSM Spectrum from (0,2)-Deformations of the Heterotic Standard Embedding},''
\textsf{\doiref{10.1007/JHEP05(2012)127}{JHEP {\bf 1205} (2012) 127}, \arxivref{1112.1097}}.

 \bibitem{Anderson:2011ns}
  L.~B.~Anderson, J.~Gray, A.~Lukas and E.~Palti,
  ``Two Hundred Heterotic Standard Models on Smooth Calabi-Yau Threefolds,''
  \textsf{\doiref{10.1103/PhysRevD.84.106005}{Phys.\ Rev.\ D {\bf 84} (2011) 106005}, \arxivref{1106.4804}}.
 
 \bibitem{Anderson:2012yf}
  L.~B.~Anderson, J.~Gray, A.~Lukas and E.~Palti,
  ``Heterotic Line Bundle Standard Models,''
  \textsf{\doiref{10.1007/JHEP06(2012)113}{JHEP {\bf 1206} (2012) 113}, \arxivref{1202.1757}}.

\bibitem{Anderson:2013xka}
L.~B.~Anderson, A.~Constantin, J.~Gray, A.~Lukas and E.~Palti,
``A Comprehensive Scan for Heterotic SU(5) GUT models,''
\textsf{\doiref{10.1007/JHEP01(2014)047}{JHEP {\bf 1401} (2014) 047}, \arxivref{1307.4787}}.

\bibitem{Strominger:1985ks}
A.~Strominger,  ``Yukawa Couplings in Superstring Compactification,''
\textsf{\doiref{10.1103/PhysRevLett.55.2547}{Phys.\ Rev.\ Lett.\  {\bf 55} (1985) 2547}}.
   
\bibitem{Candelas:1987se}
P.~Candelas,  ``Yukawa Couplings Between (2,1) Forms,''
\textsf{\doiref{10.1016/0550-3213(88)90351-3}{Nucl.\ Phys.\ B {\bf 298} (1988) 458}}.

\bibitem{Candelas:1990pi}
P.~Candelas and X.~de la Ossa,  ``Moduli Space of {Calabi-Yau} Manifolds,''
\textsf{\doiref{10.1016/0550-3213(91)90122-E}{Nucl.\ Phys.\ B {\bf 355} (1991) 455}}.

\bibitem{Greene:1986bm}
  B.~R.~Greene, K.~H.~Kirklin, P.~J.~Miron and G.~G.~Ross,
  ``A Three Generation Superstring Model. 1. Compactification and Discrete Symmetries,''
\textsf{\doiref{10.1016/0550-3213(86)90057-X}{Nucl.\ Phys.\ B {\bf 278} (1986) 667}}.

\bibitem{Greene:1986jb}
  B.~R.~Greene, K.~H.~Kirklin, P.~J.~Miron and G.~G.~Ross,
  ``A Three Generation Superstring Model. 2. Symmetry Breaking and the Low-Energy Theory,''
\textsf{\doiref{10.1016/0550-3213(87)90662-6}{Nucl.\ Phys.\ B {\bf 292} (1987) 606}}.  

\bibitem{Greene:1987xh}
  B.~R.~Greene, K.~H.~Kirklin, P.~J.~Miron and G.~G.~Ross,
  ``27**3 Yukawa Couplings for a Three Generation Superstring Model,''
\textsf{\doiref{10.1016/0370-2693(87)91151-8}{Phys.\ Lett.\ B {\bf 192} (1987) 111}}.  

\bibitem{Braun:2006me}
  V.~Braun, Y.~H.~He and B.~A.~Ovrut,
  ``Yukawa couplings in heterotic standard models,''
 \textsf{\doiref{10.1088/1126-6708/2006/04/019}{JHEP {\bf 0604} (2006) 019}, \arxivref{hep-th/0601204}}.

\bibitem{Bouchard:2006dn}
  V.~Bouchard, M.~Cvetic and R.~Donagi,
  ``Tri-linear couplings in an heterotic minimal supersymmetric standard model,''
\textsf{\doiref{10.1016/j.nuclphysb.2006.03.032}{Nucl.\ Phys.\ B {\bf 745} (2006) 62}, \arxivref{hep-th/0602096}}.
  
\bibitem{Anderson:2009ge}
  L.~B.~Anderson, J.~Gray, D.~Grayson, Y.~H.~He and A.~Lukas,
  ``Yukawa Couplings in Heterotic Compactification,''
\textsf{\doiref{10.1007/s00220-010-1033-8}{Commun.\ Math.\ Phys.\  {\bf 297} (2010) 95}, \arxivref{0904.2186}}.
  
\bibitem{Anderson:2010tc}
  L.~B.~Anderson, J.~Gray and B.~Ovrut,
  ``Yukawa Textures From Heterotic Stability Walls,''
 \textsf{\doiref{10.1007/JHEP05(2010)086}{JHEP {\bf 1005} (2010) 086}, \arxivref{1001.2317}}. 

 \bibitem{Green:1986ck}
  P.~Green and T.~Hubsch,
  ``Calabi-Yau Manifolds As Complete Intersections In Products Of Complex Projective Spaces,''
  \textsf{\doiref{10.1007/BF01205673}{Commun.\ Math.\ Phys.\ {\bf 109} (1987) 99}}.

 \bibitem{Candelas:1987kf}
  P.~Candelas, A.~M.~Dale, C.~A.~Lutken and R.~Schimmrigk,
  ``Complete Intersection Calabi-Yau Manifolds,''
  \textsf{\doiref{10.1016/0550-3213(88)90352-5}{Nucl.\ Phys.\ B {\bf 298} (1988) 493}}.
  
\bibitem{Hubsch:1992nu}
T.~Hubsch,  ``Calabi-Yau manifolds: A Bestiary for physicists,''
Singapore: World Scientific, Singapore,  (1992).  

\bibitem{Buchbinder:2013dna}
  E.~I.~Buchbinder, A.~Constantin and A.~Lukas,
  ``The Moduli Space of Heterotic Line Bundle Models: a Case Study for the Tetra-Quadric,''
  JHEP {\bf 1403} (2014) 025
  [arXiv:1311.1941 [hep-th]].


\bibitem{Buchbinder:2014qda}
  E.~I.~Buchbinder, A.~Constantin and A.~Lukas,
  ``A heterotic standard model with $B - L$ symmetry and a stable proton,''
\textsf{\doiref{10.1007/JHEP06(2014)100}{JHEP {\bf 1406} (2014) 100}, \arxivref{1404.2767}}.

\bibitem{Buchbinder:2014sya}
E.~I.~Buchbinder, A.~Constantin and A.~Lukas,  ``Non-generic Couplings in Supersymmetric Standard Models,''
\textsf{\doiref{10.1016/j.physletb.2015.07.012}{Phys.\ Lett.\ B {\bf 748} (2015) 251}, \arxivref{1409.2412}}.

\bibitem{Buchbinder:2014qca}
E.~I.~Buchbinder, A.~Constantin and A.~Lukas, ``Heterotic QCD axion,''
\textsf{\doiref{10.1103/PhysRevD.91.046010}{Phys.\ Rev.\ D {\bf 91} (2015) 4,  046010}, \arxivref{1412.8696}}.  

\bibitem{H}
D. Huybrechts, ``Complex Geometry: An Introduction", 
 \textsf{\doiref{10.1007/b137952}{Springer, Berlin (2004)}}.

\bibitem{GSW}
M. B. Green, J. H. Schwarz and E. Witten, ``Supersting Theory. Vol. 2: Loop Amplitudes, 
Anomalies and Phenomenology,"  
\textsf{\doiref{DOI:10.1002/zamm.19880680631}{Cambridge University Press 1987}}.

\bibitem{Candelas:1987is}
P.~Candelas,  ``Lectures On Complex Manifolds,''
Published in Trieste 1987, Proceedings, Superstrings 87, 1-88.

\bibitem{GH}
P. Griffiths and J. Harris, ``Principles of algebraic geometry,"  Interscience, New-York, (1978).

\end{thebibliography}
\end{document}